\documentclass[11pt]{article}
\pdfoutput=1
\usepackage{jheppub}
\usepackage[dvipsnames]{xcolor}
\usepackage{graphicx}
\usepackage{placeins}
\usepackage{wrapfig}
\usepackage{amsmath}
\usepackage{amsfonts}
\usepackage{amssymb}
\usepackage{multirow}
\usepackage{slashed}
\usepackage{physics}
\usepackage{dsfont}
\usepackage{soul}
\usepackage{ulem}
\usepackage{verbatim}
\usepackage[colorlinks=true, linkcolor=blue, citecolor=blue, urlcolor=blue]{hyperref}
\usepackage{cleveref}

\def\be{\begin{equation}}
\def\ee{\end{equation}}
\newcommand{\genprop}{\Xi_{\alpha\beta;ab}^{{\mathbf{\Gamma}}(i,j)}\left(T_f,T_i;\tau,z\right)}
\newcommand{\ratio}{R^\mu\left(\vec{p}_f,\vec{p}_i,z;T,\tau\right)}

\newcommand{\A}[1]{\mathcal{A}_{#1}\left(\nu,\xi,t,z^2\right)}

\newcommand{\amp}[1]{\mathcal{A}_{#1}}

\newcommand{\zz}{\frac{z}2}
\newcommand{\g}{\gamma}

\title{
   Towards Unpolarized GPDs from Pseudo-Distributions
}

\newcommand*{\WM}{Physics Department, William \& Mary, Williamsburg, VA 23187, USA}
\newcommand*{\JLAB}{Thomas Jefferson National Accelerator Facility, Newport News, VA 23606, USA}
\newcommand*{\ODU}{Department of Physics, Old Dominion University, Norfolk, Virginia, USA}
\newcommand*{\CNRS}{Aix Marseille Univ, Universit\'e de Toulon, CNRS, CPT, Marseille, France.}

\author[a]{Herv\'e Dutrieux,}
\author[b]{ Robert G. Edwards,}
\author[b]{ Colin Egerer,}
\author[b]{ Joseph Karpie,}
\author[a]{ Christopher Monahan,}
\author[a,b]{ Kostas Orginos,}
\author[c,b]{ Anatoly Radyushkin,}
\author[b]{ David Richards,}
\author[b]{ Eloy Romero,}
\author[d]{ Savvas Zafeiropoulos}
\author{\\(on behalf of the HadStruc Collaboration)}
\affiliation[a]{\WM}
\affiliation[b]{\JLAB}
\affiliation[c]{\ODU}
\affiliation[d]{\CNRS}

\emailAdd{hldutrieux@wm.edu}
\emailAdd{edwards@jlab.org}
\emailAdd{egerer@jlab.org}
\emailAdd{jkarpie@jlab.org}
\emailAdd{cmonahan@wm.edu}
\emailAdd{kostas@wm.edu}
\emailAdd{radyush@jlab.org}
\emailAdd{dgr@jlab.org}
\emailAdd{eromero@jlab.org}
\emailAdd{savvas.zafeiropoulos@cpt.univ-mrs.fr}

\abstract
{We present an exploration of the unpolarized isovector proton generalized parton distributions (GPDs) $H^{u-d}(x, \xi, t)$ and $E^{u-d}(x, \xi, t)$ in the pseudo-distribution formalism using distillation. Taking advantage of the large kinematic coverage made possible by this approach, we present results on the moments of GPDs up to the order $x^3$ -- including their skewness dependence -- at a pion mass $m_\pi = 358$ MeV and a lattice spacing $a = 0.094$ fm.}

\begin{document}

\date{\today}
\preprint{JLAB-THY-24-4059}
\maketitle


\section{Introduction\label{sec:intro}}

Generalized parton distributions (GPDs)~\cite{Muller:1994ses,Ji:1996nm, Radyushkin:1996nd,Radyushkin:1997ki} characterize the three-dimensional internal structure of hadrons in terms of the parton's longitudinal momentum fraction $x$, the fraction of longitudinal momentum transfer between the incoming and outgoing hadrons, or skewness $\xi$, and the invariant momentum transfer $t$. The three-dimensional nature of the distributions, in contrast to the one-dimensional parton distribution functions (PDFs), enables insights into crucial features such as the orbital angular momentum carried by the quarks and gluons \cite{Ji:1996ek, Ji:1996nm}. Generalizing the elastic form factors, GPDs can also be used to define radial profiles of energy or pressure distribution of the partonic matter \cite{Polyakov:2002yz, Polyakov:2018zvc}.  

The principal experimental means of gaining insight into GPDs is through exclusive processes, notably deeply virtual Compton scattering (DVCS) \cite{Ji:1996nm, Radyushkin:1996nd} and deeply virtual meson production (DVMP) \cite{Radyushkin:1996ru, Collins:1996fb}. There is a worldwide experimental program dedicated to these processes, including experiments at Jefferson Lab at both 6 and 12~GeV~\cite{Dudek:2012vr} and the upcoming Electron-Ion Collider (EIC)~\cite{Accardi:2012qut, AbdulKhalek:2021gbh}.  However, these experiments only give indirect access to GPDs. In the framework of collinear factorization, the amplitude of these processes depends on GPDs through a convolution with a perturbative kernel, which gives a practical access to some specific features of the GPDs (notably the diagonal region $x \approx \xi$ or some specific integrals of the GPDs). Therefore the insights they provide are primarily two-dimensional, or ``moment-like'' \cite{Qiu:2022pla, Qiu:2023mrm}, the three-dimensional behavior being obscured. A manifestation of this inverse problem is encapsulated in ``shadow GPDs''~\cite{Bertone:2021yyz}, wherein a range of GPDs can give rise to extremely similar DVCS cross-sections. In the limit of small Bjorken-$x$, corresponding to small $x$ and $\xi$, the enhanced effects of perturbative evolution may give a hope of a better control of the uncertainty propagation in this inverse problem \cite{Freund:1999xf, Dutrieux:2023qnz, Moffat:2023svr}. However, in the regime of moderate to large $x$, which precisely corresponds to the kinematical region accessible to lattice QCD, a model independent extraction of GPDs from those experimental processes seems difficult to achieve. It is especially the case in the so-called ERBL region ($|x| < |\xi|$) where the lack of theoretical constraints like positivity \cite{Radyushkin:1998es,Pire:1998nw,Diehl:2000xz,Pobylitsa:2002gw} makes the deconvolution problem particularly poorly behaved \cite{Dutrieux:2021wll}. For example, the very different results obtained when extracting gravitational form factors from DVCS data using different parametrizations \cite{Burkert:2018bqq, Kumericki:2019ddg, Dutrieux:2021nlz} demonstrate a relative lack of sensitivity of the current DVCS dataset to this quantity. In contrast, these form factors have been probed with great precision on the lattice using local composite operators~\cite{Hagler:2003jd,LHPC:2007blg,Bali:2018zgl,Alexandrou:2019ali,Hackett:2023rif}.

GPDs provide therefore an ideal setup where, at least for the time-being, only lattice QCD can provide systematically controlled information. Besides DVCS and DVMP, an increasing number of exclusive processes with enhanced sensitivity to GPDs are being studied, but their experimental exploration is only at its beginning: double DVCS \cite{Guidal:2002kt, Belitsky:2002tf}, where a recent study of measurability looks promising \cite{Deja:2023ahc}, di-photon production \cite{Pedrak:2020mfm, Grocholski:2022rqj}, photon meson pair production \cite{Boussarie:2016qop, Duplancic:2018bum}, or the general category of single diffractive hard exclusive processes \cite{Qiu:2022bpq, Qiu:2022pla}. The maturity of this new phenomenology of GPDs in the perspective of the EIC therefore complements the progress of the extraction of GPDs on the lattice. Preliminary joint lattice-experimental phenomenology of GPDs are already being explored \textit{e.g.} \cite{Guo:2022upw,Guo:2023ahv,Riberdy:2023awf}. 

\vspace{2em}

GPDs are defined as the matrix elements of operators separated along the light cone.  As in the case of the PDFs, their direct calculation is precluded on a Euclidean lattice, except for low-order moments which can be computed from local operators. New developments in the last decade have led to the possibility of computing the $x$-dependence of parton distributions using non-local operators with a space-like separation~\cite{Liu:1993cv,Detmold:2005gg,Braun:2007wv,Ji:2013dva,Ji:2014gla,Radyushkin:2017cyf,Chambers:2017dov,Ma:2017pxb,Bali:2018nde}. The application of some of those approaches to GPDs was performed in~\cite{Ji:2015qla,Liu:2019urm,Radyushkin:2019owq,Ma:2022ggj} and pioneering calculations have been conducted in~\cite{Chen:2019lcm,Lin:2020rxa,Alexandrou:2020zbe,Lin:2021brq,Alexandrou:2021bbo,Bhattacharya:2022aob,Bhattacharya:2023ays,Bhattacharya:2023jsc,Bhattacharya:2023nmv,Lin:2023gxz,Bhattacharya:2024qpp,Hannaford-Gunn:2024aix}.

\subsection*{Main phenomenological results}

\begin{figure}[b]
    \centering
    \includegraphics[width=0.7\linewidth]{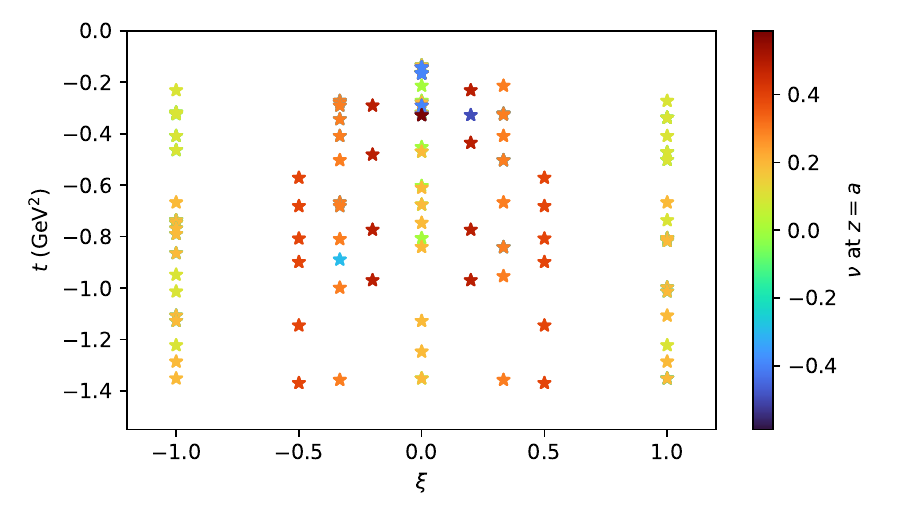}
    
    \caption{Kinematic coverage presented in this study, consisting of 186 pairs $(\vec{p}_f, \vec{p}_i)$ and 116 points in the $(\xi, t)$ plane. The values of the Ioffe time $\nu = z \cdot P$ are quoted for $z = a = 0.094$ fm. By using up to $z = 6a$, $\nu$ reaches up to 3.5 in some kinematics.}
    \label{fig:kincov}
\end{figure}

We present a formalism for the computation of pseudo-GPDs using distillation. This technique allows us to probe the GPDs on a large kinematic range in terms of skewness and momentum transfer as presented in Fig.~\ref{fig:kincov}. For numerical application, we use 186 combinations of initial and final momenta giving 116 kinematic values of $(\xi, t)$. This grants us access to Mellin moments of the GPDs sensitive to the skewness on top of those only sensitive to the forward limit $\xi = 0$. We use a lattice ensemble with pion mass $m_\pi = 358$ MeV, lattice spacing $a = 0.094$ fm, and lattice volume 32$^3\times$64. For the extraction of the GPD moments, we have used hadron momenta up to 1.4 GeV, yielding maximal Ioffe time values of $\nu \leq 0.6\,z/a$ where $z$ is the space-like separation of the non-local operator. This corresponds to $\nu_{\max} = 3.5$ using $z = 6a$. As demonstrated in the context of PDFs in \cite{Egerer:2021ymv}, we can reach momenta at least twice as large when using the distillation framework with momentum smearing, allowing for a study of the $x$-dependence of the parton distributions. However, we reserve this effort for a subsequent paper, as preliminary exploration at large momentum hinted at the necessity of a better control of excited state contamination and other lattice systematic uncertainties.

\begin{figure}[ht!]
    \centering
    \includegraphics[width=0.95\linewidth]{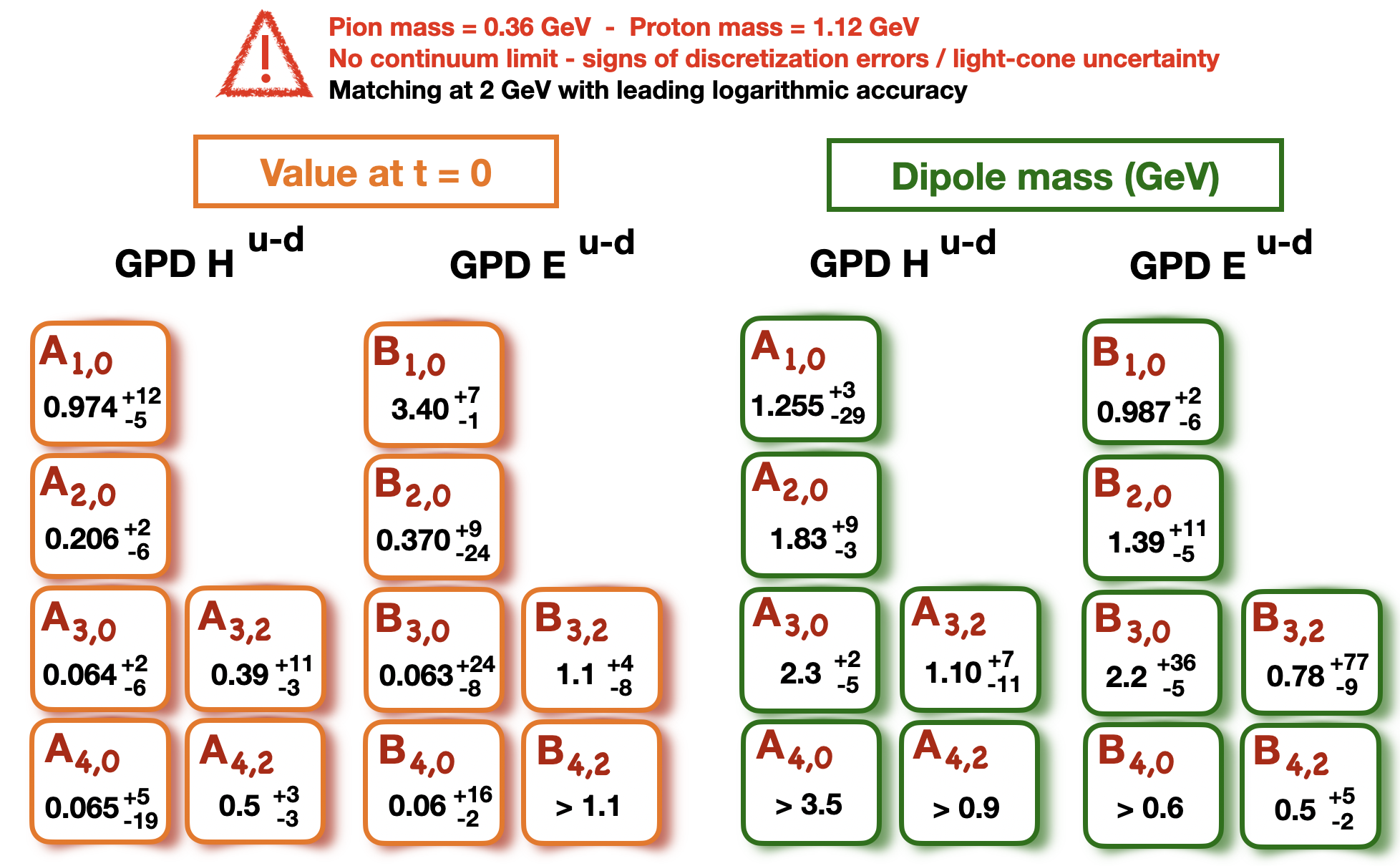}
\hspace{-20pt}\includegraphics[width=0.95\linewidth]{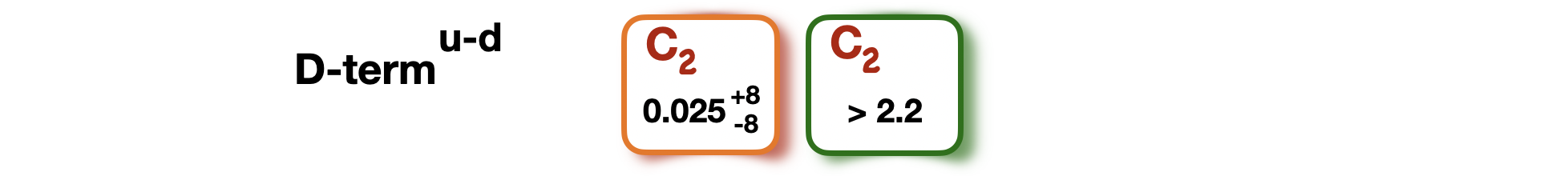}

    \caption{Short summary of the extraction of GPD Mellin moments using a dipole representation. The quoted uncertainty contains an evaluation of statistical and excited state uncertainty. More details starting from section \ref{sec:EFFs}.}
    \label{fig:summary}
\end{figure}

For the reader interested in the phenomenology of GPDs beyond the framework of lattice calculations, we summarize in Fig.~\ref{fig:summary} the main physics results of this study, in the form of moments of the isovector GPDs $H^{u-d}$ and $E^{u-d}$:
\begin{align}
    \int_{-1}^1 \mathrm{d}x\,x^{n-1} \begin{pmatrix}H^{u-d} \\ E^{u-d} \end{pmatrix}(x, \xi, t) = \sum_{k = 0\textrm{ even}}^{n-1} \begin{pmatrix}A_{n,k}(t) \\ B_{n,k} (t)\end{pmatrix} \xi^k \pm \textrm{mod}(n+1,2) \xi^n C_n(t)\,,
\end{align}
where $C_n(t)$ are moments of the isovector $D$-term with $+$ for the $H$ GPD and $-$ for the $E$ GPD. For brevity, the results are quoted at the scale $\mu = 2$ GeV in the form of a dipole fit with the value at $t=0$ and a dipole mass $\Lambda_{n,k}$ following the example:
\begin{align}
A_{n,k}(t) = A_{n,k}(t = 0) \left(1-\frac{t}{\Lambda_{n,k}^2}\right)^{-2}\,.
\end{align}

More sophisticated treatment of the data is presented starting from section \ref{sec:EFFs} of this document, in particular using a more flexible parametrization of the $t$-dependence. The quoted uncertainty is constructed by merging the 68\% confidence intervals of two different extractions to  include an evaluation of excited state contamination. This systematic uncertainty is mostly responsible for the asymmetric errors. A slight asymmetry also arises for statistical reasons for the most poorly determined moments (notably $A_{4,2}$ and $B_{4,2}$) due to the non-linearity of the dipole fit. 

These results are obtained at an unphysical pion mass of 358 MeV without an attempt at extrapolating to the physical pion mass. Likewise, as we will discuss in detail further in the document, we observe clear signs of systematic effects, either attributable to the lack of continuum limit (discretization effects) or the uncertainty in the light-cone limit of our space-like matrix elements (higher-twist effects). We use data with separations up to $z = 0.6$ fm, which presents some challenges with respect to the procedure of perturbative matching as discussed later. With these limitations in mind, we hope that the results, and in particular the novel characterization of the skewness-dependent moments $A_{3,2},B_{3,2}$ and $A_{4,2},B_{4,2}$ can provide insights for theoretical models of GPDs and extractions from the data.

The paper is organized as follows. In section \ref{sec:theory}, we review the formalism of GPDs in the short-distance factorization, notably  deriving the matching formulas for the moments sensitive to the skewness. In section \ref{sec:numerics}, we present our calculation strategy on the lattice, stressing the extensive coverage of the kinematic domain enabled by distillation. In section \ref{sec:isolateMats}, we describe our numerical analysis of the correlation functions. Then we present our results, first for the elastic form factors (section \ref{sec:EFFs}) and second for the generalized form factors (section \ref{sec:GFFs_num}).

\section{Theoretical formalism\label{sec:theory}}

\subsection{Building light-cone GPDs from space-like matrix elements\label{sec:twoA}}

GPDs parameterize off-forward matrix elements of quark and gluon operators with a light-like separation. In the convention of~\cite{Ji:1998pc}, the leading twist-2 vector quark GPDs of the nucleon are defined according to:
\begin{align}
    F^q\left(x,p_f,p_i\right)&=\frac{1}{2}\int\frac{{\rm d}z^-}{2\pi}e^{ixP^+z^-} \label{eq:def-GPDs-HE_int}\\
    & \quad \times\bra{N\left(p_f,\lambda_f\right)}\bar{\psi}^q\left(-\frac{z}{2}\right)\gamma^+\hat{W}\left(-\frac{z}{2},\frac{z}{2};A\right)\psi^q\left(\frac{z}{2}\right)\ket{N\left(p_i,\lambda_i\right)}\mid_{z^+=0,\mathbf{z}_\perp=\mathbf{0}_\perp}\,, \nonumber \\
    &=\frac{1}{2P^+}\overline{u}\left(p_f,\lambda_f\right)\left[\gamma^+ H^q\left(x,\xi,t\right)+\frac{i\sigma^{+\nu}q_\nu}{2m} E^q\left(x,\xi,t\right)\right]u\left(p_i,\lambda_i\right)\,,\label{eq:def-GPDs-HE}
\end{align}
where $\hat{W}$ is the Wilson line in the fundamental representation, $m$ the nucleon mass, $\lambda_i, \lambda_f$ are the initial and final state helicities, and $p_i, p_f$ are the initial and final state momenta that define the variables
\begin{equation}
    P \equiv \frac{1}{2}(p+p')\,,\ \ q \equiv p'-p\,,\ \ t \equiv q^2\,,\ \ \xi \equiv -\frac{q^+}{2P^+}\,.\label{eq:var_def}
\end{equation}
We use the spinor normalization $\bar{u}(p,\lambda)u(p,\lambda')=2m\delta_{\lambda\lambda'}$. 

For a Euclidean lattice QCD calculation, the light-cone definition is unsuitable. Since real-time is inaccessible, we use instead equal-time non-local operators with a space-like separation. Let us consider more general fundamental matrix elements defining GPDs of spin-1/2 nucleon without restriction on $z^\mu$:
\be
M^\mu\left(p_f,p_i,z\right)=\bra{N\left(p_f,\lambda_f\right)}\mathcal{O}^\mu\left(z;A\right)\ket{N\left(p_i,\lambda_i\right)}.\label{eq:mat-to-match}
\ee
This matrix element has multiplicative ultraviolet divergences~\cite{Ishikawa:2017faj} that only depend on the length of the Wilson line, so it is useful to use instead the ratio:
\be
\mathcal{M}^\mu\left(p_f,p_i,z\right) = \frac{M^\mu\left(p_f,p_i,z\right)}{M^0\left(0,0,z\right)}\,,\label{eq:reducedmatelem}
\ee
which has a finite continuum limit. In fact, this ratio is renormalization group invariant (RGI), which means that it is independent of the scheme that is used to define the renormalized quantities in the numerator and denominator. This denominator is the same matrix element used in the analogous ratio of the forward matrix elements in~\cite{Egerer:2021ymv}. In the case of isovector unpolarized quarks, the denominator also directly normalizes the Dirac elastic form factor at 1 when $t = 0$.

Introducing the Ioffe time $\nu = z \cdot P$, the light-cone GPD definition \eqref{eq:def-GPDs-HE_int} can be expressed as:
\begin{align}
    F^q\left(x,p_f,p_i\right)&= \lim_{z \rightarrow (0,z^-,0_\perp)} \frac{1}{2P^+}\int\frac{{\rm d}\nu}{2\pi}e^{ix\nu} \mathcal{M}^+(p_f,p_i,z)\,,\\
    &=\lim_{z^2 \rightarrow 0} \frac{1}{2}\int\frac{{\rm d}\nu}{2\pi\nu}e^{ix\nu} z_\mu \mathcal{M}^\mu(p_f,p_i,z)\,.\label{eq:lorentzinv}
\end{align}
This expression is particularly convenient as it is directly Lorentz invariant. It should be noted that the limit in Eq.~\eqref{eq:lorentzinv} suffers from collinear divergences, but perturbation theory allows us to predict the behavior in $z^2$ of the leading twist contribution when $z$ is small enough compared to hadronic scales. However, since all our calculations are performed at $z^2 < 0$, there exists an uncertainty associated with the light-cone limit $z^2 \rightarrow 0$.

The matrix element admits the  Lorentz decomposition~\cite{Bhattacharya:2022aob}:
\begin{align}  \mathcal{M}^\mu\left(p_f,p_i,z\right)&=\langle\langle\gamma^\mu\rangle\rangle\mathcal{A}_1\left(\nu,\xi,t,z^2\right)+z^\mu\langle\langle\mathds{1}\rangle\rangle\mathcal{A}_2\left(\nu,\xi,t,z^2\right)+i\langle\langle\sigma^{\mu z}\rangle\rangle\mathcal{A}_3\left(\nu,\xi,t,z^2\right)\nonumber \\
    &\qquad\qquad+\frac{i}{2m}\langle\langle\sigma^{\mu q}\rangle\rangle\mathcal{A}_4\left(\nu,\xi,t,z^2\right)+\frac{q^\mu}{2m}\langle\langle\mathds{1}\rangle\rangle\mathcal{A}_5\left(\nu,\xi,t,z^2\right)\nonumber \\
    &\quad+\frac{i}{2m}\langle\langle\sigma^{zq}\rangle\rangle\left[P^\mu\mathcal{A}_6\left(\nu,\xi,t,z^2\right)+q^\mu\mathcal{A}_7\left(\nu,\xi,t,z^2\right)+z^\mu\mathcal{A}_8\left(\nu,\xi,t,z^2\right)\right]\label{eq:decomp}\,.
\end{align}
The Lorentz structures preceding the amplitudes employ the following abbreviations: $\sigma^{\mu z}\equiv\sigma^{\mu\rho}z_\rho$, $\sigma^{\mu q}\equiv\sigma^{\mu\rho}q_\rho$, $\sigma^{zq}\equiv\sigma^{\rho\lambda}z_\rho q_\lambda$, and $\langle\langle\Gamma\rangle\rangle \equiv \overline{u}\left(p_f,\lambda_f\right) \Gamma u\left(p_i,\lambda_i\right)$. $\xi$ may be generalized to space-like separations via:
\begin{equation}
    \xi = -\frac{q\cdot z}{ 2 P\cdot z} = - \frac{q\cdot z}{2 \nu}\,.
\end{equation}
This expression coincides exactly with the definition of $\xi$ in the light-cone limit \eqref{eq:var_def}, but interestingly the kinematic bound satisfied by the light-cone $\xi$
\begin{equation}
    |\xi_{\textrm{light-cone}}|\leq \frac{\sqrt{-t}}{\sqrt{-t+4m^2}} \leq 1\,,
\end{equation}
no longer holds for space-like $z$. In fact, with this definition of $\xi$, one can choose kinematics such that $\nu =0$ and $\xi$ is infinite. This is, however, not an issue as all amplitudes and matching relations appear naturally as functions of $\bar{\nu} = \nu\xi = -q\cdot z/2$, which is always finite. The decomposition of the GPD matrix elements in terms of $\nu$ and $\bar{\nu}$ emerges clearly in the double distribution (DD) framework~\cite{Radyushkin:2023ref}, as shown in the Lorentz invariant relationship Eq.~\eqref{eq:pitd-dd_eq}.

The matrix element contracted with $z_\mu$ that enters the light-cone limit, Eq.~\eqref{eq:lorentzinv}, reads:
\begin{align}  z_\mu\mathcal{M}^\mu\left(p_f,p_i,z\right)&=\langle\langle\slashed{z}\rangle\rangle\mathcal{A}_1\left(\nu,\xi,t,z^2\right)+z^2\langle\langle\mathds{1}\rangle\rangle\mathcal{A}_2\left(\nu,\xi,t,z^2\right)+0\times\mathcal{A}_3\left(\nu,\xi,t,z^2\right)\nonumber \\
    &\qquad\qquad+\frac{i}{2m}\langle\langle\sigma^{z q}\rangle\rangle\mathcal{A}_4\left(\nu,\xi,t,z^2\right)-\frac{\nu\xi}{m}\langle\langle\mathds{1}\rangle\rangle\mathcal{A}_5\left(\nu,\xi,t,z^2\right)\nonumber \\
    &\quad+\frac{i}{2m}\langle\langle\sigma^{zq}\rangle\rangle\left[\nu\mathcal{A}_6\left(\nu,\xi,t,z^2\right)-2\xi\nu\mathcal{A}_7\left(\nu,\xi,t,z^2\right)+z^2\mathcal{A}_8\left(\nu,\xi,t,z^2\right)\right]\,,
\end{align}
where we have used that $\sigma^{\mu\nu}z_\mu z_\nu = 0$. One can immediately identify that $\mathcal{A}_3$ disappears in the contraction and $\mathcal{A}_{2,8}$ are $\mathcal{O}(z^2)$ contaminations since  in QCD $\mathcal{A}_{2,8}$ can only  have logarithmic colinear divergences in the limit of $z^2\rightarrow 0$. If we introduce Ioffe-time GPDs as:
\begin{align}
    \begin{pmatrix} H^q \\ E^q \end{pmatrix} (x, \xi, t) = \int \frac{\mathrm{d}\nu}{2\pi} e^{ix\nu} \begin{pmatrix} H^q \\ E^q \end{pmatrix}(\nu, \xi, t)\,,
\end{align}
then we can identify which amplitudes contribute to the lightcone limit by comparison with Eq.~\eqref{eq:def-GPDs-HE}:
\begin{align}
    H^q(\nu, \xi, t) &= \lim_{z^2 \rightarrow 0} \bigg[\mathcal{A}_1 - \xi A_5\bigg]\,, \label{eq:defH}\\
    E^q(\nu, \xi, t) &= \lim_{z^2 \rightarrow 0} \bigg[\mathcal{A}_4 + \nu \mathcal{A}_6 - 2\xi\nu \mathcal{A}_7 + \xi A_5\bigg]\,,\label{eq:defE}
\end{align}
where scale dependence from regulating the collinear divergences at the $z^2 \to 0$ limit has been suppressed. 
We have used the Gordon identity to distribute the contribution of $\mathcal{A}_5$ between $H$ and $E$. 
This relation is equivalent to the one quoted in~\cite{Bhattacharya:2022aob} up to choices of conventions. The discussion in \cite{Radyushkin:2023ref} using the DD formalism \cite{Radyushkin:1996nd,Radyushkin:1996ru,Radyushkin:1997ki} demonstrates that the $\mathcal{A}_5$ amplitude is sensitive to the $D$-term in the limit $z^2 \rightarrow 0$, while the other amplitudes $\mathcal{A}_{1,4,6,7}$ build up the double distributions in a similar fashion as here. The right-hand sides of Eqs. \eqref{eq:defH} and \eqref{eq:defE} are very advantageous since they can be evaluated immediately at any value of $z^2$ with the assurance that they converge towards the correct light-cone limit for arbitrarily small values of $z^2$. However, there is clearly no unique way to construct a Lorentz invariant object at $z^2 < 0$ which converges analytically towards the correct light-cone limit.

In the case of unpolarized quark PDFs, a single amplitude contributes leading twist information at lowest powers of $z^2$ and $\alpha_s$. For helicity distributions a linear combination of two is required~\cite{HadStruc:2022nay,Balitsky:2021cwr,HadStruc:2022yaw}, not dissimilar to Eqs.~\eqref{eq:defH} and ~\eqref{eq:defE}. In the case of GPDs, different choices of the quasi-GPD were compared in terms of the Lorentz invariant amplitudes~\cite{Bhattacharya:2022aob}. In the light-cone limit $z^2\to0$ at fixed $\nu,\xi$, these definitions converge to the same final GPD. The practical differences between the definitions for realistic scales would be dominated by twist-3 contributions~\cite{Braun:2023alc,Radyushkin:2023ref} in the amplitudes that come from terms proportional to the transverse momentum transfer $\Delta_\mu^T = q^\mu + 2 \xi P^\mu $. When considering the light-cone matrix element in Eq.~\eqref{eq:lorentzinv} with a generic $\gamma^\mu$, when $\mu\neq +$ the dominant deviations from the leading twist matrix elements again occur in the OPE from twist-3 operators that are total derivatives of leading twist operators or from genuine quark-gluon-antiquark correlations~\cite{Anikin:2000em,Penttinen:2000dg, Belitsky:2000vx,Radyushkin:2000jy,Radyushkin:2000ap,Kivel:2000cn}. In~\cite{Braun:2023alc}, the effect of such twist-3 contributions is to partially restore the translational symmetry of the off-light-cone matrix element, which is broken in a leading twist calculation. As shown in~\cite{Radyushkin:2023ref} adding the $\amp{6,7}$ terms to Eq.~\eqref{eq:defE} is explicitly canceling the contribution to two such twist-3 contributions. If one could determine higher twist combinations, of any order and from the same or a different calculation, then they could be added similarly to Eq.~\eqref{eq:defH} and~\eqref{eq:defE} to improve the light-cone limit $z^2\to0$, even if not all twist-3 contribution are separable using Lorentz invariant functions. An evaluation of the plausible size and analytical properties of the genuine higher-twist power corrections is carried out in \cite{Braun:2024snf}, where the corrections are found to be fairly mild when using the ratio normalization of Eq.~\eqref{eq:reducedmatelem} in the range of Ioffe time relevant for this study.

\subsection{Perturbative matching}

Extending the pseudo-PDF's one-loop matching~\cite{Radyushkin:2018cvn,Zhang:2018ggy,Izubuchi:2018srq}, the matching of unpolarized non-singlet Ioffe-time GPDs in the $\overline{\rm MS}$ scheme to $z^2$-dependent pseudo-distributions in coordinate space was first presented at one-loop accuracy in~\cite{Radyushkin:2019owq} for the pion's analog of $\amp{1}$. For clarity, we denote with a bar the $\overline{\rm MS}$ quantity. The matching is given by:
\be
    F\left(\nu,\xi,t,z^2\right)=\overline{F}\left(\nu,\xi,t,\mu^2\right)-\frac{\alpha_sC_F}{2\pi}\left\lbrace\ln\left(-\frac{e^{2\gamma_E+1}}{4}z^2\mu^2\right)B+L\right\rbrace\otimes\overline{F}+\mathcal{O}\left(z^2\Lambda_{\rm QCD}^2\right) + \mathcal{O}\left(z^2 t\right)
  \label{eq:gitd-pgitd-matching},
\ee
where:
\begin{align}
B\otimes\overline{F}&=\int_0^1{\rm d}\alpha\ \left[\left(\frac{2\alpha}{1-\alpha}\right)_+\cos\left(\bar{\alpha}\xi\nu\right)+\frac{\sin\left(\bar{\alpha}\xi\nu\right)}{\xi\nu}-\frac{\delta\left(1-\alpha\right)}{2}\right]\overline{F}\left(\alpha\nu,\xi,t,\mu^2\right)\label{eq:BCrossM} \\
L\otimes\overline{F}&=\int_0^1{\rm d}\alpha\ \left[4\left(\frac{\ln\left(1-\alpha\right)}{1-\alpha}\right)_+\cos\left(\bar{\alpha}\xi\nu\right)-2\frac{\sin\left(\bar{\alpha}\xi\nu\right)}{\xi\nu}+\delta\left(1-\alpha\right)\right]\overline{F}\left(\alpha\nu,\xi,t,\mu^2\right),\label{eq:LCrossM}
\end{align}
with  $\bar{\alpha}=1-\alpha$. The standard {\it plus-prescription} is defined as \begin{equation}\int_0^1{\rm d}\alpha\ G\left(\alpha\right)_+f\left(\alpha x\right)=\int_0^1{\rm d}\alpha\ G\left(\alpha\right)\left[f\left(\alpha x\right)-f\left(x\right)\right]\,.\end{equation}

Here $B$ is simply the ordinary leading-order evolution kernel of non-singlet GPDs, except expressed in Ioffe time instead of $x$. Compared to its expression in $x$-space, the evolution operator is much simpler. For instance, it is analytical in $\xi\nu$, whereas the $x$-space evolution operator has non-analytic structures in $\xi/x$. Furthermore, the evolution/matching of Ioffe-time GPDs shares with Ioffe-time PDFs the characteristic of only involving smaller Ioffe times (the $\alpha$ variable in the integral runs from 0 to 1). Thus, we do not need to know the entire Ioffe-time dependence of the GPD for evolution. In contrast, the $x$-space evolution of GPDs will typically require knowledge of the entire $x$-dependence of the GPD. It is only when $|x| > |\xi|$ that a partial knowledge of the $x$-dependence of GPDs is enough to evolve them\footnote{Depending on which kinematic region in $(x, \xi)$ of the GPDs one exactly knows, it is possible to imagine a somewhat cumbersome strategy where one would use double distributions as an intermediate step before performing the evolution \cite{DallOlio:2024vjv}. It is however potentially numerically challenging, and certainly less straightforward than the evolution of a Ioffe-time GPD.}. As a consequence, it seems that Ioffe-time GPDs are generally speaking a much friendlier object to manipulate than their $x$-dependent counterpart, even though they contain the same information if the entire range in $\nu$ and $x$ is known.

\begin{figure}[ht!]
    \centering
    \includegraphics[width=0.9\linewidth]{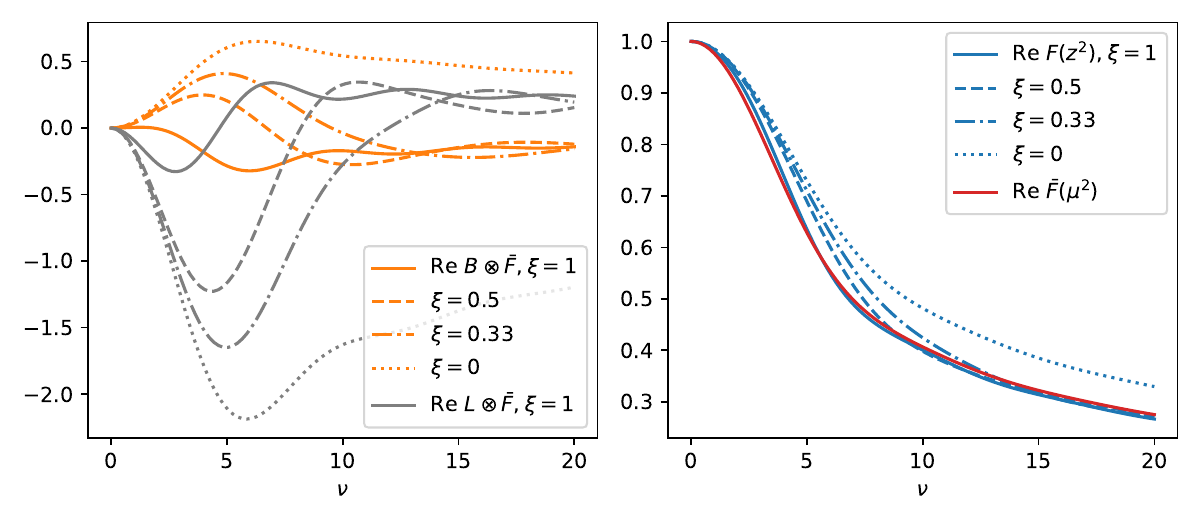}
    \caption{(left) Convolution of $\mathrm{Re}\,\overline{F}\left(\nu,\xi,t,\mu^2\right)$ obtained from Eq.~\eqref{eq:testgpd} with the evolution kernel $B$ and the matching kernel $L$ for various values of $\xi$.  
    (right) The original Ioffe-time GPD (red) and the result of the matching for $\mu = 2$ GeV, $\alpha_s = 0.28$, $z = 0.094$ fm and various values of $\xi$.}
    \label{fig:exampleMatchingKernels-real}
\end{figure}

\begin{figure}[ht!]
    \centering
    \includegraphics[width=0.9\linewidth]{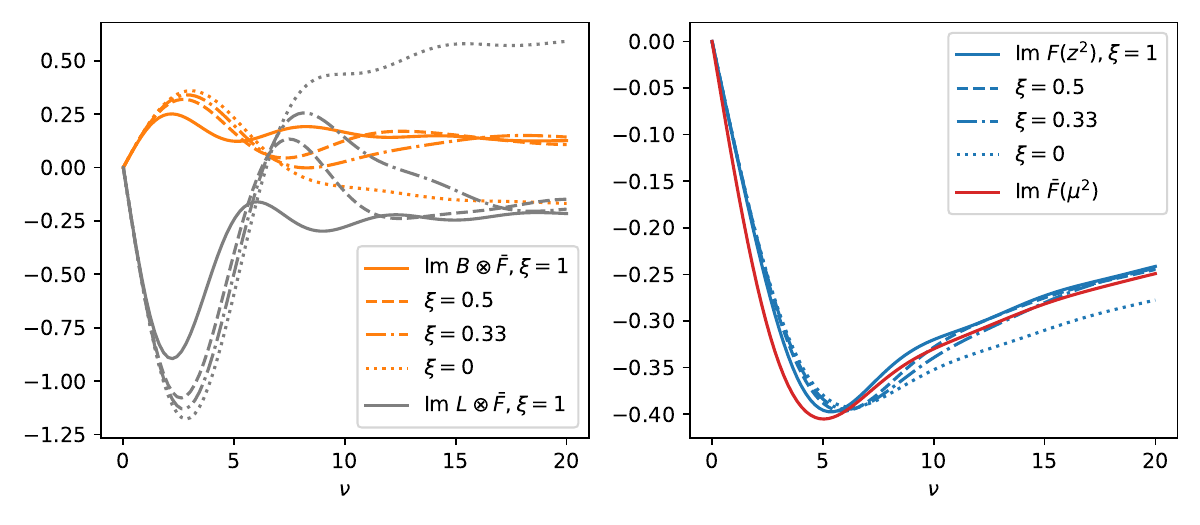}
    \caption{Same as Fig. \ref{fig:exampleMatchingKernels-real}, but for the imaginary part.}
    \label{fig:exampleMatchingKernels-imag}
\end{figure}

The kernels governing the evolution, Eq.~\eqref{eq:BCrossM}, and scheme matching, Eq.~\eqref{eq:LCrossM}, connect the scales $\mu^2$ in $\overline{\rm{MS}}$ and $z^2$ at fixed values of Ioffe time, skewness, and momentum transfer. In the limit $\xi \rightarrow 0$, the kernels reduce to the PDF case. It is interesting to observe the effects of non-zero skewness. To isolate the skewness dependence attached purely to the perturbative matching, we introduce a $\xi$-independent GPD resembling a normalized isovector PDF:
\begin{equation}
\overline{F}\left(x,\xi,t,\mu^2\right)=\frac{15}{16}\,x^{-0.5}\left(1-x\right)^2\,.\label{eq:testgpd}
\end{equation}
The associated Ioffe-time GPD
\begin{equation}
\overline{F}\left(\nu,\xi,t,\mu^2\right)=\int_{-1}^1 \mathrm{d}x\,e^{-ix\nu}\overline{F}\left(x,\xi,t,\mu^2\right)\label{eq:testitgpd}   
\end{equation}
is presented as the red curve in the right panels of Fig.~\ref{fig:exampleMatchingKernels-real} for the real part and Fig.~\ref{fig:exampleMatchingKernels-imag} for the imaginary part. The blue curves on the same plots represent the matched pseudo-GPD $F(\nu, \xi, t, z^2)$ where we have assumed that $\mu = 2$ GeV, $\alpha_s = 0.28$ and $z = a = 0.094$ fm. This yields $\ln(-z^2\mu^2 e^{2\gamma_E+1}/4) \approx 0.67$. It can be observed that the matching effects are quite small, especially for larger values of $\xi$. 

To understand how evolution and matching separately contribute to the effects of non-zero skewness, we show the separate convolutions $B \otimes \overline{F}$ and $L \otimes \overline{F}$ on the left panels of Figs.~\ref{fig:exampleMatchingKernels-real} and~\ref{fig:exampleMatchingKernels-imag}. A striking feature is that the curves for various skewnesses seem to become similar as $\nu$ gets larger, but the smaller the skewness, the longer they take to fall towards this ``universal'' limit. Although it is difficult to assess precisely, a general rule can be observed from the plots that the universal limit is reached when $\nu \approx 5/\xi$, and it is unclear whether the curve for $\xi = 0$ participates in this universal limit at all. We have checked that this remains consistently true for much larger values of $\nu$.

In fact, this is not surprising. The $x$-space evolution kernel has a cusp when $x = \xi$, which leaves us to expect some kind of trouble in Ioffe time when $\nu = \mathcal{O}(1/\xi)$. As we have noticed before, the relevant variable that appears consistently in our calculations is not $\xi$, but rather $\nu\xi$. A similar study is conducted in $x$-space in \cite{Bertone:2022frx} for the evolution kernel. It demonstrates the formation of non-analytic behavior at $x = \xi$ through the evolution operator and shows how evolution effects are considerably increased in this region (see Fig.~3 in that paper, for instance). We observe here the same effect in Ioffe-time space, although in the range of $\nu$ available for realistic calculations, the $\xi$ dependence of the evolution kernels is quite benign.

Additionally, one can notice in Fig.~\ref{fig:exampleMatchingKernels-imag} that the small $\nu$ behavior of the imaginary part of the convolutions is independent of the skewness. As we will see in the next section, this stems from the fact that the first derivative with respect to $\nu$ of the imaginary part of the non-singlet Ioffe-time GPD at $\nu =0$ is independent of $\xi$ due to the polynomiality property of GPDs. This also explains why the small $\nu$ behavior of the real part of the convolutions in Fig.~\ref{fig:exampleMatchingKernels-real} is of order $\mathcal{O}(\nu^2)$, but this time dependent on $\xi$.

Finally, a feature of the matching already noticed before in the case of PDFs or Good Lattice Cross Sections (\textit{e.g.}~\cite{Joo:2019jct, Sufian:2020vzb, Dutrieux:2023zpy}) is the fact that the evolution and matching convolutions have generally opposite signs in the regime of $\nu$ where data exists. The result is a reduction or even near cancellation of the dominant perturbative effect. Of course, we did not include any initial skewness dependence in the model, so such results are qualitative at best. It will be interesting to study if these canceling effects continue to hold with higher orders in perturbation theory and more sophisticated resummation techniques.

In~\cite{Dutrieux:2023zpy}, the perturbatively calculated kernel for the resummed $z^2$ PDF step-scaling from one scale to another was studied and compared to DGLAP evolution. The kernel for the forward pseudo-PDF was found to be smaller than the $\overline{\rm MS}$ DGLAP evolution kernel. Further comparisons should be made in future studies to other scheme-dependent methods, such as RI-Mom renormalization, where the scale-independent structure is different.

\subsection{Moments of GPDs} \label{sec:GFFs}

By differentiating the defining relationship between the Ioffe-time distribution and the $x$-space distribution, Eq.~\eqref{eq:testitgpd}, one finds that:
\begin{equation}
\frac{d^n}{d\nu^n}F(\nu, \xi, t, z^2)\bigg|_{\nu = 0} = (-i)^n \int_{-1}^1 dx\,x^n F(x, \xi, t, z^2)\,.
\end{equation}
We introduce the short-hand notation:
\begin{equation}
    F_n(\xi, t, z^2) \equiv \int_{-1}^1 dx\,x^{n-1}F(x, \xi, t, z^2)\,,
\end{equation}
where $n{-}1$ is a widely encountered convention. Then:
\begin{equation}
F(\nu, \xi, t, z^2) = \sum_{n=0}^\infty \frac{(-i\nu)^n}{n!} F_{n+1}(\xi, t, z^2)\,.
\end{equation}
Introducing the matching relation:
\begin{equation}
    F(\nu, \xi, t, z^2) = \int_0^1 d\alpha\,K(\alpha, \xi\nu, \alpha_s, z^2\mu^2) \overline{F}(\alpha\nu, \xi, t, \mu^2) \,,
\end{equation} 
and differentiating it $n$ times yields:
\begin{align}
    (-i)^n F_{n+1}(\xi, t, z^2) &= \sum_{k = 0}^n \begin{pmatrix}n \\ k\end{pmatrix}\int_0^1 d\alpha\,\frac{d^{n-k}}{d\nu^{n-k}}K(\alpha, \xi\nu, \alpha_s, z^2\mu^2) \bigg|_{\nu = 0}\frac{d^k}{d\nu^k}\overline{F}(\alpha\nu, \xi, t, \mu^2)\bigg|_{\nu = 0} \,,\\
    &= \sum_{k = 0}^n  \overline{F}_{k+1}(\xi, t, \mu^2)  \begin{pmatrix}n \\ k\end{pmatrix}\int_0^1 d\alpha\,(-i\alpha)^k\frac{d^{n-k}}{d\nu^{n-k}}K(\alpha, \xi\nu, \alpha_s, z^2\mu^2)\bigg|_{\nu = 0} \\
    &\equiv (-i)^n \sum_{k = 0}^n  \overline{F}_{k+1}(\xi, t, \mu^2) \xi^{n-k} c_{n+1,n-k}(\alpha_s, z^2\mu^2) \,,\label{eq:wdsg}
\end{align}
where
\begin{equation}
    c_{n+1,k}(\alpha_s, z^2\mu^2) \equiv i^k \begin{pmatrix}n \\ k\end{pmatrix}\int_0^1 d\alpha\,\alpha^{n-k}\frac{d^{k}}{d(\xi\nu)^{k}}K(\alpha, \xi\nu, \alpha_s, z^2\mu^2)\bigg|_{\nu = 0} \,.
\end{equation}
In this paper we are concerned with the spin-1/2 unpolarized GPDs, so $K(\alpha, \xi\nu)$ is even in $\xi\nu$, and thus only the $c_{n+1,k}$ where $k$ is even are non-zero. In particular, $c_{n+1,k}$ is always real. 

We now list the first few values of $c_{n+1,k}$ at order $\mathcal{O}(\alpha_s)$, using the convention that
\begin{equation}
    L = \ln\left(-\frac{e^{2\gamma_E+1}}{4}z^2\mu^2\right)\,:
\end{equation}

\begin{center}
\bgroup
\def\arraystretch{2}
\begin{tabular}{@{\hspace{1em}}c@{\hspace{1em}}@{\hspace{1em}}c@{\hspace{1em}}}
\hline
\hline
skewness independent & quadratic skewness\\
\hline
   $c_{1,0} = 1$ & \\ 
   $c_{2,0} = 1- \frac{\alpha_sC_F}{2\pi}\left(-\frac{4}{3} L +\frac{14}{3}\right)$ &  \\ 
   $c_{3,0} = 1- \frac{\alpha_sC_F}{2\pi}\left(-\frac{25}{12}L +\frac{47}{6}\right)$ & $c_{3,2} = -\frac{\alpha_sC_F}{2\pi}\left(\frac{5}{12}L - \frac{7}{6}\right)$ \\
   $c_{4,0} = 1- \frac{\alpha_sC_F}{2\pi}\left(-\frac{157}{60}L +\frac{931}{90}\right)$ & $c_{4,2} = -\frac{\alpha_sC_F}{2\pi}\left(\frac{11}{20}L - \frac{53}{30}\right)$ \\
   \hline
   \hline
\end{tabular}
\egroup
\end{center}

\vspace{2em}
The coefficient of $-\alpha_s C_FL/2\pi $ in $c_{n+1,0}$ is simply the leading-order DGLAP anomalous dimension. For $\xi \neq 0$, the GPD Mellin moments mix among themselves through the coefficients $c_{n+1, k}$ with $k > 0$. It is convenient to write the matching in matrix form:
\begin{equation}
\begin{pmatrix}
F_1(\xi, t, z^2)\\F_2(\xi, t, z^2)\\F_3(\xi, t, z^2)\\F_4(\xi, t, z^2)\\\cdots
\end{pmatrix} = \begin{pmatrix}
   c_{1,0} & 0 & 0 & 0 &\\
   0 & c_{2,0} & 0 & 0 &\\
   \xi^2c_{3,2} & 0 & c_{3,0} & 0 & \\
   0 & \xi^2 c_{4,2} & 0 & c_{4,0} & \\ &  &  &  & \ddots
\end{pmatrix}\times \begin{pmatrix}
\overline{F}_1(\xi, t, \mu^2)\\\overline{F}_2(\xi, t, \mu^2)\\\overline{F}_3(\xi, t, \mu^2)\\
\overline{F}_4(\xi, t, \mu^2)\\\cdots
\end{pmatrix}\,.
\end{equation}
Then, by isolating the logarithmic part of the kernel, one can introduce an expression resumed at leading-logarithmic accuracy which generalizes the well-known PDF case (for the LL resummation applied to the pseudo-PDF matching, see \cite{Gao:2021hxl}) by using the appropriate set of moments diagonalizing the GPD evolution \cite{Ohrndorf:1981qv, CHASE1980109}:
\begin{equation}
    \begin{pmatrix}
\overline{F}_1(\xi, t, \mu^2)\\\overline{F}_2(\xi, t, \mu^2)\\\overline{F}_3(\xi, t, \mu^2)\\\overline{F}_4(\xi, t, \mu^2) \\ \cdots
\end{pmatrix} = G^{-1}(\xi) A(\mu^2,z^2) G(\xi) \times \bigg(I - B(\xi, z^2)\bigg) \times \begin{pmatrix}
F_1(\xi, t, z^2)\\F_2(\xi, t, z^2)\\F_3(\xi, t, z^2)\\F_4(\xi, t, z^2)\\\cdots
\end{pmatrix}\,, \label{eq:matching_numerical}
\end{equation}
where $A$ is the matrix of the leading-logarithmic resummed DGLAP evolution
\begin{equation}
A(\mu^2, z^2) = \begin{pmatrix}
   1 & 0 & 0 & 0 &\\
   0 & r^{-4/3} & 0 & 0 &\\
   0 & 0 & r^{-25/12} & 0 &\\
   0 & 0 & 0 & r^{-157/60} &\\
    &  &  &  & \ddots
\end{pmatrix}\textrm{ with } r = \left(\frac{\alpha_s(-1/z^2)}{\alpha_s(\mu^2)}\right)^{C_F/(2\pi\beta_0)}\,.
\end{equation}
Here $\beta_0 = (33-2n_f)/(12\pi) \approx 0.716$, $G$ is the matrix diagonalizing the evolution of moments of GPDs (a specific type of Gegenbauer moments for leading-order evolution),
\begin{equation}
    G(\xi) = \begin{pmatrix}
1 & 0 & 0 & 0 &\\
0 & 1 & 0 & 0 &\\
-\xi^2/5 & 0 & 1 & 0 &\\
0 & -3\xi^2/7 & 0 & 1 &\\
 &  &  &  & \ddots
    \end{pmatrix}\,,
\end{equation}
and $I$ is the identity matrix of the relevant dimension. The non-logarithmic part of the matching kernel is
\begin{equation} 
B(\xi, z^2) = \begin{pmatrix}
   0 & 0 & 0 & 0 &\\
   0 & -a_s \left(-\frac{4}{3}l+\frac{14}{3}\right) & 0 & 0 &\\
   -\xi^2a_s\left(\frac{5}{12}l-\frac{7}{6}\right) & 0 & -a_s \left(-\frac{25}{12}l+\frac{47}{6}\right) & 0 &\\
   0 & -\xi^2a_s\left(\frac{11}{20}l-\frac{53}{30}\right) & 0 & -a_s \left(-\frac{157}{60}l+\frac{931}{30}\right) &\\
    &  &  &  & \ddots\\
\end{pmatrix}\,,
\end{equation}
where $a_s = \alpha_s(-1/z^2) C_F / (2\pi)$ and $l = \ln(e^{2\gamma_E+1}/4) \approx 0.768$.

We have taken the ``natural'' scale of non-local lattice calculations to be $-1/z^2$, but one can easily choose to vary this scale. It is important to notice that the expression of Eq.~\eqref{eq:matching_numerical} is far from unique, and several sensible perturbative choices would result in different formulations, differing by contributions of order $\mathcal{O}(\alpha_s^k L^m)$ where $k \geq 2$ and $m \leq k-1$. In the simpler context of PDFs, more complicated choices of perturbative approach and some of their properties were explored in \cite{Dutrieux:2023zpy}, where the existence of this ambiguity motivates the desire to compute at least the logarithmic part of the kernel directly on the lattice.

Now we introduce the polynomiality property of GPDs. It is useful here to separate in the analysis the $D$-term from the rest of the GPD. As we have already mentioned, the $D$-term is contained in a separate Lorentz amplitude $\mathcal{A}_5$ \cite{Radyushkin:2023ref}, while the rest of the $H$ and $E$ GPDs is constructed out of a combination of the $\mathcal{A}_{1,4,6,7}$ amplitudes. Furthermore, the matching does not mix the $D$-term with the rest of the GPD. To see it, we notice that the polynomiality property writes:
\begin{equation}
    \overline{F}_{n+1}(\xi, t, \mu^2) = \sum_{\substack{k=0\\ k\textrm{ even}}}^n \xi^k A_{n+1,k}(t, \mu^2) + \textrm{mod}(n,2) \xi^{n+1} C_{n+1}(t, \mu^2)\,. \label{eq:polynomiality}
\end{equation}
where
\begin{equation}
    C_{n+1}(t, \mu^2) = \int_{-1}^1 d\alpha\,\alpha^n D(\alpha, t, \mu^2)\,
\end{equation}
are the moments of the $D$-term, while the $A_{n,k}$ do not depend on the $D$-term. Then, the matching formula of Eq. \eqref{eq:wdsg} yields:
\begin{align}
&\sum_{k = 0}^n \bigg[\sum_{\substack{i=0\\ i\textrm{ even}}}^k \xi^i A_{k+1,i}(t, \mu^2) + \textrm{mod}(k,2) \xi^{k+1} C_{k+1}(t, \mu^2) \bigg] \xi^{n-k} c_{n+1, n-k} \nonumber\\
&\hspace{50pt}=\sum_{k = 0}^n\bigg[\sum_{\substack{i=0\\ i\textrm{ even}}}^k \xi^{n+i-k} A_{k+1,i}(t, \mu^2)c_{n+1, n-k}\bigg] + \xi^{n+1}\textrm{mod}(n,2)\sum_{k=0}^n  C_{k+1}(t, \mu^2) c_{n+1, n-k}\,. \label{eq:matching_Dterm}
\end{align}
The first term only contains powers of $\xi$ up to $\xi^n$, while the second term has the same form as a $D$-term. This means that the $z^2$ dependent pseudo-GPD obeys a similar polynomiality property as the $\overline{MS}$ GPD, and that the $D$-term of the pseudo-GPD can be matched to the $D$-term of the $\overline{MS}$ GPD independently from the rest of the GPD. It is convenient to write the polynomiality property of the rest of the GPD, noted here as $\overline{F}_{n+1,\textrm{ no D-term}}$ as a simple matrix multiplication:
\begin{equation}
\begin{pmatrix}
\overline{F}_{1,\textrm{ no D-term}}(\xi, t, \mu^2)\\\overline{F}_{2,\textrm{ no D-term}}(\xi, t, \mu^2)\\\overline{F}_{3,\textrm{ no D-term}}(\xi, t, \mu^2)\\\overline{F}_{4,\textrm{ no D-term}}(\xi, t, \mu^2) \\ \cdots
\end{pmatrix} = \begin{pmatrix}
1 & 0 & 0 & 0 & 0 & 0 & \\
0 & 1 & 0 & 0 & 0 & 0 & \\
0 & 0 & 1 & \xi^2 & 0 & 0 & \\
0 & 0 & 0 & 0 & 1 & \xi^2 & \\
 &  &  &  &  &  & \ddots
\end{pmatrix}\times \begin{pmatrix}
    A_{1,0}(t,\mu^2) \\ A_{2,0}(t,\mu^2) \\ A_{3,0}(t,\mu^2) \\ A_{3,2}(t,\mu^2) \\ A_{4,0}(t, \mu^2) \\ A_{4,2}(t, \mu^2) \\ \cdots
\end{pmatrix}\,,
\end{equation}
and the independent matching of this piece is obtained by inserting this expression into Eq.~\eqref{eq:matching_numerical}, and evaluating the result for a sufficient range of different values of $\xi$ to obtain invertibility (one value of $\xi$ for $F_{1,2}$ and two values for $F_{3,4}$ are enough). 

As for the $D$-term, from:
\begin{align}
\overline{H}(x, \xi, t, \mu^2) = \overline{H}_{\textrm{ no D-term}}(x, \xi, t, \mu^2) + \textrm{sgn}(\xi) \overline{D}\left(\frac{x}{\xi}, t, \mu^2\right) \Theta(|\xi|-|x|)\,,
\end{align}
one gets:
\begin{align}
\overline{H}(\nu, \xi, t, \mu^2) = \overline{H}_{\textrm{ no D-term}}(\nu, \xi, t, \mu^2) + \xi \int_{-1}^1 d\alpha\,e^{-i \alpha \xi \nu} \overline{D}(\alpha, t, \mu^2)
\end{align}
and comparing with Eq. \eqref{eq:defH} gives that:
\begin{align}
\lim_{z^2 \rightarrow 0} \mathcal{A}_5(\nu, \xi, t, z^2) &= -\int_{-1}^1 d\alpha\,e^{-i \alpha \xi \nu} D(\alpha, t, z^2)\,,\\
&= i \xi \nu \int_{-1}^1 d\alpha\,\alpha D(\alpha, t, z^2) - i (\xi \nu)^3 \int_{-1}^1 d\alpha\,\alpha^3 D(\alpha, t, z^2) + ... \,,
\end{align}
where we have used the oddness in $\alpha$ of the $D$-term which implies that $\mathcal{A}_5$ should be purely imaginary in the limit $z^2 \rightarrow 0$. Deviations from this might arise both because of lattice systematics and higher twist contaminations due to the non-zero value of $z^2$. Finally, the introduction of the matching yields:
\begin{align}
&\lim_{z^2 \rightarrow 0} \mathcal{A}_5(\nu, \xi, t, z^2) = i \xi \nu C_2(t, \mu^2) c_{2,0}(\alpha_s, z^2\mu^2) \nonumber \\
&\hspace{70pt}- i (\xi \nu)^3 \bigg[ C_4(t, \mu^2) c_{4,0}(\alpha_s, z^2\mu^2) + C_2(t, \mu^2) c_{4, 2}(\alpha_s, z^2\mu^2) \bigg] + ... \label{eq:Dtermdecomp}
\end{align}

\subsection{Discrete and Lorentz symmetries of Ioffe-time GPDs}\label{sec:spacetime}

Let us first understand the symmetries of the amplitudes $\mathcal{A}_k$ entering the Lorentz decomposition of Eq.~\eqref{eq:decomp} with respect to $\nu\to-\nu$ and $\xi\to-\xi$. In this section, for brevity we choose a gauge where the link connecting the quark and anti-quark fields is unity. Since it is a gauge invariant quantity the discrete symmetries will not be different in other gauges. The parity (P), charge conjugation (C), and time reversal (T) properties of the operator $O^\mu(z)=\bar{\psi}(-\zz)\g^\mu\psi(\zz)$ in Eq.~\eqref{eq:def-GPDs-HE_int} and its (anti)-hermitian combination $O^\mu_\pm(z) =\frac12 \big(O^\mu(z) \pm O^\mu(-z)\big)$ are given by:
\begin{align}
    PO^\mu(z)P =& (-1)^\mu \bar{\psi}\left (\zz \right )\g^\mu\psi \left (-\zz \right) =& (-1)^\mu O^\mu(-z) \quad&\to\quad PO_\pm^\mu(z)P =& \pm(-1)^\mu O_\pm^\mu(z)\nonumber\\
    CO^\mu(z)C =&-\bar{\psi}\left (\zz\right )\g^\mu\psi(-\zz)=& -O^\mu(-z)\quad&\to\quad  CO_\pm^\mu(z)C =&\mp O_\pm^\mu(z)\nonumber\\
    TO^\mu(z)T =&(-1)^\mu \bar{\psi}\left (-\zz\right )\g^\mu\psi(\zz)=& (-1)^\mu O^\mu(z) \quad&\to\quad TO^\mu_\pm(z)T =& (-1)^\mu O^\mu_\pm(z)\,,
\end{align}
where $(-1)^\mu = (1, -1, -1, -1)$ and the operators have the following properties with respect to hermitian conjugation:
\begin{equation}
O^\mu(z)^\dagger = \bar{\psi}\left (\zz \right) \gamma^\mu \psi \left (-\zz\right) =  O^\mu(-z)   \quad\to\quad O^\mu_\pm(z)^\dagger = \pm O^\mu_\pm(z) \,.
\end{equation}
By using the combined $PT$ symmetries, the matrix element in Eq.~\eqref{eq:mat-to-match}, written now with explicit helicity arguments, obeys
\begin{equation}
    M^\mu(p_f,p_i, z, \lambda_f, \lambda_i) = M^\mu(p_i,p_f, z, -\lambda_i, -\lambda_f)\,, \label{eq:pt_matelem}
\end{equation}
which swaps the sign of $\xi$. The relation between amplitudes with $\pm \xi$ can be determined term by term from this relationship once the changes in the kinematic factors in the Lorentz decomposition of Eq.~\eqref{eq:decomp} are known. With the spinors employed in this study, the bilinears $\bar{u}(p_f,\lambda_f) \Gamma u(p_i,\lambda_i)$ equal $\lambda_f\lambda_i [\bar{u}(p_f,-\lambda_f) \Gamma u(p_i,-\lambda_i)]^\ast$ for $\Gamma=1, \gamma^\mu, \sigma^{\mu\nu}$. Note that all structures with the tensor bilinear are accompanied by an additional factor of $i$ that generates an additional sign when complex conjugated. Under these transformations, we can identify the following relation obeyed by the kinematic factors labeled using $k \in \{1,..,8\}$:  
\begin{equation}
K_k^\mu(p_f,p_i, z, \lambda_f, \lambda_i) = \lambda_f\lambda_i Z^{PT}_k K_k^\mu(p_i,p_f, z, -\lambda_i, -\lambda_f)\,,
\end{equation} 
where $Z^{PT}_k $ is the parity of the amplitudes $\amp{k}$ under $\xi\to-\xi$, which takes the values $Z^{PT}_{3,5,7}=-1$ and $Z^{PT}_{k\neq3,5,7}=+1$.

Next, consider the impact of hermiticity:
\begin{equation}
    M^\mu(p_f,p_i,z,\lambda_f, \lambda_i)^\ast = M^\mu(p_i,p_f,-z,\lambda_i, \lambda_f) \,,
\end{equation}
which swaps the sign of $\xi$ and $\nu$ simultaneously. The parity of the kinematic factors under hermiticity is given by $Z^{H}_{2,5,6,8}=-1$ and $Z^{H}_{k\neq2,5,6,8}=+1$. Therefore, the amplitudes are related to their complex conjugates as:
\begin{equation}
    \amp{k}(\nu,\xi) = Z^H_k \amp{k}(-\nu,-\xi)^\ast\,.
\end{equation}
Combining this result with the previous rule, we find that the real part of the amplitudes changes with the sign of $\nu$ following $Z^\nu_k=Z^H_k Z^{PT}_k$ which is $Z^\nu_k=-1$ for $k=2,3,6,7,8$ and $Z^\nu_k=+1$ otherwise. The opposite is true for the imaginary parts.

To understand the implications of this behavior, consider the (anti-)hermitian operators $O_\pm^\mu(z)$, whose matrix elements are purely real or imaginary in the forward case, $(p_f, \lambda_f) = (p_i, \lambda_i)$. In the off-forward case the matrix elements $M_\pm^\mu(p_f,p_i, z)=\frac12 \big(M^\mu(p_f,p_i, z) \pm M^\mu(p_f,p_i, -z)\big)$ are given by
\begin{eqnarray}
    M_\pm^\mu(p_f,p_i, z) &= \frac 12\sum_k \left[K_k^\mu(p_f,p_i, z)\mathcal{A}_k(\nu,\xi,t,z^2) \pm K_k^\mu(p_f,p_i, -z) \mathcal{A}_k(-\nu,\xi,t,z^2)\right] \\
    &=\frac 12 \sum_k K_k^\mu(p_f,p_i, z)\left[\mathcal{A}_k(\nu,\xi,t,z^2) \pm Z^z_k \mathcal{A}_k(\nu,\xi,t,z^2)^\ast \right] \,,
\end{eqnarray}
where $Z^z_k$ is the parity of the kinematic factors in the sign of $z$. As can be seen, the hermitian and anti-hermitian operators expose the real and imaginary components of the amplitudes, but the relation depends on the amplitude. The phases in the matrix elements are entirely controlled by the kinematic factors and choices in defining spinors within the calculations. 

Finally, let us observe how these discrete symmetries interact with the Lorentz symmetry, which materializes through the polynomiality property, Eq.~\eqref{eq:polynomiality}. The feature of polynomiality is an intricate relationship in the $\xi$ dependence of the moments of GPDs. It arises naturally in the double distribution representation, which describes GPDs in $x$-space as (we use only the $D$-term independent part here):
\begin{equation}
    H(x,\xi) = \int_{-1}^1 \mathrm{d}\beta \int_{-1+|\beta|}^{1-|\beta|} \mathrm{d}\alpha\, \delta(x-\beta - \alpha \xi) h(\beta,\alpha) \,.
\end{equation}
We have omitted the $t$-dependence and the scale of the double distribution and GPD.
It is easy to observe that the parity in $\alpha$ and $\beta$ of the double distribution controls respectively the $\xi$ and $x$ parity of the GPD.
The relation between the $(\nu,\xi)$ space of Ioffe-time distributions and $(\beta,\alpha)$ of double distributions is given by integrals of the form:
\begin{equation}
    \mathcal{A}(\nu,\bar{\nu}=\xi\nu) = \int_{-1}^1 \mathrm{d}\beta\, e^{i\nu\beta} \int_{-1+|\beta|}^{1-|\beta|} \mathrm{d}\alpha\, e^{i\bar{\nu}\alpha} h(\beta,\alpha)\,.\label{eq:pitd-dd_eq}
\end{equation}
Therefore, definite parity in $\alpha$ of the double distribution is equivalent to the same definite parity in $\xi$ of the amplitude $\mathcal{A}$. Provided the double distribution $h$ is real, evenness in $\beta$ is equivalent to $\mathcal{A}$ even in $\nu$ and real, whereas oddness in $\beta$ is equivalent to $\mathcal{A}$ odd in $\nu$ and imaginary. 

\section{Numerical Implementation}
\label{sec:numerics}
Previous calculations from the HadStruc collaboration isolated the isovector twist-2 quark PDFs~\cite{Orginos:2017kos,Karpie:2018zaz,Joo:2019jct, Joo:2019bzr,Joo:2020spy,Egerer:2021ymv,HadStruc:2022nay,HadStruc:2021qdf} and unpolarized and helicity gluon PDFs~\cite{HadStruc:2021wmh,HadStruc:2022yaw} of the nucleon using a combination of distillation~\cite{HadronSpectrum:2009krc} and the pseudo-distribution formalism. We use the same gauge ensemble, featuring an $m_\pi=358\ {\rm MeV}$ pion mass within a $32^3\times64$ lattice volume with an $a=0.094\ {\rm fm}$ lattice spacing--we denote this ensemble as {\tt a094m358}. The {\tt a094m358} ensemble is an isotropic $2+1$ flavor Wilson clover fermion ensemble generated by the JLab/W\&M/LANL/MIT collaboration~\cite{jlab-wm-lanl}, with the strange quark fixed to its physical value and Wilson clover fermion sea quarks. A total of $348$ configurations with four distillation ``sources" evenly separated in the time extent per configuration (detailed below) are utilized to determine all matrix elements.  Aspects of the {\tt a094m358} ensemble are summarized in Table~\ref{tab:ensem}--further details can be found in Refs.~\cite{Yoon:2016dij,Yoon:2016jzj}.
\begin{table}[t]
    \setlength{\tabcolsep}{8pt}
    \renewcommand{\arraystretch}{1.5}
    \centering
    \begin{tabular}{ccccccccc}
    \hline\hline
    ID & $a$ (fm) & $m_\pi$ (MeV) & $\beta$ & $m_\pi L$ & $L^3\times N_T$ & $N_{\rm cfg}$ & $N_{\rm srcs}$ & ${\rm rk}\left(\mathcal{D}\right)$ \\
    \hline
    {\tt a094m358} & $0.094(1)$ & $358(3)$ & $6.3$ & $5.4$ & $32^3\times64$ & $348$ & $4$ & $64$ \\
    \hline\hline
    \end{tabular}
    \caption{Lattice spacing, pion mass, inverse coupling, control parameter $m_\pi L$ for finite-volume effects, lattice volume. The total number of configurations $N_{\rm cfg}$, temporal sources $N_{\rm srcs}$ per configuration, and rank of the distillation space ${\rm rk}\left(\mathcal{D}\right)$ are also given.}
    \label{tab:ensem}
\end{table}

To isolate the bare isovector matrix elements of the space-like quark bilinear in Eq.~\eqref{eq:def-GPDs-HE_int} we require two-point functions and
connected three-point functions with the insertion of the operator $\overline{\psi}\left(-\frac z2\right)\gamma^\mu W^{(f)}\left(-\frac z2,\frac z2;A\right)\frac{\tau^3}{2}\psi\left(\frac z2\right)$, where the flavor isovector projector $\tau^3/2$ is included. This projector eliminates statistically noisy disconnected diagrams and gives the difference of the $u$ and $d$ quark distributions. 
Placing the operator symmetric about the origin, with quark and anti-quark fields at $(z/2)$ and $(-z/2)$, ensures the operator has transparent $C,P,T$ and hermiticity properties, reviewed in Section~\ref{sec:spacetime}. 

Inserting complete sets of energy eigenstates, the spectral content of the two- and three-point correlation functions reads
\begin{align}
    C_{2{\rm pt}}\left(\vec{p},T\right)&=\langle\mathcal{N}\left(-\vec{p},T\right)\overline{\mathcal{N}}\left(\vec{p},0\right)\rangle=\sum_n\frac{\left|\mathcal{Z}_n\left(\vec{p}\right)\right|^2}{2E_n\left(\vec{p}\right)}e^{-E_n\left(\vec{p}\right)T}\label{eq:2pt} \\
    C_{3{\rm pt}}^{\left[\gamma^\mu\right]}\left(\vec{p}_f,\vec{p}_i,T;z,\tau\right)&=\langle\mathcal{N}\left(-\vec{p}_f,T\right)\overline{\psi}\left(-\frac z2,\tau\right)\gamma^\mu W^{(f)}\left(-\frac z2,\frac z2;A\right)\psi\left(\frac z2,\tau\right)\overline{\mathcal{N}}\left(\vec{p}_i,0\right)\rangle \nonumber\\
    &=\sum_{n',n}\frac{\mathcal{Z}_{n'}\left(\vec{p}_f\right)\mathcal{Z}_n^\dagger\left(\vec{p}_i\right)}{4E_{n'}\left(\vec{p}_f\right)E_n\left(\vec{p}_i\right)}\bra{n'}\mathring{\mathcal{O}}^\mu\left(-\frac z2,\frac z2;A\right)\ket{n}e^{-E_{n'}\left(\vec{p}_f\right)\left(T-\tau\right)}e^{-E_n\left(\vec{p}_i\right)\tau}\,,\label{eq:3pt}
\end{align}
with the source and sink interpolating fields $\mathcal{N}$ separated by a Euclidean time $T$, momentum-dependent interpolator-state overlaps given by $\mathcal{Z}_n\left(\vec{p}_{i,f}\right)$. The bare quark bilinear, abbreviated by $\mathring{\mathcal{O}}^\mu\left(z,\tau;A\right)$, is introduced between the source and sink interpolators for Euclidean times $1\leq\tau\leq T-1$. It is the ground state, or $n,n'=0$ in Eqs.~\eqref{eq:3pt}, that corresponds to the physical nucleon state of interest.

Our computation of the correlation functions in Eqs.~\eqref{eq:2pt} and~\eqref{eq:3pt} using distillation~\cite{HadronSpectrum:2009krc} provides potential cost-saving benefits when mapping $x$-dependent GPDs because distillation enables us to efficiently calculate many combinations of source and sink momenta while reusing the most expensive components of the calculation. The Wick contracted quark fields smeared with distillation at both source and sink reduce Eqs.~\eqref{eq:2pt} and~\eqref{eq:3pt} into traces over products of distinct, reusable computational units in the distillation space. The independently generated hadron elementals and perambulators encode the operator construction at source/sink and quark propagation between time slices, respectively. The elementals and perambulators are shared between both two-point and three-point correlation functions (see~\cite{Egerer:2021ymv} Fig.~1). The color and spatial degrees of freedom (a $N_c \times L^3$ dimensional space for quarks) are projected into the lowest modes of the Laplacian (a $N_{\textrm{vec}}$ dimensional space). Since hadronic physics is dominated by the long-distance modes of QCD, which the lowest modes of the Laplacian are a proxy for, the number of modes required to reproduce a ground state hadronic correlation function is much smaller than triple the volume of modern lattices. 
However, correlators computed in distillation effectively sample the whole source and sink timeslices and therefore are equivalent to a large number of smeared point-source propagators. As a result, our computation with four time sources on 348 configurations results in much more precise matrix elements than traditional approaches. 

A substantial amount of the computational resources for this calculation are expended in computing generalized perambulators, or {\it genprops} for short:
\begin{equation}
\genprop=\sum_{\vec{y}}\xi_a^{(i)\dagger}\left(T_f\right)D^{-1}_{\alpha\sigma;ac}\left(T_f;\tau,\vec{y}\right)\mathbf{\Gamma}\left(\tau\right)D^{-1}_{\rho\beta;db}\left(\tau,\vec{x};T_i\right)\xi_b^{(j)}\left(T_i\right)\,,
\label{eq:genprop}
\end{equation}
where color (Lorentz) indices are denoted by Latin (Greek) letters, $D^{-1}$ are inverses of the Wilson-clover fermion operator, and the Dirac structure of the insertion is abbreviated $\mathbf{\Gamma}=\gamma_{\sigma\rho}^\mu e^{-i\vec{q}\cdot\vec{y}}\delta_{\vec{y},\vec{x}+z\hat{z}}W_{cd}\left(\vec{x}+z\hat{z},\vec{x}\right)$. The $\xi^{(k)}$ are eigenvectors of the three-dimensional gauge-covariant Laplacian at a time slice $T$,
\begin{equation}
-\nabla^2\left(T\right)\xi^{(k)}\left(T\right)=\lambda^{(k)}\left(T\right)\xi^{(k)}\left(T\right)\,,
\label{eq:lap}
\end{equation}
that defines the distillation operator. In practice, we compute inversions against eigenvectors at both source and sink and use $\gamma_5$-hermiticity to assemble each genprop $\genprop$. Within the distillation approach, we only need to produce genprops with all desired Dirac matrices, Wilson line lengths, and $3$-momentum transfers $\lbrace\vec{q}\rbrace$. We are then free to choose any off-forward matrix element, defined by final (initial) momenta $\vec{p}_f$ ($\vec{p}_i$), that is consistent with the pre-determined set of momentum transfers $\lbrace\vec{q}\rbrace$, and compute that matrix element via comparatively cheap tensor contractions.  

The distillation framework affords two particular advantages over other methods generally employed.  The first is that it enables a momentum projection to be performed at each point in our two- and three-point functions.  Thus it provides a far better sampling of the gauge configurations.  Second, the factorization of correlation functions into the convolution of the perambulators describing the quark propagation and the elementals representing the interpolating operators admits the use of a variational method; we defer the use of this for later work.

In this manner, distillation provides straight-forward access to what has been called in the literature ``symmetric'' and ``asymmetric'' matrix elements~\cite{Bhattacharya:2022aob} without repeating nearly identical inversions when varying the sources/sinks, which are otherwise required for standard Gaussian-smeared correlation functions computed with the help of sequential inversion methods~\cite{Bernard1986,Kilcup:1985fq}.

\subsection{Interpolator Construction\label{sec:interpConstruction}}
Mass eigenstates in the continuum are labeled according to angular momentum and parity quantum numbers $J^P$, each of which corresponds to an irreducible representation (irrep) of the continuous rotation group $O\left(3\right)$. Fermionic and bosonic states are subsequently described by double and single-valued representations, respectively. The projection of $J$ along a standard axis, typically taken to be $J_z$, provides a further quantum number and is used to label the rows of the representation.

On a finite isotropic lattice the continuum rotational symmetry reduces to the finite-dimensional cubic group or its dual, the octahedral group $O_h$. Continuum mass eigenstates in this environment are characterized by their patterns of subduction across the finite number of irreps $\Lambda$ of $O_h$. As the nucleon is our numerical arena in this study, we will be concerned with irreps of the double-cover octahedral group $O_h^D$. Due to this implied many-to-one mapping, the spectrum of the ground-state nucleon will be contaminated not only by excited states of the same continuum $J^P$ quantum numbers but by higher spin states as well.

For continuum states in motion, $J_z$ ceases to be a good quantum number and is replaced by helicity $\lambda$; $J^\lambda$ then labels the irreps of the little group, or subgroup, $O\left(3\right)$ that leaves the momentum vector invariant. Unlike the continuum little group, which is independent of momentum direction, the $O_h^D$ symmetry is broken further into little groups that depend on $^*\left(\vec{p}\right)$, the group of rotations that leave $\vec{p}$ invariant~\cite{Moore:2005dw}, further compounding the mixing of (continuum) mass eigenstates.

The reduced symmetry of an isotropic cubic lattice and the mixing of distinct continuum $J^P$ states motivates our construction of nucleon interpolators that transform irreducibly under the irreps of $O_h^D$ and its little groups. Our strategy for isolating the $1/2^+$ ground-state nucleon with distillation hinges on first building continuum interpolators $\mathcal{O}^{J,P,M}(\vec{p}=\vec{0})$ that possess definite $J^P$ and flavor quantum numbers~\cite{Edwards:2011jj,Dudek:2012ag}. This starting point reads symbolically
\be
\bra{\vec{p}=\vec{0};J',P',M'} \left[\mathcal{O}^{J,P,M}(\vec{p}=\vec{0})\right]^\dagger \ket{ \vphantom{ \vec{p}=\vec{0};J',P',M'} 0} =Z_J\delta_{JJ'}\delta_{PP'}\delta_{MM'}\,,
\ee
where the overlap of a continuum spin-$J'$ state with the continuum interpolator $\mathcal{O}^{J,P,M}$ is denoted by $Z_J$. To reliably isolate nucleons with a momentum $\vec{p}$ within our isotropic lattice, which breaks $O_h^D$ into its little groups, we implement an algorithm~\cite{Thomas:2011rh} first envisioned for mesons in flight. The starting point is continuum helicity operators, which are obtained by boosting our continuum operators at rest by an amount $\left|\vec{p}\right|$ along their quantization axis $\hat{z}$, followed by a series of active rotations that enforce a change of basis from $J_z$ to $\lambda$. This process reads symbolically as:
\be
\left[\mathbb{O}^{J,P,\lambda}\left(\vec{p}\right)\right]^\dagger=\sum_M\mathcal{D}_{M\lambda}^{(J)}\left(R\right)\left[\mathcal{O}^{J,P,M}\left(\left|\vec{p}\right|\hat{z}\right)\right]^\dagger\,,\label{eq:interpolate_roterpolate}
\ee
where a continuum helicity operator $\mathbb{O}^{J,P,\lambda}\left(\vec{p}\right)$ is obtained via application of a Wigner-$\mathcal{D}$ matrix $\mathcal{D}_{M\lambda}^{(J)}\left(R\right)$ onto the boosted $J_z$-quantized operator $\mathcal{O}^{J,P,M}\left(\left|\vec{p}\right|\hat{z}\right)$. Ref.~\cite{Thomas:2011rh} highlights the ambiguity within a finite cubic volume in rotating from $\left|\vec{p}\right|\hat{z}$ to $\vec{p}$ in this manner. A consistent convention is established by splitting the rotation $R$ into two distinct rotations,
\be
R=R_{\rm lat}R_{\rm ref}\,,
\ee
where $R_{\rm ref}$ rotates $\left|\vec{p}\right|\hat{z}$ to a reference vector $\vec{p}_{\rm ref}$\footnote{Each $O_h^D$ little group defined by $*\left(\vec{p}\right)$ is assigned a reference direction such that momenta within the same little group (\textit{e.g.}~$\left(0,1,0\right)$ and $\left(-2,0,0\right)$) can be treated consistently. For example, we assign momenta within the ${\rm Dic}_2$ little group, such as $\left(1,0,1\right)$, the reference direction $\vec{p}_{\rm ref}=\left(0,1,1\right)$. Euler angles within the $zyz$-convention then rotate $\left|\vec{p}\right|\hat{z}$ to $\vec{p}_{\rm ref}$.}, and $R_{\rm lat}$ is a lattice rotation that rotates $\vec{p}_{\rm ref}$ to $\vec{p}$. In general, $R_{\rm ref}$ may not be allowed within a cubic volume, however, its action is permissible given that it merely produces the aforementioned basis change. We have validated this procedure for all possible momenta within a finite cubic volume, with the exception of momenta of the form $\vec{p}=\left(n,m,l\right)$, with distinct $n,m,l\in\mathbb{Z}\backslash\lbrace0\rbrace$, which are left invariant under the $C_2$ little group of $O_h^D$. Hence, we will exclude $C_2$-type momenta from this study. The resulting continuum helicity operators respect the following overlap criterion:
\begin{equation}
\bra{\vec{p};J',P',\lambda'\vphantom{\left[\mathbb{O}^{J,P,\lambda}\left(\vec{p}\right)\right]^\dagger}}\left[\mathbb{O}^{J,P,\lambda}\left(\vec{p}\right)\right]^\dagger\ket{\vphantom{\vec{p};J'^{P'},\lambda'}\!\ 0}=Z_{J'J;P'P;\lambda}\delta_{\lambda\lambda'}\,.
\end{equation}
In motion, parity $P$ and spin $J$ are no longer conserved quantities, because their symmetries are broken by the explicit direction $\vec p$, while helicity remains conserved in flight. The construction of our interpolators, or equivalently subduced helicity operators, is finalized by specifying how a continuum helicity $\lambda$ subduces into a little group irrep $\Lambda$ with rows $\mu\in\lbrace1,\cdots,{\rm dim}\left(\Lambda\right)\rbrace$:
\begin{equation}
\left[\mathbb{O}^{[J,P,\left|\lambda\right|]}_{\Lambda,\mu}\left(\vec{p}\right)\right]^\dagger=\sum_{\hat{\lambda}=\pm\lambda}\left[S_{\Lambda,\mu}^{J,\hat{\lambda}}\right]^\ast\left[\mathbb{O}^{J,P,\hat{\lambda}}\left(\vec{p}\right)\right]^\dagger\,, \label{eq:subduction_duction_whats_your_function}
\end{equation}
where the subduced helicity operators used in our study are linear combinations of each value of continuum helicity, the precise combinations of which are set by so-called subduction coefficients $S_{\Lambda,\mu}^{J,\hat{\lambda}}$. As a consequence, the continuum energy eigenstates as part of the spectral content of the three-point functions in Eq.~\eqref{eq:3pt} are instead subduced eigenstates of helicity.

Interpolators constructed in this manner are assigned a compact spectroscopic notation $X^{2S+1}L_\mathcal{P}J^P$, with $X$ denoting the hadronic state, $S$ the overall Dirac spin, $L$ the angular momentum induced by covariant derivatives with permutation symmetry $\mathcal{P}$, and $J^P$ the continuum spin/parity quantum numbers. We select the simplest non-relativistic and spatially-local operator, denoted $N^2S_s\frac{1}{2}^+$, to interpolate the ground-state nucleon from the vacuum. A follow-on study could make use of an expanded operator basis (such as in Refs.~\cite{HadStruc:2021wmh,Egerer:2018xgu,HadStruc:2022yaw}) to improve excited state control within each $\vec{p}_i/\vec{p}_f$ momentum channel.

\subsection{Computing the kinematic matrix\label{sec:pGITD-extraction}}

The connection between the physically relevant matrix elements and the subduced matrix elements we extract from the three-point correlation functions was established in~\cite{HadStruc:2022nay} for the isovector quark helicity PDFs of the nucleon. In the interest of self-containment, we highlight the points that are key to the extraction of the GPDs.

To isolate the Lorentz invariant amplitudes, the subduced matrix elements can be related to the helicity matrix elements, which requires inverting the effect of the subduction coefficients $S^{J,\hat{\lambda}}_{\Lambda,\mu}$, Eq.~\eqref{eq:subduction_duction_whats_your_function}, that describe the way the different helicity $\lambda$ states are related (subduced) into the various rows $\mu\in\lbrace1,{\rm dim}\left(\Lambda\right)\rbrace$ of the irreps $\Lambda$. Following the conventions of~\cite{Thomas:2011rh}, the creation operators require the complex conjugate of the coefficient. After, the Lorentz decomposition can be applied to the continuum matrix element.
Finally, the subduced matrix elements are given by a linear combination of invariant amplitudes: 
\begin{equation}
\bra{\vec{p},\Lambda,\mu_f}\mathcal{J}^{\Lambda_\Gamma,\mu_\Gamma}\ket{\vec{p},\Lambda,\mu_i}=\sum_k\sum_{\lambda_f,\lambda_\Gamma,\lambda_i}S^{\Lambda,J}_{\mu_f,\lambda_f}\left[S_{\mu_i,\lambda_i}^{\Lambda,J}\right]^*\mathcal{K}_k\left(\lambda_f\left[J,\vec{p}\right];\lambda_i\left[J,\vec{p}\right]\right)\mathcal{A}_k\left(\nu,z^2\right)\,.
\label{eq:subduced-matelem}
\end{equation}
Here each term is weighted by a helicity-dependent kinematic prefactor $\mathcal{K}_k\left(\lambda_f\left[J,\vec{p}\right];\lambda_i\left[J,\vec{p}\right]\right)$, given as the Lorentz covariant coefficients of $\amp{k}$ in Eq.~\eqref{eq:decomp}, whose irreducibility under the relevant cubic (sub)group is encoded via subduction coefficients.

Following the procedure in~\cite{HadStruc:2022nay}, one need not relate the subduced matrix element to the helicity matrix element as an intermediate step. Instead one can calculate the kinematic matrices,
\begin{equation}
    K_k (\mu_f, [\Lambda, \vec{p}]) =\sum_{\lambda_f,\lambda_\Gamma,\lambda_i}S^{\Lambda,J}_{\mu_f,\lambda_f}\left[S_{\mu_i,\lambda_i}^{\Lambda,J}\right]^*\mathcal{K}_k\left(\lambda_f\left[J,\vec{p}\right];\lambda_i\left[J,\vec{p}\right]\right)\,,
\end{equation}
and invert the relationship to obtain the amplitudes directly. As discussed in~\cite{HadStruc:2022nay}, this viewpoint exposes the ever-present potential issues with mixing between states of different parity or continuum spin, all of which belong to the same finite volume irrep $\Lambda$. In QCD, the spectrum is ordered in spin, and ultimately the ground state which will be the desired is the $\frac 12^+$ nucleon. These other states contribute to excited state contamination that can be modeled and removed. As such, the kinematic matrices will use only spin-$1/2$ spinors.

The construction of the subduced spinors was outlined in detail in~\cite{HadStruc:2022nay}. The algorithm begins with constructing a continuum spinor with spin and momentum in the $z$ direction. This spinor is then rotated, as the interpolating field was in Eq.~\eqref{eq:interpolate_roterpolate}, to align with the correct lattice-allowed momenta. Finally, the spinor is subduced into the finite volume $\Lambda$ irrep as the interpolator was in Eq.~\eqref{eq:subduction_duction_whats_your_function}. The convenience of parameterizing a rotation with Euler angles is that these rotations are independent of the spin representation of the object being considered. Armed with subduced spinors, we can construct the Lorentz covariant coefficients of each of the desired invariant amplitudes in their relation to the subduced matrix elements. 

\subsection{From the matrix element to the amplitudes \label{sec:SVD}}

To isolate each amplitude individually we need to consider many combinations of initial/final rows as well as $\Gamma$ structures. We expand upon the strategy of~\cite{HadStruc:2022nay} to address this challenge.

In the case of the unpolarized quark forward matrix element, the nucleon matrix element $M^0\left(p,p,z\right)$ was sufficient to isolate $\mathcal{A}_1(\nu, \xi=0, t, z^2)$ and obtain a PDF. With so many amplitudes in the off-forward case, Eq.~\eqref{eq:decomp}, this is impossible, just as it was in the case of helicity quark and gluon matrix elements~\cite{HadStruc:2022nay,Balitsky:2021cwr,HadStruc:2022yaw}. While fixing the momenta and separation, the matrix element is considered in the four possible initial and final spin combinations\footnote{Regardless of source or sink momentum, each cubic irrep relevant for the (continuum) $J^P=1/2^+$ nucleon is two-dimensional.} and for each of the four $\gamma^\mu$ matrices. These combined sixteen matrix elements can be used to isolate the eight unknown $\amp{k}$ given the matrix relation
\begin{equation}
\underset{16\times1}{\mathbf{M}}=\underset{16\times8}{\mathbf{K}}\times\underset{8\times1}{\mathbf{A}}\,,
\label{eq:sysEqns}
\end{equation}
where we define a column vector of fitted subduced matrix elements by
\begin{equation}
\underset{16\times1}{\mathbf{M}}=
\begin{bmatrix}
\begin{pmatrix}
M^0\left(p_f,p_i,z\right)_{11} \\
M^0\left(p_f,p_i,z\right)_{12} \\
M^0\left(p_f,p_i,z\right)_{21} \\
M^0\left(p_f,p_i,z\right)_{22}
\end{pmatrix}^T
\begin{pmatrix}
M^1\left(p_f,p_i,z\right)_{11} \\
M^1\left(p_f,p_i,z\right)_{12} \\
M^1\left(p_f,p_i,z\right)_{21} \\
M^1\left(p_f,p_i,z\right)_{22}
\end{pmatrix}^T
\begin{pmatrix}
M^2\left(p_f,p_i,z\right)_{11} \\
M^2\left(p_f,p_i,z\right)_{12} \\
M^2\left(p_f,p_i,z\right)_{21} \\
M^2\left(p_f,p_i,z\right)_{22}
\end{pmatrix}^T\begin{pmatrix}
M^3\left(p_f,p_i,z\right)_{11} \\
M^3\left(p_f,p_i,z\right)_{12} \\
M^3\left(p_f,p_i,z\right)_{21} \\
M^3\left(p_f,p_i,z\right)_{22}
\end{pmatrix}^T
\end{bmatrix}^T\,.
\label{eq:fittedMats}
\end{equation}
The kinematic matrix is built from each Lorentz covariant structure in Eq.~\eqref{eq:decomp} sandwiched between initial/final-state subduced spinors: 
\begin{equation}
\underset{16\times8}{\mathbf{K}}=
    \begin{bmatrix}
    \langle\langle\gamma^0\rangle\rangle_{11} & z^0 \langle\langle\mathds{1}\rangle\rangle_{11} & i\langle\langle\sigma^{0 z}\rangle\rangle_{11} & \cdots & \frac{i}{2m}z^0\langle\langle\sigma^{z\Delta}\rangle\rangle_{11}  \\
    {\LARGE{\color{red}\vdots}} & {\LARGE{\color{red}\vdots}} & {\LARGE{\color{red}\vdots}} & 
    & 
    {\LARGE{\color{red}\vdots}}  \\
    
    \langle\langle\gamma^1\rangle\rangle_{11} & z^1\langle\langle\mathds{1}\rangle\rangle_{11} & i\langle\langle\sigma^{1 z}\rangle\rangle_{11} & \cdots & \frac{i}{2m}z^1\langle\langle\sigma^{z\Delta}\rangle\rangle_{11}  \\
    {\LARGE{\color{red}\vdots}} & {\LARGE{\color{red}\vdots}} & {\LARGE{\color{red}\vdots}} & 
    & 
    {\LARGE{\color{red}\vdots}} \\
        
    \langle\langle\gamma^2\rangle\rangle_{11} & z^2\langle\langle\mathds{1}\rangle\rangle_{11} & i\langle\langle\sigma^{2 z}\rangle\rangle_{11} & \cdots & \frac{i}{2m}z^2\langle\langle\sigma^{z\Delta}\rangle\rangle_{11}  \\
    {\LARGE{\color{red}\vdots}} & {\LARGE{\color{red}\vdots}} & {\LARGE{\color{red}\vdots}} & 
    & 
    {\LARGE{\color{red}\vdots}} \\
        
    \langle\langle\gamma^3\rangle\rangle_{11} & z^3\langle\langle\mathds{1}\rangle\rangle_{11} & i\langle\langle\sigma^{3 z}\rangle\rangle_{11} & \cdots & \frac{i}{2m}z^3\langle\langle\sigma^{z\Delta}\rangle\rangle_{11}  \\
    {\LARGE{\color{red}\vdots}} & {\LARGE{\color{red}\vdots}} & {\LARGE{\color{red}\vdots}} & 
    & 
    {\LARGE{\color{red}\vdots}} \\
    
    \end{bmatrix} \,,
\label{eq:kinmat}
\end{equation}
and the vector
\begin{eqnarray}
\underset{8\times1}{\mathbf{A}}=
\begin{bmatrix}
    \amp{1} & \amp{2} & \amp{3} & \amp{4} & \amp{5} & \amp{6} & \amp{7} & \amp{8}
\end{bmatrix}^T
\label{eq:ampVec}
\end{eqnarray}
contains the required amplitudes with omitted $(\nu,\xi,t,z^2)$ arguments. The subscripts in each entry of Eqs.~\eqref{eq:fittedMats} and~\eqref{eq:kinmat} are integer pairs designating the sink/source irrep row combination $\left(\mu_f \mu_i\right)$; red dots are employed for brevity to indicate repeated Lorentz covariant structures sandwiched between distinct $\left(\mu_f\mu_i\right)$ combinations. 
 
Given the linear system of sixteen equations for eight amplitudes \eqref{eq:sysEqns}, our goal is to obtain the most probable $\amp{k}$. The traditional solution to an over-constrained linear system of equations is to find the minimum of $\chi^2=||\mathbf{M} - \mathbf{K} \mathbf{A}||_2^2$ where $||.||_2$ denotes the $l^2$ norm. The solution is given by $\bar{\mathbf{A}} = \mathbf{K}^+ \mathbf{M}$, where $\mathbf{K}^+$ is the pseudoinverse of $\mathbf{K}$~\cite{penrose_1956}. The pseudoinverse can be calculated via singular value decomposition (SVD). Numerically, we use a cutoff on the relative size of the smallest to the largest singular value of $\epsilon=10^{-15}$ in the \verb+linalg.pinv+ function of the publicly available \verb+NumPy+ package. This cutoff formally shifts the matrix from the original definition of the Moore-Penrose pseudoinverse but increases the numerical stability.

Half of the equations in Eq.~\eqref{eq:sysEqns} are related to the other half by various symmetries. Ultimately the four helicity combinations only give two independent equations. For most kinematics, this is enough to reliably reconstruct the eight amplitudes $\mathcal{A}_k$. A few specific situations arise however:

\begin{enumerate}
\item If $z = 0$, the kinematic factors of all amplitudes except $\mathcal{A}_{1,4,5}$ cancel. Those are therefore the only amplitudes that we can hope to extract. We will discuss the results at $z = 0$ specifically in section \ref{sec:EFFs}.

\item

If the spatial momentum transfer $\vec{q} = \vec{p}_f - \vec{p}_i$ is exactly along the $z$ axis, several issues may appear. For some kinematic pairs and $z \neq 0$ (respectively $z = 0$), there are fewer than eight (respectively three) singular values in the kinematic matrix, meaning that there exist exact (up to $\epsilon=10^{-15}$) linear relations between some amplitudes that cannot be separated from one another. In other cases, in spite of eight different singular values, the unitary matrices entering the SVD present large similarities in their columns and result in a very bad case of uncertainty propagation. As we observe in a consistent fashion that these kinematics behave poorly, we remove them altogether. They are not included in the summary plot of Fig.~\ref{fig:kincov}.

\end{enumerate}

If $z \neq 0$ and $\vec{q}$ is not along the $z$ axis ($q_1$ or $q_2 \neq 0$), then we observe that, due to our choice of $z^\mu = (0, 0, 0, z_3)$, only the $M^3(p_f, p_i, z)$ matrix elements are sensitive to $\mathcal{A}_2$ and $\mathcal{A}_8$. In practice, this means that the $M^3$ matrix elements are entirely and exclusively used to characterize those two amplitudes. Although it has been argued that using $M^3$ introduces a finite mixing \cite{Constantinou:2017sej}, using it does not change the extraction of any other amplitude, and therefore does not impact our light-cone GPDs, which do not involve either $\mathcal{A}_2$ or $\mathcal{A}_8$.

When $z = 0$, the $\mathcal{A}_2$ and $\mathcal{A}_8$ amplitudes are absent from the decomposition and the $M^3$ matrix elements have a constraining power on the amplitudes $\mathcal{A}_{1,4,5}$. We observe, however, that the difference is typically of the order of a few percent at most on $\mathcal{A}_{1,4}$, which contain the elastic form factors (EFFs). As we will see later, we perform the extraction of the EFFs using both the $z = 0$ and $z \neq 0$ data, and obtain very similar results up to a $\mathcal{O}(z^2)$ contamination which cannot be attributed to the $M^3$ data (since it plays no role when $z \neq 0$). We give more details on the question of the use of $M^3$ in appendix \ref{app:to_gammaz}.

As an extension to this approach, one could consider the effect of additional momentum combinations that select the same values of $\nu$, $\xi$, and $t$ to produce an even more constrained system of equations. This could especially help in reducing lattice systematic effects, such as discretization errors of $\mathcal{O}\left(aq\right)$. Using the relationships in section~\ref{sec:spacetime} between amplitudes of $\pm\nu$ and $\pm\xi$, even more combinations can be included. In this paper, we limit ourselves to averaging the $\pm z_3$ separations and perform an external amplitude extraction for each pair $(\vec{p}_f, \vec{p}_i)$. Nonetheless, we have a certain redundancy, because we have data for $\pm \xi$ and various $\nu$ at the same or very nearby $t$, which will enables us to reduce some of the lattice artifacts.

An alternative possibility to perform the extraction is to remove or average directly the trivial duplicate constraints and form a true matrix inverse. The solution of this linear inverse would exactly reproduce the analog of the analytic formulas presented in~\cite{Bhattacharya:2022aob} to extract the amplitudes from the matrix elements. In our calculation, the spinors required to calculate $\langle\langle\Gamma\rangle\rangle_{\mu_f \mu_i}$ are complicated functions of the momenta, depending upon its magnitude, the specific finite volume irreducible representation, and choices of convention. With many momenta combinations, solving the analytic formulas would require a numerical algebra implementation for practical use. The approach outlined in this section replaces the step of re-deriving these analytic solutions for varying considerations such as off-axis separations or new momentum frames with a generic linear algebra problem.

Finally, we have cross-checked the stability of our SVD extraction by using an $l^1$-norm minimization of $||\mathbf{M} - \mathbf{K} \mathbf{A}||_1$, also known as least absolute deviation (LAD) estimator \cite{LAD}. This procedure has disadvantages compared to SVD because it requires an iterative minimization for each jack-knife sample of the data, which is hundreds of times 
computationally more expensive than a simple matrix operation. The results are consistent, as one can check in appendix \ref{ap:fullset}, where we present the extraction of a full set of amplitudes in some kinematics.

\section{Analysis of correlation functions\label{sec:isolateMats}}

\begin{figure}[t!]
    \centering
    \includegraphics[width=0.9\linewidth]{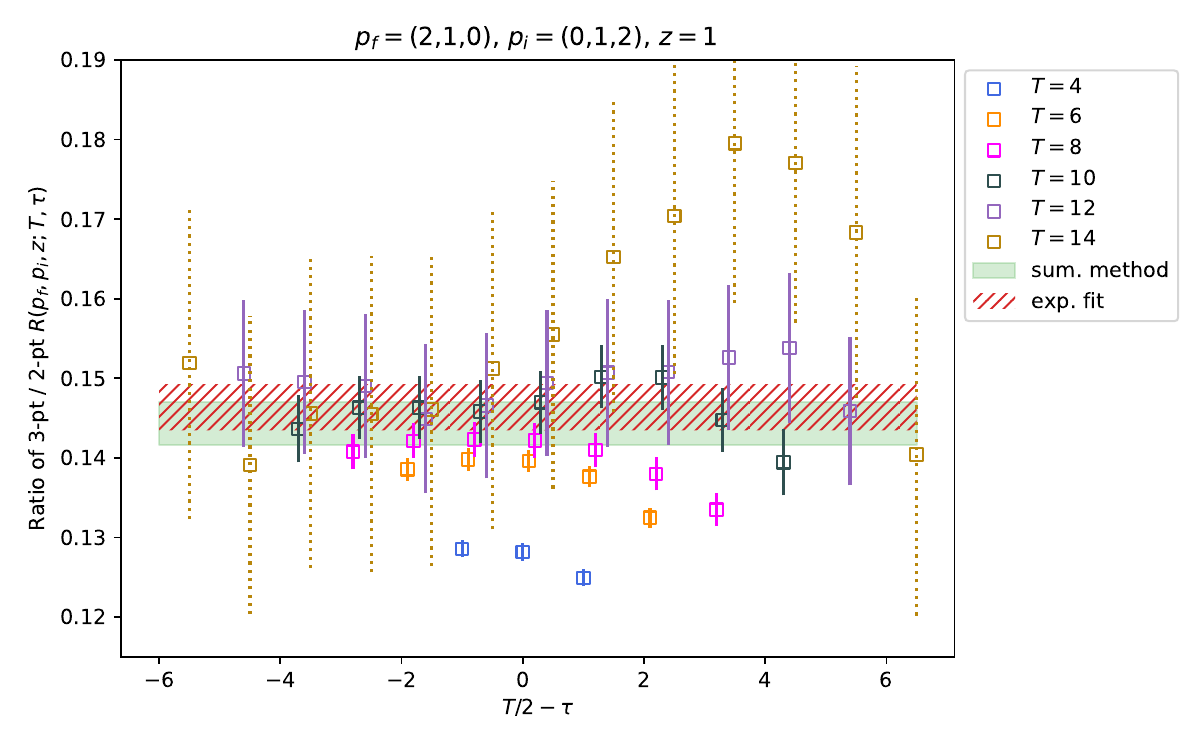}

    \caption{Ratio $R^0(\vec{p}_f, \vec{p}_i, z = a; T, \tau)$ for $\vec{p}_f = (2,1,0)$ and $\vec{p}_i = (0, 1, 2)$ (lattice units) with proton rows $1,1$ for various values of $T$ and $\tau$. We show the result of the matrix element fit using either the summation method or an exponential fit with one excited state, with the exclusion of all $\tau$ and $T-\tau < 3$. The data at $T = 14$ is dotted for legibility purposes.}
    \label{fig:ratio}
\end{figure}

We isolate the desired bare matrix element $M^\mu\left(\vec{p}_f,\vec{p}_i,z\right)$ using two strategies to evaluate possible systematics linked to excited state contamination. Our first procedure is the {\it summation method}~\cite{Maiani:1987by,Capitani:2012gj}. We form an optimized ratio of three-point and two-point correlation functions,
\begin{equation}
\ratio=\frac{C^\mu_{3{\rm pt}}\left(\vec{p}_f,\vec{p}_i,z;T,\tau\right)}{C_{2{\rm pt}}\left(\vec{p}_f;T\right)}\sqrt{\frac{C_{2{\rm pt}}\left(\vec{p}_i;T-\tau\right)C_{2{\rm pt}}\left(\vec{p}_f;\tau\right)C_{2{\rm pt}}\left(\vec{p}_f;T\right)}{C_{2{\rm pt}}\left(\vec{p}_f;T-\tau\right)C_{2{\rm pt}}\left(\vec{p}_i;\tau\right)C_{2{\rm pt}}\left(\vec{p}_i;T\right)}}\,,
\label{eq:ratio}
\end{equation}
where $T$ is the temporal separation between the source/sink interpolating fields and $0\leq\tau\leq T$ is the current insertion time slice. The three-point correlation functions are computed for the values of $T \in \{4,6,8,10,12,14\}$ and every $1 \leq \tau \leq T-1$. For asymptotically large temporal separations ($0\ll\tau\ll T$), the ratio $\ratio$ will become proportional to the desired matrix element,
\be
\ratio\overset{\tau,T\rightarrow\infty}{\longrightarrow}\frac{1}{\sqrt{2}}\times\frac{1}{\sqrt{4E_0\left(\vec{p}_f\right)E_0\left(\vec{p}_i\right)}}M^\mu\left(p_f,p_i,z\right)\,.
\ee
The additional factor of $\sqrt{2}$ is due to the normalization of the isovector vector current.\footnote{
The electromagnetic current in the light-quark sector reads $\mathcal{J}^\mu=\frac{2}{3}\overline{u}\gamma^\mu u-\frac{1}{3}\overline{d}\gamma^\mu d$, which can be expressed in terms of isovector $\rho^\mu$ and isoscalar $\omega^\mu$ components according to $\mathcal{J}^\mu=\frac{\rho^\mu}{\sqrt{2}}+\frac{\omega^\mu}{3\sqrt{2}}$, where $\rho^\mu=\left(\overline{u}\gamma^\mu u-\overline{d}\gamma^\mu d\right)/\sqrt{2}$ and $\omega^\mu=\left(\overline{u}\gamma^\mu u+\overline{d}\gamma^\mu d\right)/\sqrt{2}$. We use the current $\rho^\mu$ to probe the isovector flavor structure of the off-forward nucleon matrix elements.} We form $\ratio$ by treating both three-point and all two-point correlators as complex-valued functions, as the construction of the subduced helicity operators we use as interpolators, detailed in section~\ref{sec:interpConstruction}, involves a delicate interplay of complex phases. A depiction of the ratio for a specific choice of $\vec{p}_f$ and $\vec{p}_i$ is given in Fig. \ref{fig:ratio}.

\begin{figure}[ht!]
    \centering
    \includegraphics[width=0.54\linewidth]{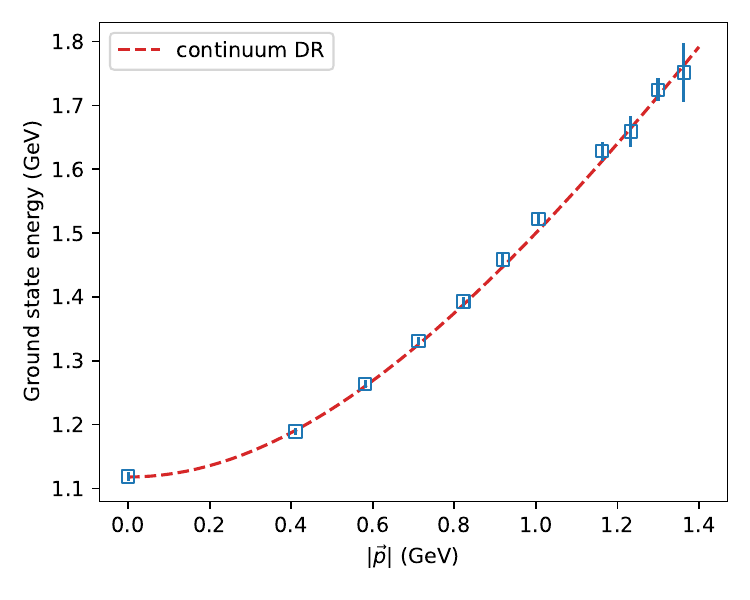} \includegraphics[width=0.44\linewidth]{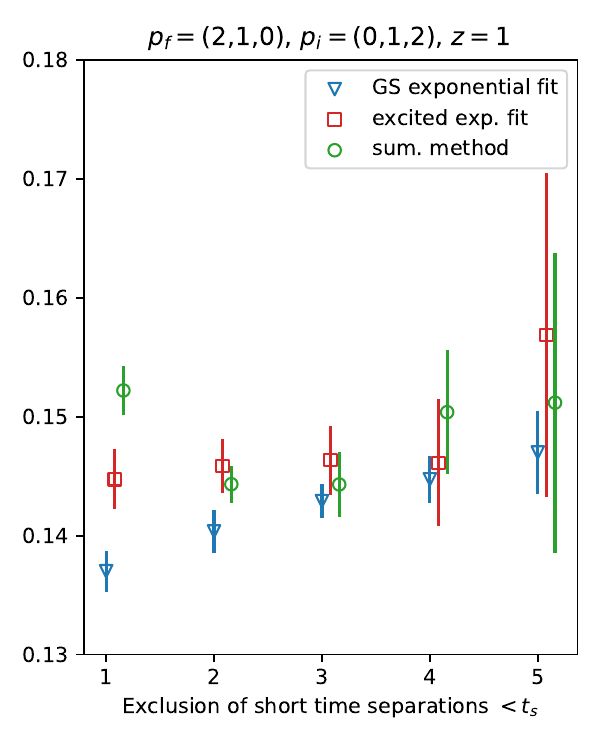}

    \caption{(left) Dispersion relation $E_0\left(|\vec{p}|\right)$ obtained from non-linear fits of the ground state and first excited state on the two-point correlation functions. The dotted line is the continuum dispersion relation.  -- (right) Impact of the exclusion of short-time separations below $t_s$ on the extraction of the same bare matrix element as in Fig. \ref{fig:ratio}. We show the result of a ground state exponential fit (blue), exponential fit with one excited state (red), and ratio method (green). The standard result in this analysis will be using both the red point at $t_s = 3$ and at $t_s = 4$ to evaluate excited state uncertainty.}
    \label{fig:compare_ts}
\end{figure}

Then the matrix elements are extracted by summing the ratio $R^\mu\left(\vec{p}_f,\vec{p}_i,z;T,\tau\right)$ over the insertion time slice $\tau$, excluding all time separations below a threshold $t_s$,
\begin{equation}
\mathcal{R}_{t_s}^\mu\left(\vec{p}_f,\vec{p}_i,z;T\right)\equiv\sum_{\tau/a=t_s}^{T-t_s}R^\mu\left(\vec{p}_f,\vec{p}_i,z;T,\tau\right)\,.
\label{eq:summedRatio}
\end{equation}
The summed ratio $\mathcal{R}_{t_s}^\mu\left(\vec{p}_f,\vec{p}_i,z;T\right)$ is a geometric series that depends linearly on the bare matrix element $M^\mu\left(p_f,p_i,z\right)$. We apply the functional form
\begin{equation}
\mathcal{R}^\mu_{\rm fit}\left(\vec{p}_f,\vec{p}_i,z;T\right)=C+M^\mu\left(p_f,p_i,z\right)T
\label{eq:SRfit}
\end{equation}
to extract the bare matrix element. This procedure should yield errors of the order $\mathcal{O}\left(e^{-\Delta ET}\right)$, where $\Delta E$ is the energy gap between the lowest-lying effective excited state and the ground state. We show in the right panel of Fig.~\ref{fig:compare_ts} the impact of the exclusion of short time separations, both in the ratio method (green points) and in the exponential fits that we will describe now.

Our second procedure extracts the bare matrix elements using exponential fits, following the spectral decompositions in Eqs.~\eqref{eq:2pt} and \eqref{eq:3pt}. 

We first perform a non-linear fit of the ground state and first excited state on the two-point correlation functions, excluding small source-sink separations. For each value of $|\vec{p}|$, we use several rotationally equivalent two-point functions to increase the robustness of our fit. As a result, we obtain for each jack-knife sample the fitted values of  $\mathcal{Z}_n\left(\vec{p}\right)$ and $E_n\left(\vec{p}\right)$ for $n = 0$ and $1$. The dispersion relation of the ground state is shown on the left panel of Fig.~\ref{fig:compare_ts}. 

Then we use the fitted values of $\mathcal{Z}_n\left(\vec{p}\right)$ and $E_n\left(\vec{p}\right)$ in the decomposition of the three-point correlation function. We first fit the three-point function assuming pure ground state dominance, that is, using only the $n = 0$ fits of the two-point functions. This gives the blue points in the right panel of Fig.~\ref{fig:compare_ts}, where we sequentially exclude the small separations $\tau$ and $T-\tau < t_s$. It is clear that the points follow a non-constant trend, a sign of sizeable excited state contamination. Based on these results, we use both the results for $n = 0$ and $n=1$ (red points in the right panel of Fig.~\ref{fig:compare_ts}), meaning that we fit the three-point function with four free parameters corresponding, in principle, to $\langle 0|\mathcal{O} |0\rangle$, $\langle 0|\mathcal{O} |1\rangle$, $\langle 1|\mathcal{O} |0\rangle$ and $\langle 1|\mathcal{O} |1\rangle$. The results are now generally in very good agreement with the ratio fits. In general, we observe that the exponential fit with one excited state seems less prone to fluctuations compared to the ratio method when increasing the cut in $t_s$. Therefore, we will use the excited state exponential fit with a cut $t_s = 3$ and $t_s = 4$ to include a measure of excited state contamination. 

Finally, we form the ratio with the matrix element for $\vec{p}_f = \vec{p}_i = 0$, $\Lambda_f = \Lambda_i = 1$ and $\mu = 0$ to form the RGI reduced matrix element, Eq.~\eqref{eq:reducedmatelem}.

The process of subduction detailed in section~\ref{sec:interpConstruction} leads to ${\rm dim}\left(\Lambda_f\right)\times{\rm dim}\left(\Lambda_i\right)$ determinations of each matrix element $M^\mu\left(\vec{p}_f,\vec{p}_i,z\right)$, where $\Lambda_f$ and $\Lambda_i$ are the irreps at sink and source, respectively. The fitted matrix elements will be determined for each combination of initial and final interpolator rows $\mu_i/\mu_f$.

\section{Elastic form factors \label{sec:EFFs}}

\subsection{Local matrix elements}

Let us first consider the local limit $z=0$. Then, Eq.~\eqref{eq:decomp} reduces to a decomposition resembling the standard form factors parametrizing an electromagnetic interaction with a spin-1/2 nucleon:
\begin{align}
M^\mu\left(p_f,p_i,0\right)&=\langle\langle\gamma^\mu\rangle\rangle\mathcal{A}_1\left(0,\xi,t,0\right)+\frac{i}{2m}\langle\langle\sigma^{\mu q}\rangle\rangle\mathcal{A}_4\left(0,\xi,t,0\right)+\frac{q^\mu}{2m}\langle\langle\mathds{1}\rangle\rangle\mathcal{A}_5\left(0,\xi,t,0\right),\label{eq:F1F2-withDeltaMu} \\
&=\langle\langle\gamma^\mu\rangle\rangle F_1\left(t\right)+\frac{i}{2m}\langle\langle\sigma^{\mu q}\rangle\rangle F_2\left(t\right)+\frac{q^\mu}{2m}\langle\langle\mathds{1}\rangle\rangle\mathcal{A}_5\left(0,\xi,t,0\right)\label{eq:dirac-pauli-FFs}\,,
\end{align}
where $F_1\left(t\right)$ and $F_2\left(t\right)$ are the familiar Dirac and Pauli form factors of the nucleon. The Ward identity $q_\mu M^\mu=0$ requires that $\amp{5}$ vanishes in the local limit.\footnote{The terms $q_\mu\langle\langle\gamma^\mu\rangle\rangle$ and $q_\mu\langle\langle\sigma^{\mu q}\rangle\rangle$ evaluate to zero via the Dirac equation, while the term $q_\mu q^\mu\langle\langle\mathds{1}\rangle\rangle$ does not vanish.} Any departure of $\A{5}$ from zero in the local limit can thus be interpreted as the degree to which lattice artifacts systematically affect a calculation of Eq.~\eqref{eq:decomp} since we use the local vector current which is not the conserved vector current on the lattice.

\begin{figure}[t!]
    \centering
    \includegraphics[width=0.9\linewidth]{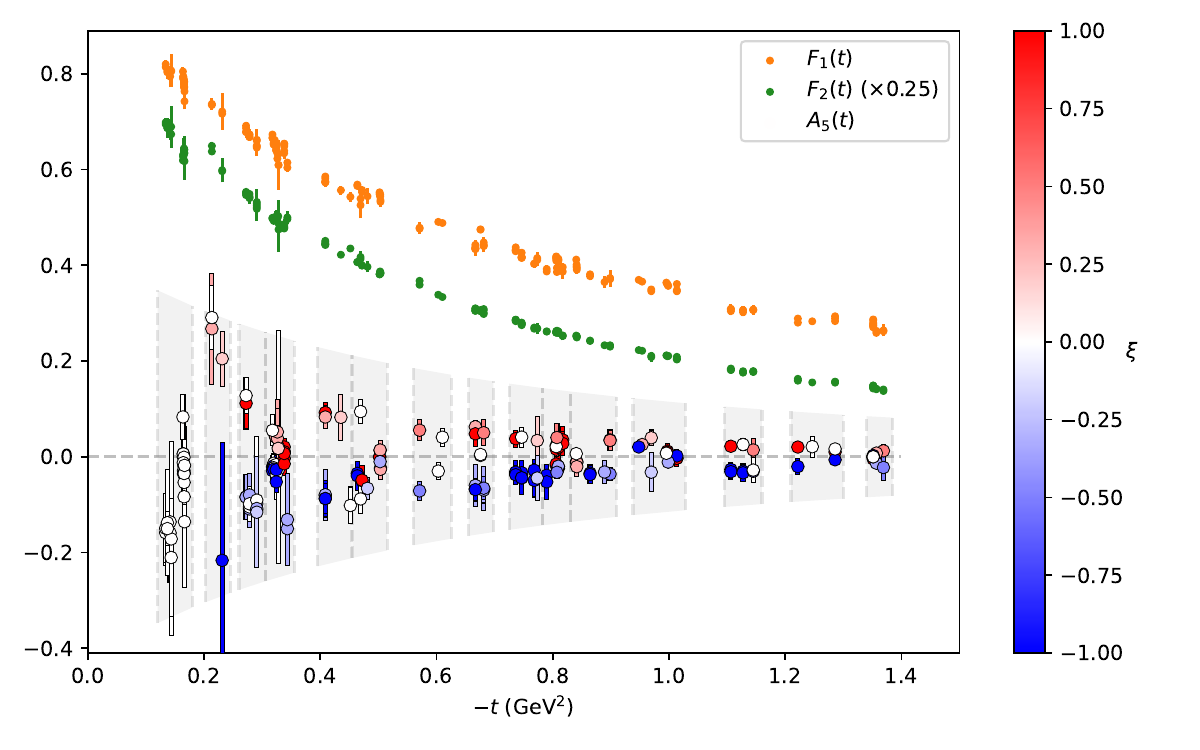}
    
    \caption{Elastic form factors $F_1(t)$, $F_2(t)$ and $\mathcal{A}_5(t)$ extracted from the local matrix elements using all skewness data. The $\mathcal{A}_5(t)$ data are plotted with a color scheme that reflects the skewness at each point. The 15 grey areas represent the 15 bins in $t$ which will come into play later on.}
    \label{fig:EFFs}
\end{figure}

We present in Fig.~\ref{fig:EFFs} the results of the elastic form factors on the 186 pairs of momenta $(\vec{p}_f, \vec{p}_i)$ whose kinematic coverage is displayed on Fig.~\ref{fig:kincov}. We have merged the errorbars provided by the cut of lattice time separation $t_s = 3$ and 4 (see section \ref{sec:isolateMats}). As expected, $F_1(t)$ and $F_2(t)$ are globally independent of $\xi$. The amplitude $\mathcal{A}_5(0, \xi, t, 0)$ is generally at most a few sigmas away from zero but there seems to be a non-vanishing signal, loosely odd in $\xi$ as expected using the discrete symmetries of section~\ref{sec:spacetime}. The equivalent term in~\cite{Bhattacharya:2022aob} was found to be small, but nonetheless non-zero for $z=0$. App.~\ref{app:to_gammaz} describes the effect on this systematic error from the inclusion or exclusion of $M^3$ data. 

When using the extraction of bare matrix elements from the summation of ratios instead of the exponential fit, an even starker signal of non-vanishing $\mathcal{A}_5(t)$ appears. The fact that $\mathcal{A}_5(t)$ at $z = 0$ presents an enhanced sensitivity to the treatment of excited states suggests that some underestimation of excited state contamination may play a role in addition to the discretization errors in the non-vanishing value of $\mathcal{A}_5(t)$. On the other hand, $F_1(t)$ and $F_2(t)$ seem generally unaltered by the behavior of $\mathcal{A}_5(t)$ and its skewness dependence. In fact, performing the SVD while ignoring altogether the $\mathcal{A}_5(t)$ contribution does not produce a significant change in the extraction of $F_1(t)$ and $F_2(t)$.

\begin{figure}[ht!]
    \centering
    \includegraphics[width=0.9\linewidth]{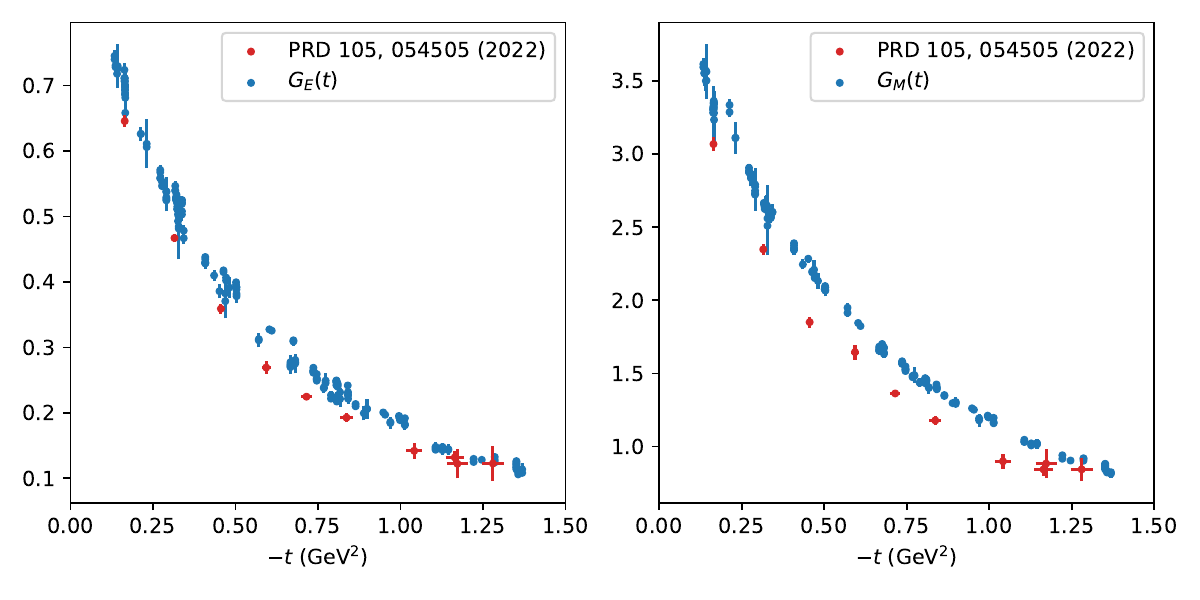}
    
    \caption{Comparison between our extraction of the $G_E$ and $G_M$ form factors from local matrix elements and the determination in~\cite{Park:2021ypf} on a slightly lighter Wilson clover fermion ensemble (pion mass of 270 Mev vs. 358 MeV in this work).}
    \label{fig:GEGM}
\end{figure}

A recent expansive calculation of nucleon form factors by the Nucleon Matrix Elements Collaboration~\cite{Park:2021ypf} included determinations of the electric and magnetic Sachs form factors 
\be
G_E\left(t\right)=F_1\left(t\right)+\frac{t}{4m^2}F_2\left(t\right)\,,\qquad G_M\left(t\right)=F_1\left(t\right)+F_2\left(t\right)\,,
\label{eq:sachs-ffs}
\ee
on a Wilson clover fermion ensemble characterized by a pion mass ($m_\pi\sim270$ MeV) that is slightly lighter than the {\tt a094m358} ensemble we consider in this work. The result is shown in Fig. \ref{fig:GEGM}. As expected, the extraction with a lighter pion mass has a more pronounced decrease in $t$. 

The next step of our analysis consists in forming bins of data in $t$, as indicated by the grey regions in Fig.~\ref{fig:EFFs}. This is beneficial for several reasons. First, by averaging over data with various $(\vec{p}_f, \vec{p}_i)$ in a single bin, we hope to mitigate some lattice systematic uncertainty. For instance, we have many cases of symmetry $\xi \longleftrightarrow -\xi$ in our data, but also different magnitudes of $(\vec{p}_f, \vec{p}_i)$ in the same bin producing different excited state behavior. Furthermore, the binning results in a decrease in the number of degrees of freedom for the final fit of the $t$-dependence, meaning that the empirical covariance matrix is a better estimate of the true covariance of the data. Indeed, to estimate the covariance matrix, one should typically have many more measurements than random variables. Here, with 186 kinematic values for 348 gauge configurations, the stability of the empirical covariance matrix is reduced. Trimming down the 186 kinematic values to fifteen bins improves the confidence in this estimation. Finally, the binning gives us access to a significant number of values of $(\nu, \xi\nu)$ in each bin when $z \neq 0$, which will allow for the extraction of generalized form factors in the next section.

We perform the binning in two steps. First, we fit each amplitude by a constant within each bin. Then we take the fifteen data points obtained in this fashion, and fit an overall $t$-dependence using a dipole fit,
\begin{equation}
     F(t) = A \left(1 - \frac{t}{\Lambda^2}\right)^{-2}\,.\label{eq:dipole}
 \end{equation}8
Next, we displace the points in each bin towards the center of their bin using the $t$-slope indicated by the dipole fit. With this method, we expect to correct the intrabin $t$-dependence of order $\mathcal{O}(t-t_{\textrm{center}})$, where $t_{\textrm{center}}$ is the value at the center of the bin to which each point belongs. In effect, the modification is generally very small as $|t - t_{\textrm{center}}|$ is typically below 0.02 GeV$^2$. A few reach 0.04 GeV$^2$, but only for larger values of $t$, where the $t$ dependence is reduced. The only noticeable effect of intrabin $t$-correction arises in the first bin, where the slope in $t$ is pronounced and the data very numerous, and we consider this therefore as a negligible effect in the overall uncertainty budget. For this reason, we will not quote an uncertainty in $t$ in our results.

\begin{figure}[ht!]
    \centering
    \includegraphics[width=0.99\linewidth]{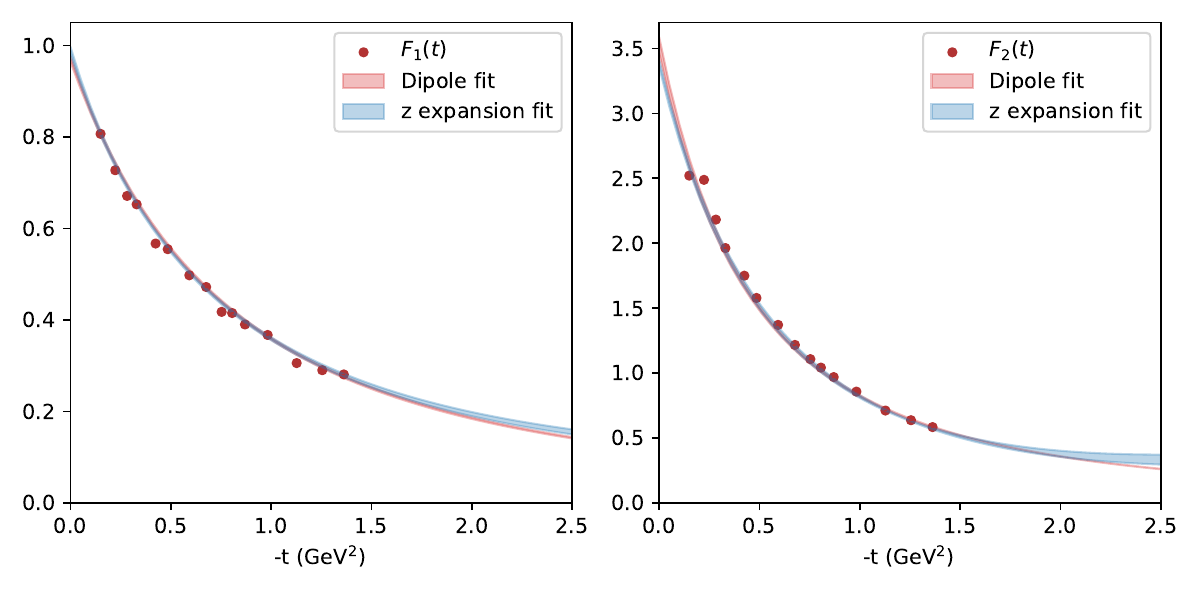}
    
    \caption{Fit of the elastic form factors from local matrix elements after binning into 15 values. The uncertainty bands are not only obtained from the statistical uncertainty of the data but also by merging the fit results using two cuts in lattice time separation to evaluate excited state contaminations. The results of the fit are produced in the main body.}
    \label{fig:localEFF}
\end{figure}

We fit the elastic form factor using either the dipole form of Eq.~\eqref{eq:dipole} or a $z$-expansion \cite{Lee:2015jqa} with three free-parameters,
\begin{equation}
    F(t) = \sum_{k = 0}^2 a_k z(t)^k\,,\ \ z(t) = \frac{\sqrt{t_{\textrm{cut}} - t} - \sqrt{t_{\textrm{cut}} - t_0}}{\sqrt{t_{\textrm{cut}} - t} + \sqrt{t_{\textrm{cut}} - t_0}}\,, \label{eq:zexp}
\end{equation}
where $t_{\textrm{cut}} = 4 m_\pi^2$ and $t_0 = t_{\textrm{cut}} (1-\sqrt{1-t_{\max} / t_{\textrm{cut}}})$. We use $t_{\max} = -1.4$ GeV$^2$. We observe that the specific value of $t_{\textrm{cut}}$ and $t_{\max}$ does not produce a significant effect on the extraction. The final fit results are produced in Table \ref{resEFF} and displayed in Fig.~\ref{fig:localEFF}. We quote in Table \ref{resEFF} both a statistical uncertainty and an excited state uncertainty obtained by comparing the results with a cut of $t_s = 3$ and 4 on the correlation functions. The uncertainty band in Fig.~\ref{fig:localEFF} is constructed as the union of the bands obtained using the two cuts. As one can observe, both the statistical and excited state uncertainties are extremely narrow for the elastic form factor, to the point that the model dependence in the extrapolation to $t = 0$ becomes a significant source of uncertainty as well. The $\chi^2$ per degree of freedom of those fits is bad, between 6 and 30. We attribute that to the fact that the precision of the measurements is such that lattice spacing errors become very noticeable and make it apparent that the Lorentz decomposition is violated on the lattice. As a result, the points do not fall exactly on a single curve, and increasing the statistics would probably only result in even worse fits.  

 \setlength{\tabcolsep}{0.5em}
 \begin{table}
 \begin{tabular}{c|c|c|c}
      &  $A$ (dipole at $t=0$) & $\Lambda$ (MeV) & $z$-exp order 2 at $t = 0$ \\ \hline \hline
  $F_1(t)$ local  & $0.97 \pm 0.004 \pm 0.006$ & $1246 \pm 3 \pm 6$ &  $0.989 \pm 0.004 \pm 0.008$ \\ \hline
  $F_1(t)$ non-local & $0.969 \pm 0.005 \pm 0.010$ & $1248 \pm 4 \pm 13$& $0.988 \pm 0.005 \pm 0.012$ \\ \hline \hline
  $F_2(t)$ local  & $3.53 \pm 0.01 \pm 0.06$ & $966 \pm  2 \pm 7$ & $ 3.38 \pm 0.01 \pm 0.03 $ \\ \hline 
  $F_2(t)$ non-local & $3.44 \pm 0.01 \pm 0.03$ & $982 \pm 2 \pm 4$ & $3.30 \pm 0.01 \pm 0.02$
 \end{tabular}
 \caption{Values of the fit of the $F_1$ and $F_2$ form factors both from local and non-local matrix elements with a dipole and a $z$-expansion up to order 2 (three free parameters). The statistical and an estimate of the systematic uncertainty linked to excited state contamination are provided. \label{resEFF}}
 \end{table}

\subsection{Elastic form factors from non-local data\label{nonloceff}}

\begin{figure}[ht!]
    \centering
    \includegraphics[width=0.99\linewidth]{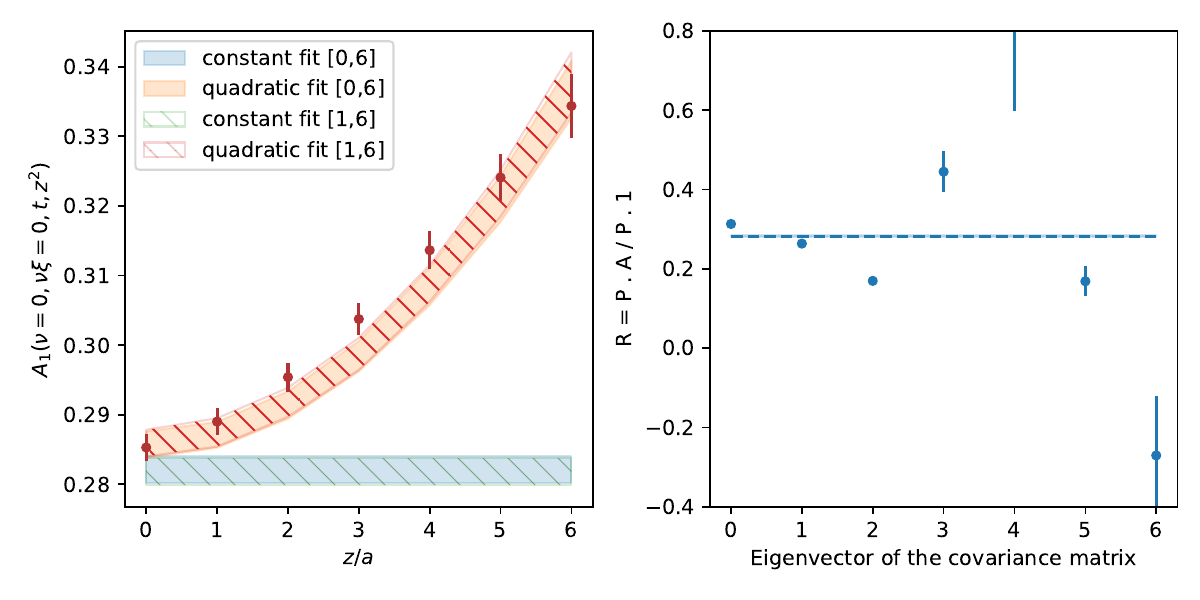}
    
    \caption{(left) Value of $\mathcal{A}_1(\nu = 0, \nu\xi = 0, -1.35\,\textrm{GeV}^2, z^2)$ for the pair of momenta $\vec{p}_f = (1, -1, 0)$ and $\vec{p}_i = (-1, 1, 0)$ in lattice units, using the cut $t_s = 3$. We show both fits by a constant (blue and green) and a quadratic model $\alpha + \beta (z/a)^2$ (orange and red), using either all data points (solid area) or excluding the local data $z = 0$ (hatches). -- (right) To understand why the constant fit ends up below every single point, we show the full set of measurements in the diagonalized space of the covariance matrix. See explanations in the text.} \label{fig:driftEFF}
    
\end{figure}

Although elastic form factors are typically computed using the local operator $z = 0$, they can be accessed using non-local operators as well. In fact, some kinematics $(\vec{p}_f, \vec{p}_i)$ are particularly convenient for that purpose: if $p_{f,3} = p_{i,3} = 0$ and $z = (0, 0, 0, z_3)$, then both $\nu$ and $\nu \xi$ are identically zero whatever the value of $z^2$, and one has:
\begin{align}
    F_1(t) = \lim_{z^2 \rightarrow 0} \mathcal{A}_1(\nu = 0, \nu\xi = 0, t, z^2)\,,\\
    F_2(t) = \lim_{z^2 \rightarrow 0} \mathcal{A}_4(\nu = 0, \nu\xi = 0, t, z^2)\,.    
\end{align}
Using non-local operators exposes us to the risk of introducing additional uncertainty due to the evaluation of the right-hand sides at $z^2 < 0$. We plot on the left panel of Fig.~\ref{fig:driftEFF} the value of $\mathcal{A}_1(\nu = 0, \nu\xi=0, t = -1.35$ GeV$^2, z^2)$ using $\vec{p}_f = (1, -1, 0)$ and $\vec{p}_i = (-1, 1, 0)$ for every value of $z$ from 0 to $6a$ (we average the $\pm z$ data) using the cut $t_s = 3$. Evidently, a noticeable $z$ dependence can be observed.

It is important to notice that the various points on the plot are highly correlated with one another, as they all stem from the same external proton states computed on the same gauge configurations. The correlation matrix of the data is given by:
\begin{equation}
    \textrm{Corr} = \begin{pmatrix}
1 &   0.96 & 0.93 & 0.86 & 0.76 & 0.64 & 0.51 \\
0.96 & 1  &  0.98 & 0.93 & 0.84 & 0.73  & 0.60 \\
0.93 & 0.98 & 1  &  0.98 & 0.91 & 0.82 & 0.70 \\
0.86 & 0.93 & 0.98 & 1 &   0.98 & 0.91 & 0.80 \\
0.76 & 0.84 & 0.91 & 0.98 & 1   & 0.97 & 0.90\\
0.64 & 0.73 & 0.82 & 0.91 & 0.97 & 1  &   0.97\\
0.51 & 0.60 & 0.70 & 0.80 & 0.90 & 0.97 & 1
    \end{pmatrix}.
\end{equation}
Neighboring points on the plot are correlated by more than 96\%, and the furthest are still over 50\%. Those circumstances are common in lattice QCD plots where the axis variable is the non-local separation $z$. This can produce counter-intuitive results.

For pedagogical purpose on the counter-intuitive effect of very high correlation, despite the clear trend in $z$ that appears visually, one could wish to fit our dataset by a constant. Intuitively, one would expect the constant to fall somewhere in the middle of the plot, around $F_1 \approx 0.31$. Such would indeed be the case if our points were uncorrelated. But the best fit by a constant is in fact $F_1 \approx 0.28$, strangely close to the result at $z = 0$, and in good agreement with the projection to $z = 0$ of the quadratic fit by $\alpha + \beta (z/a)^2$. Excluding the local data at $z = 0$ produces exactly the same fit results, either for the constant or quadratic fit. However, now the constant fit is incompatible with every single point of the fitted dataset. The remarkable agreement of the constant fit with the quadratic one at $z = 0$ is quite surprising and happens in a general fashion in other kinematics as well. Although we will use the quadratic fit for the non-local elastic form factor, it is interesting to try to understand why a value that is significantly below every single data point can be the best constant fit of the dataset. This illustrates how such plots can be quite misleading if the very high degree of correlation is not taken into account.

One needs to remove correlations from the data. First, we diagonalize the covariance matrix $\textrm{Cov} = P^{-1} D P$ where $D$ is the diagonal matrix of eigenvalues. Then the vector
\begin{equation}
    P \times \begin{pmatrix}
        \mathcal{A}_1(z = 0) \\ \cdots \\ \mathcal{A}_1(z = 6a)
    \end{pmatrix}
\end{equation}
is an uncorrelated random vector. However, at this stage, its central values are difficult to interpret. In our case, it is interesting to compare these uncorrelated data with what would happen if $\mathcal{A}_1$ was indeed independent of $z$. Therefore, we form:
\begin{equation}
R = P \times \begin{pmatrix}
        \mathcal{A}_1(z = 0) \\ \cdots \\ \mathcal{A}_1(z = 6a)
    \end{pmatrix} \bigg/ P \times \begin{pmatrix}
        1 \\ \cdots \\ 1
    \end{pmatrix}\,,
\end{equation}
where the division is understood as a term-by-term ratio of the elements in both vectors. If $\mathcal{A}_1$ was indeed constant in $z$, its uncorrelated version $R$ would remain constant with the same value. The values of the seven uncorrelated random variables contained in R are shown on the right panel of Fig.~\ref{fig:driftEFF}. 

Now the origin of our problem appears clearly. There exist uncorrelated linear combinations of our measurements which favor very small constant fits. In fact, it is easy to see from this example that, by carefully crafting the correlations between the data, the best fit from a given model can be arbitrarily far from any of the correlated measurements.

\begin{figure}[ht!]
    \centering
    \includegraphics[width=0.9\linewidth]{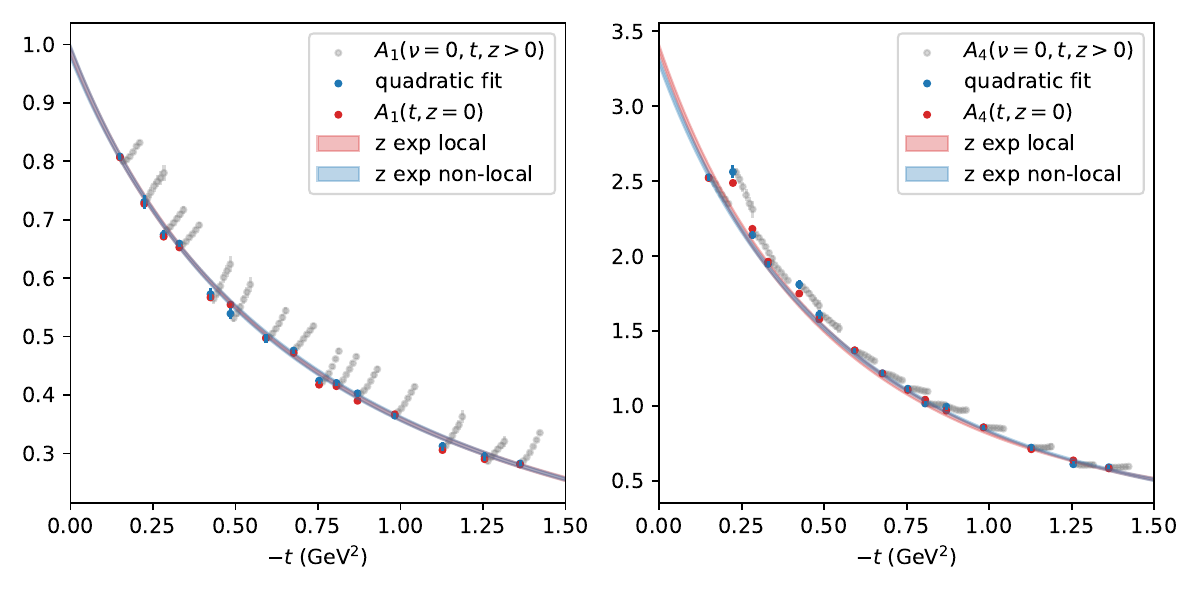}
    
    \caption{In grey, the series of 6 points $\mathcal{A}_{1,4}(\nu = 0, t, z^2)$ for $z \in \{a, ..., 6a\}$, in blue a quadratic (correlated) fit of the grey points, in red the data at $z = 0$. The grey points are shifted gradually in $t$ to allow us to observe their trend. For instance, at large $t$, there is no $z^2$ drift in $F_2$ on the right panel. The fit by a $z$-expansion \eqref{eq:zexp} is very similar whether one uses the local or non-local extraction. We merge the two cuts in lattice time separation.}
    \label{fig:nonlocalEFF}
\end{figure}

With this in mind, we can extract the elastic form factor using only non-local matrix elements, and performing quadratic fits over $z \in \{a, ..., 6a\}$. When using kinematics that are not the special $p_{f,3} = p_{i,3} = 0$, since $\nu \neq 0$ and $\xi \neq 0$, the matrix element contains more than just the elastic form factor. But when the higher-order moments are fitted jointly, as we will describe in the next section, the entire kinematic dataset can be used to constrain the elastic form factor using the non-local data. The result is shown in Fig.~\ref{fig:nonlocalEFF}. In light grey, the drifting value of $\mathcal{A}_{1,4}$ with increasing $z$ is depicted, while its value at $z = 0$ is represented by the red dots. The fit of the grey dots by a quadratic is depicted as the blue dots, barely distinguishable from the $z = 0$ data, or the result of a constant fit. The final fit by a $z$ expansion of either the local (red) or non-local (blue) elastic form factor extraction is completely compatible, although the $\chi^2$ per d.o.f. are typically significantly smaller when using the non-local extraction. We report the results of the non-local extraction of elastic form factors in Table \ref{resEFF} as well, and the dipole result serves as the standard result reported at the beginning of this document in Fig.~\ref{fig:summary}. One will notice that the pattern of $z^2$ dependence is clearly different between the $F_1$ and $F_2$ elastic form factors.

\section{Generalized Form Factors\label{sec:GFFs_num}}

\begin{figure}[ht!]
    \centering
    \includegraphics[width=0.99\linewidth]{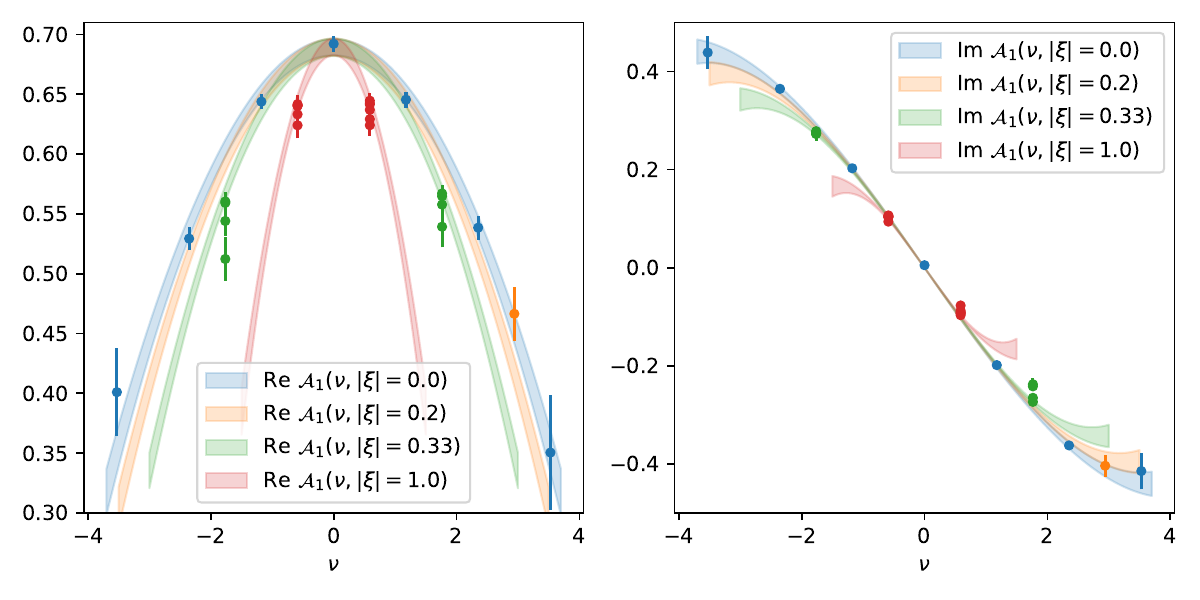}

\caption{The values of $\mathcal{A}_1(\nu, |\xi|, t \approx -0.33$ GeV$^2, z = 6a)$ with a cut $t_s = 3$ inside a single bin for the real (left) and imaginary (right) parts, along with the results of fits by the functional forms of Eqs. \eqref{eq:refit} and \eqref{eq:imfit}. We take the absolute value of $\xi$ to simplify the plot as we expect $\mathcal{A}_1$ to be even in $\xi$.\label{fig:inside_bin}}
\end{figure}

From the discussion in section \ref{sec:GFFs}, we expect that
\begin{align}
    \textrm{Re}\,\mathcal{A}_1(\nu, \xi, t, z^2) &= F_1(t) - \frac{\nu^2}{2} \bigg(A_{3,0}(t, z^2) + \xi^2 A_{3,2}(t, z^2)\bigg) + \mathcal{O}(\nu^4)+ \mathcal{O}(\Lambda_{\rm QCD}^2 z^2, t z^2)\,,\label{eq:refit}\\
    \textrm{Im}\,\mathcal{A}_1(\nu, \xi, t, z^2) &= -\nu A_{2,0}(t, z^2) + \frac{\nu^3}{6} \bigg(A_{4,0}(t, z^2) + \xi^2 A_{4,2}(t, z^2)\bigg) + \mathcal{O}(\nu^5) + \mathcal{O}(\Lambda_{\rm QCD}^2 z^2, t z^2)\,, \label{eq:imfit}
\end{align}
up to power corrections and lattice spacing systematic uncertainties. A similar relation is obtained by substituting the appropriate combination to form the $E$ GPD $\mathcal{A}_4 + \nu \mathcal{A}_6 - 2\xi\nu\mathcal{A}_7$ and its $B_{n,k}$ generalized form factors. The question of the $D$-term extracted from $\mathcal{A}_5$ is discussed in section \ref{Dtermanalysis}. We fit these relation at a fixed value of $z$ in each bin, with an intrabin $t$ correction proceeding in the same fashion as what we have described for the elastic form factors. The result using all points in the bin $t \approx -0.33$ GeV$^2$, $z = 6a$, and the cut $t_s = 3$ is displayed in Fig. \ref{fig:inside_bin}. 

\begin{figure}[ht!]
    \centering
    \includegraphics[width=0.49\linewidth]{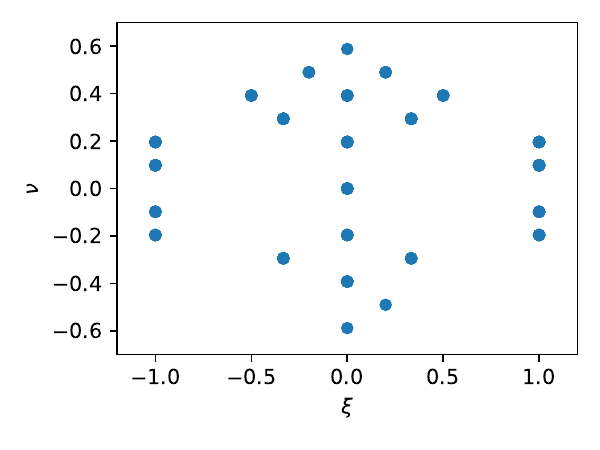}

\caption{The distribution of available values of $(\xi, \nu)$ where $\nu$ is presented at $z = a$.\label{fig:nu_xi}}
\end{figure}

The $\chi^2$ per d.o.f. of this fit of 28 points with 3 parameters is close to 2 both for the real and imaginary parts. Note that although it may seem that the special kinematic point at $\nu = 0$ corresponding to $\vec{p}_f = (0,0,0)$ and $\vec{p}_i = (1,1,0)$ is dragging the value of $F_1(t)$ upwards, this is not the case. Excluding this point from the fit would give $F_1(t, z=6a) = 0.687\pm 0.008$, whereas with it included, we find the very similar $F_1(t, z=6a) = 0.689\pm 0.007$. As we have already discussed, the data favor a value of $F_1(t)$ at $z = 6a$ that is noticeably larger than the data at $z = 0$ prefers, namely $0.652 \pm 0.007$, that is $5\sigma$ below. One will also notice that data at larger $\xi$ exists at smaller values of $\nu$ since part of the average $P_3$ momentum is sacrificed to create the momentum transfer. The coverage in $(\xi, \nu)$ available for our study is depicted in Fig.~\ref{fig:nu_xi}, with the exception of a few points where $p_{f,3} = -p_{i,3} \neq 0$, corresponding to $\nu = 0$ and $\xi$ infinite (but $\nu\xi \neq 0$). As one can see from Eqs.~\eqref{eq:refit} and \eqref{eq:imfit}, those kinematics probe exclusively the elastic form factor and the skewness-dependent generalized form factors, so they are particularly useful.

\subsection{Matching}

\begin{figure}[ht!]
    \centering
    \includegraphics[width=0.8\linewidth]{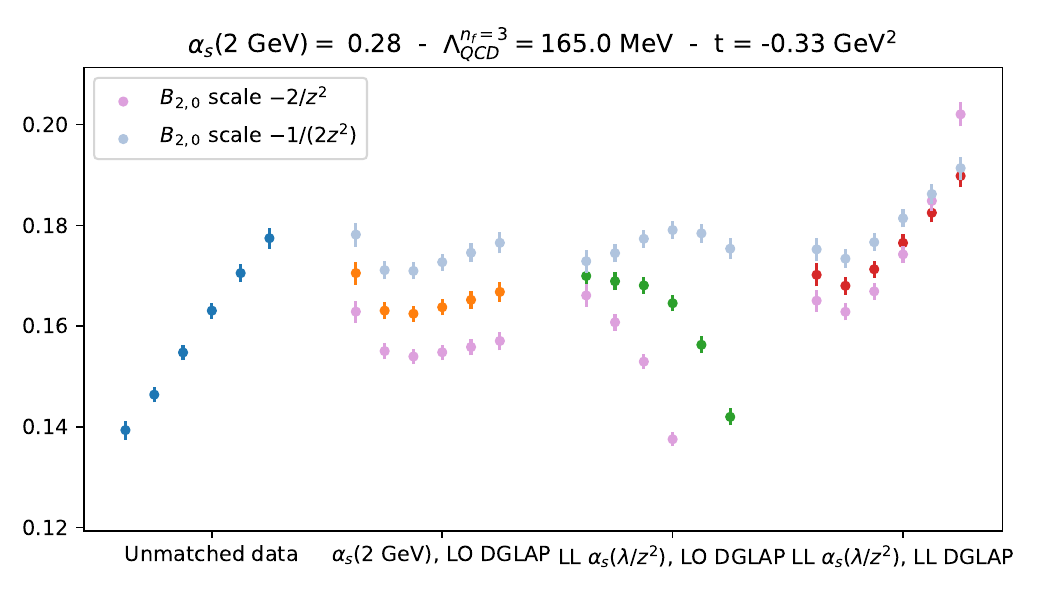}

\caption{The gravitational form factor of the $H$ GPD at $t = -0.33$ GeV$^2$ (cut $t_s = 3$) is an increasing function of the separation for $z \in \{a, ..., 6a\}$ (blue points). We show three different attempts at matching it to an $\overline{MS}$ scale of 2 GeV, using $\alpha_s(2 \textrm{ GeV}) = 0.28$ according to the three matching formulas \eqref{eq:matching1} - \eqref{eq:matching3}. The greyish points represent an account of scale variation, which is considerable compared to the statistical uncertainty in this well-constrained bin. \label{fig:matching_GFF}}
\end{figure}

At higher order and in contrast to the elastic form factors, the generalized form factors are scale-dependent, and we have derived in section \ref{sec:GFFs} a perturbative prediction for their $z^2$ dependence. The matching of the gravitational form factors $A_{2,0}, B_{2,0}$ is particularly interesting to observe because our data are very precise. Let us use again the same bin as in the previous discussion. We try three different leading-order/leading-logarithmic matching procedures:
\begin{enumerate}
\item The procedure that arises immediately from the fixed order matching in Eq.~\eqref{eq:gitd-pgitd-matching}:
\begin{align}
A_{2,0}(t, \mu^2) &= A_{2,0}(t, z^2) \bigg[1+\frac{\alpha_s(\mu^2) C_F}{2\pi} \left(\gamma_1 L + d_1\right)\bigg]\,, \label{eq:matching1}
\end{align}
where $\gamma_1 = -4/3$ is the DGLAP anomalous dimension, $d_1 = 14/3$ the moment of the non-logarithmic part of the matching kernel, and $L = \ln(e^{2\gamma_E+1}/4 \times (z^2/\lambda) \times \mu^2)$. Here $\lambda$ should be close to $-1$ and will be varied in the interval $[-1/2, -2]$ as is usual to evaluate uncertainty in scale fixing.
\item A variant in which we evaluate $\alpha_s$ using its leading-logarithmic 3-flavor evolution at the ``natural'' scale of the data $\lambda / z^2$:
\begin{align}
A_{2,0}(t, \mu^2) &= A_{2,0}(t, z^2) \bigg[1+\frac{\alpha_s(\lambda / z^2) C_F}{2\pi} \left(\gamma_1 L + d_1\right)\bigg]\,.
\end{align}
\item A full leading-logarithmic matching as derived in section \ref{sec:GFFs}:
\begin{align}
A_{2,0}(t, \mu^2) &= A_{2,0}(t, z^2) \left(\frac{\alpha_s(\lambda / z^2)}{\alpha_s(\mu^2)}\right)^{\gamma_1 C_F/(2\pi\beta_0)} \bigg[1+\frac{\alpha_s(\lambda/z^2) C_F}{2\pi} \left(\gamma_1\ln\left(\frac{e^{2\gamma_E+1}}{4}\right) + d_1\right)\bigg]\,.
\label{eq:matching3}
\end{align}
\end{enumerate}
We use the $\overline{MS}$ scale $\mu = 2$ GeV and present the result of the three matching procedures in Fig.~\ref{fig:matching_GFF}. We use $\alpha_s(\mu^2) = 0.28$, a value compatible with the usual $\overline{MS}$ value at that scale. The relative arbitrariness of this choice is balanced by the possibility of performing scale variations. 

By construction, the first matching procedure does not present any divergence except for $-z^2 \rightarrow \infty$ and can be applied to as large $-z^2$ as desired. We observe some inconsistency at small $z$, followed by a stabilization in the large $z$ regime. The second matching procedure, on the other hand, cannot be applied once as $\lambda / z^2$ reaches the $\Lambda_{QCD}$ divergence. With $\alpha_s(2$~GeV$)=0.28$ and three active flavors, $\Lambda_{QCD} = 165$ MeV, which allows us nonetheless to apply the matching to all our values of $z^2$. However, the large values result in an extremely large-scale fixing uncertainty, as one would expect. 

\begin{figure}[ht!]
    \centering
    \includegraphics[width=0.8\linewidth]{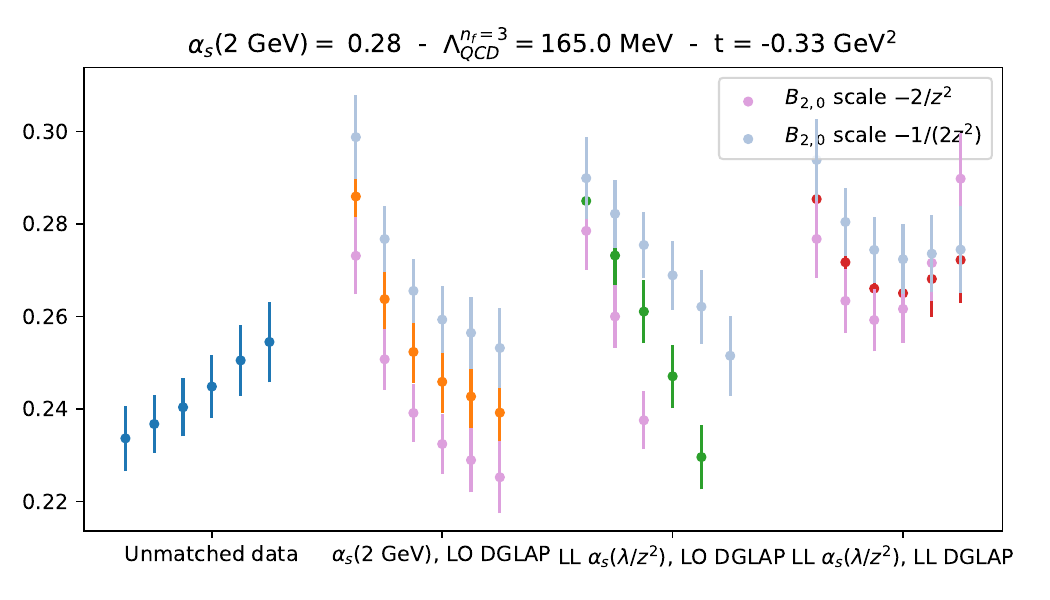}

\caption{Same as Fig.~\ref{fig:matching_GFF}, except for the $E$ GPD. The statistical uncertainty is worse than the $H$ GPD. \label{eq:mathcingE}}
\end{figure}

The full leading-logarithm calculation exhibits the smallest scale fixing uncertainties of the three matching procedures. Although $\alpha_s(\lambda/z^2)$ increases considerably at large $z$, since $\gamma_1$ is negative, the ratio $(\alpha_s(\lambda/z^2) / \alpha_s(\mu^2))^{\gamma_1 C_F/(2\pi\beta_0)}$ decreases. On the other hand, the final factor of Eq. \eqref{eq:matching3} increases at large $z$, and both variations happen in quite similar proportions. The procedure produces relatively constant results up to $z = 0.3$ fm, after which it is no longer constant in this bin. In principle, we do not have expectations that a perturbative matching explains the $z^2$ dependence of the data when $-1/z^2 \ll 1$ GeV$^2$. Therefore, this situation is not particularly concerning. In fact, when applied to the $E$ GPD in the same bin, the full leading-logarithmic matching still has the smallest scale fixing uncertainties, and overall the best behavior as seen in Fig.~\ref{eq:mathcingE}. At this stage, considering that different behaviors are observed for various bins, no strong conclusion can be drawn, especially in view of the $z^2$ dependence observed in the elastic form factors.

Only a few bins of the generalized form factor $A_{2,0}$ are precise enough in terms of statistics and systematic difference between the cut $t_s = 3$ and 4 that scale variation in the leading-logarithmic matching produces a visible difference. Overall, scale variation does not produce a significant contribution to the uncertainty budget given the other uncertainties once the $t$-dependence is reconstructed in the final fit. We will therefore stick to the full leading-logarithmic matching with $\lambda = -1$ in the following. Even if this is not a practical matter at the current stage of our calculations, we foresee that applying perturbation theory at large $z$ is not good practice when all uncertainties are under control. On the other hand, in the event that our framework retains a strong sensitivity to the leading-twist operator even at scales where perturbation theory becomes dubious, we have advocated that evolution operators in $z^2$ should be computed on the lattice in \cite{Dutrieux:2023zpy}. However, this program requires a higher confidence in the suppression of power corrections and other lattice systematic uncertainties that we do not believe to have been achieved in the present calculation.

\begin{figure}[ht!]
    \centering
    \includegraphics[width=0.89\linewidth]{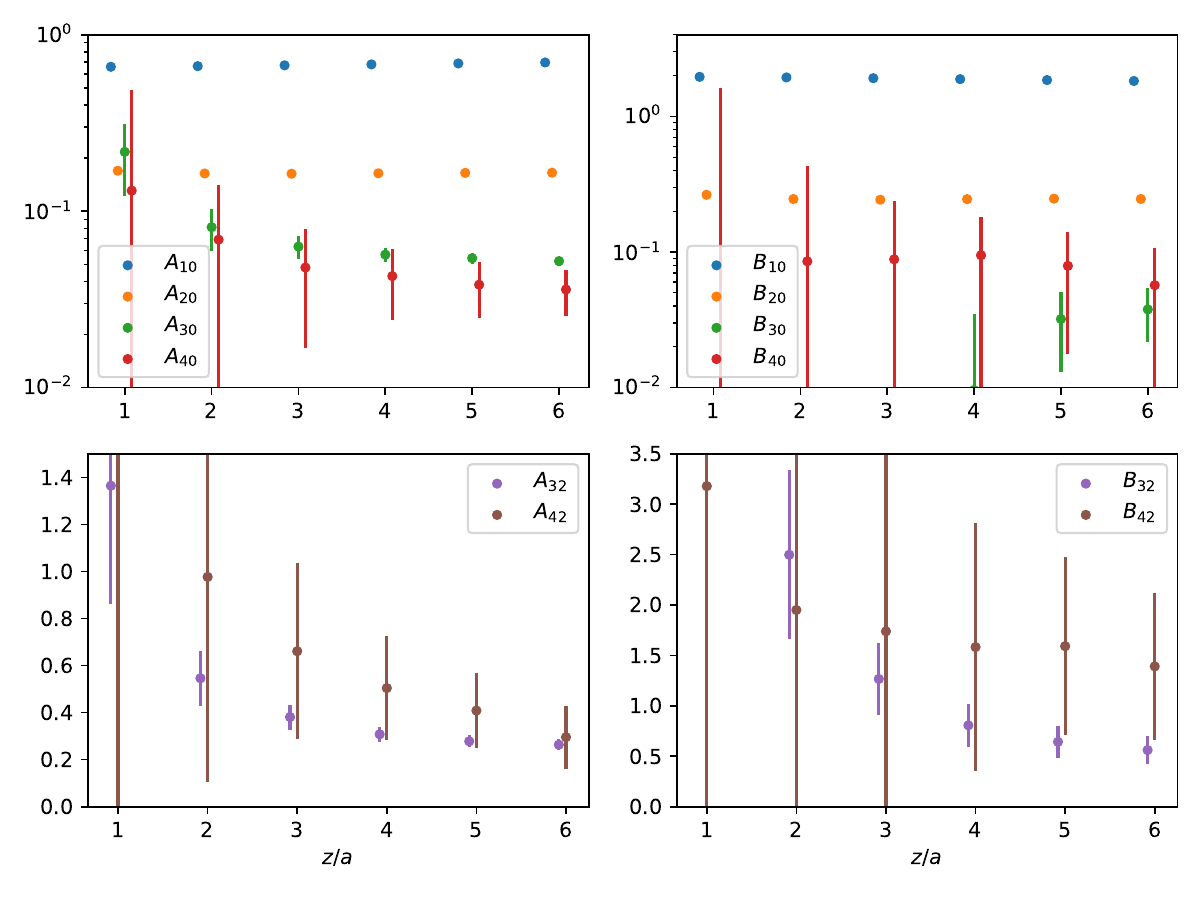}

\caption{The moments of the GPDs $H^{u-d}$ (left) and $E^{u-d}$ (right) at $t = -0.33$ GeV$^2$ with leading-logarithmic matching to $\mu = 2$ GeV as a function of $z/a$ (cut $t_s = 4$). In the upper row, we show the skewness-independent moments, and in the lower row the skewness-dependent moments. \label{fig:matched_moments}}
\end{figure}

We present in Fig.~\ref{fig:matched_moments} an overall picture of the matched moments in one bin in $t$. The importance of the large $z$ data (and therefore large Ioffe time) to constrain the higher moments is obvious.  

\subsection{$t$-dependent generalized form factors}

\begin{figure}[ht!]
    \centering
    \includegraphics[width=0.9\linewidth]{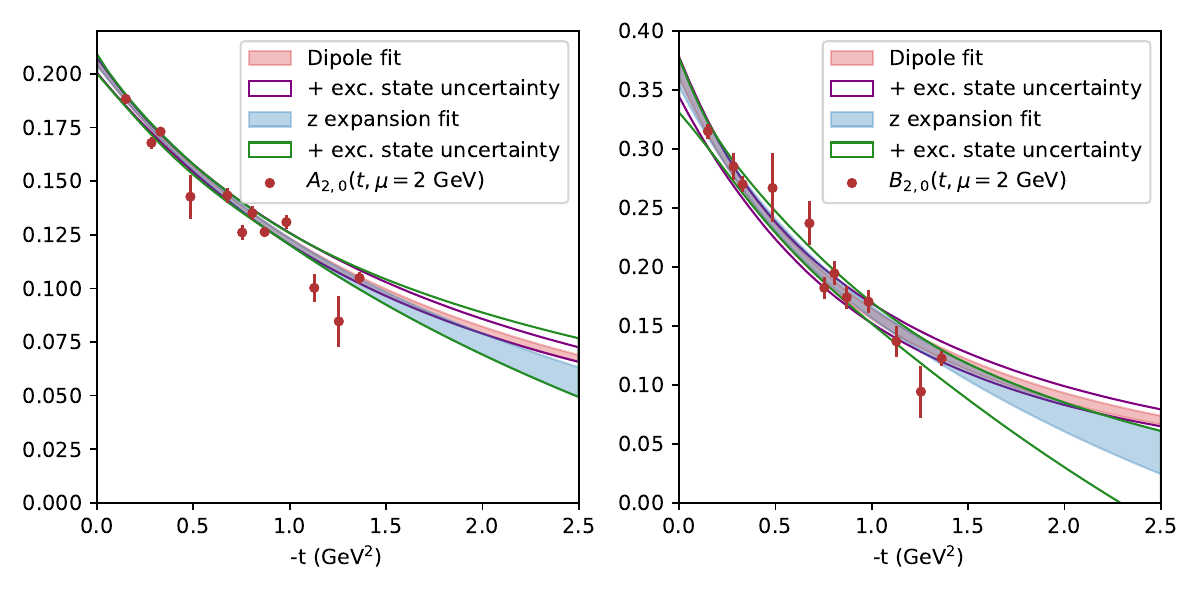}
\caption{The generalized form factors $A_{2,0}$ (left) and $B_{2,0}$ (right) as a function of $t$. The data points are obtained for a time separation cut of three in the analysis of correlation functions. The fit of those data points, either by a dipole or a $z$ expansion, is represented by the uniformly colored areas. When merging this uncertainty with the one using a time separation cut of four, one obtains the wider uncertainties, which contain information about excited state contamination.\label{fig:A20}}
    
\end{figure}

\begin{figure}[ht!]
    \centering
\includegraphics[width=0.9\linewidth]{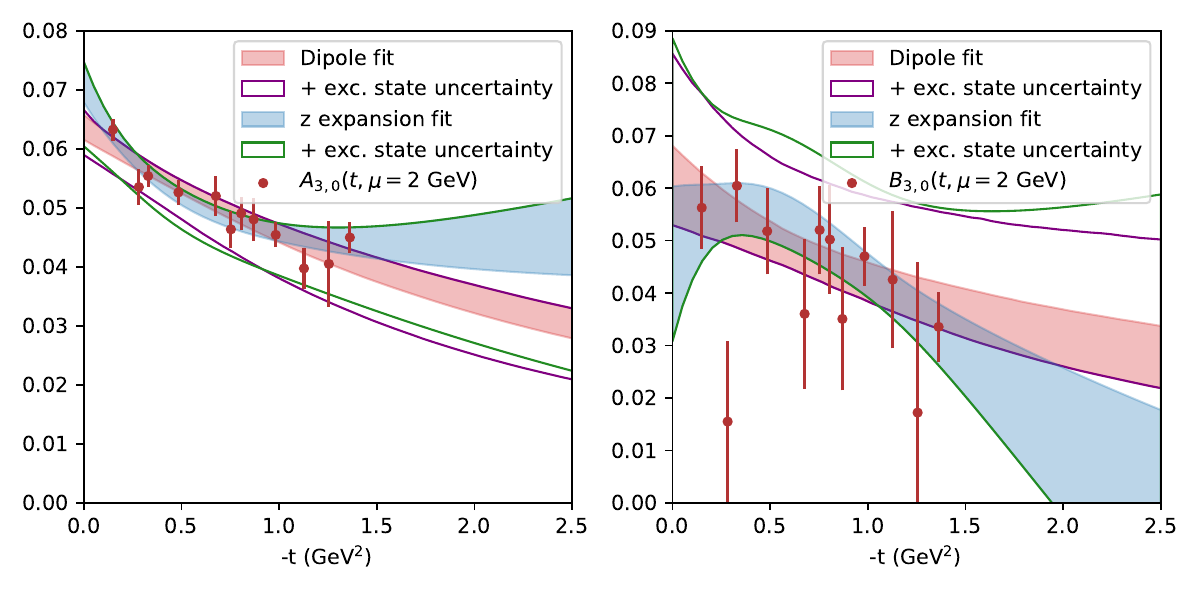}
\caption{The generalized form factors $A_{3,0}$ (left) and $B_{3,0}$ (right) (see caption of Fig. \ref{fig:A20}) \label{fig:A30}}
    
\end{figure}

\begin{figure}[ht!]
    \centering
\includegraphics[width=0.9\linewidth]{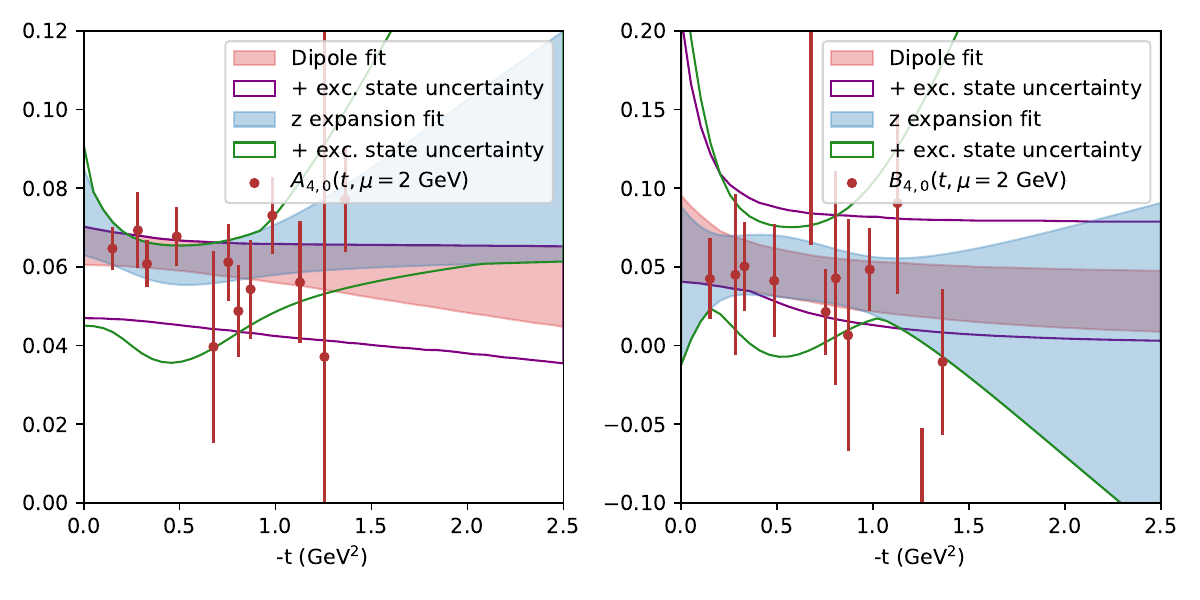}
\caption{The generalized form factors $A_{4,0}$ (left) and $B_{4,0}$ (right) (see caption of Fig. \ref{fig:A20}) \label{fig:A40}}
    
\end{figure}

\begin{figure}[ht!]
    \centering
    \includegraphics[width=0.9\linewidth]{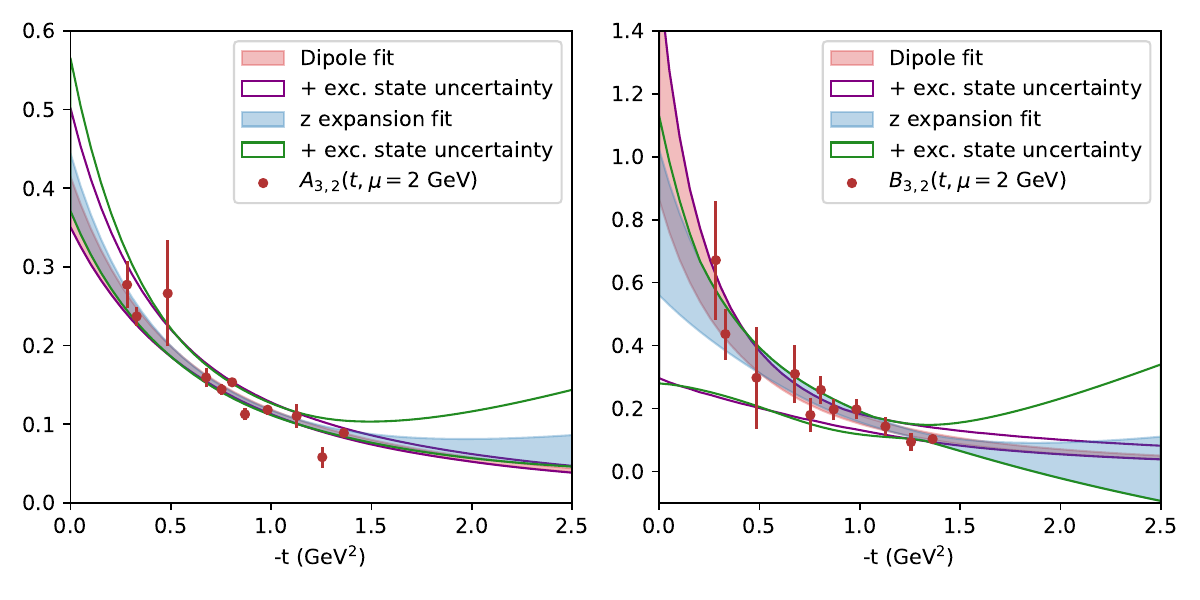}
\includegraphics[width=0.9\linewidth]{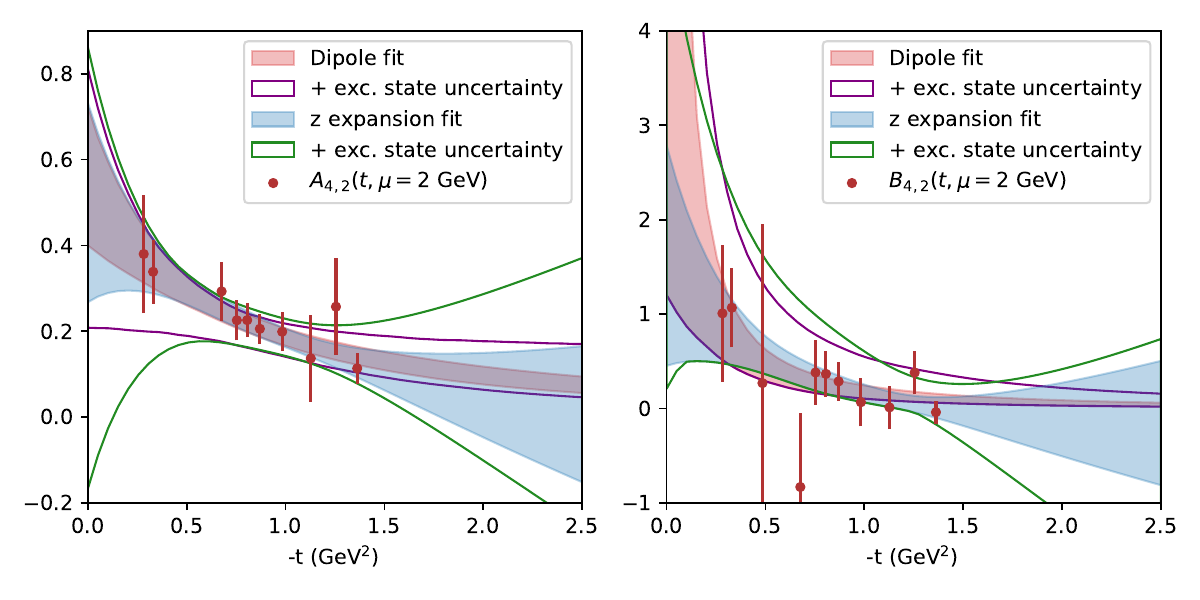}

    \caption{The skewness-dependent generalized form factors $A_{3,2}$ (top left), $B_{3,2}$ (top right), $A_{4,2}$ (bottom left) and $B_{4,2}$ (bottom right)  (see caption of Fig. \ref{fig:A20}). \label{fig:skewdep}}
    
\end{figure}

The results of the fits of the $t$-dependence of moments of GPDs are presented in Figs.~\ref{fig:A20} to \ref{fig:A40} for the generalized form factors $A_{n,0}$ and $B_{n,0}$ that are independent of $\xi$, and in Fig.~\ref{fig:skewdep} for the $A_{n,2}$ and $B_{n,2}$ that are dependent on $\xi$. Since beyond the elastic form factor, the trend in $z$ of the data is not discernible in a consistent fashion, we use a constant fit on $z \in \{a, ..., 6a\}$ after having matched the data to $\overline{MS}$ at 2 GeV using the leading-logarithmic matching derived in section \ref{sec:GFFs}. We demonstrate in Appendix \ref{differentz} that the correlated constant fit gives very robust results on a diverse subset of separations, for instance using only $z \in \{a, ..., 4a\}$. The colored bands are 68\% confidence intervals obtained by dropping the lower and upper 16\% of the best-fit results on the data with a cut in lattice time separation of three. This makes practically no difference compared to a simple $1\sigma$ uncertainty band, except for the most poorly determined moments $B_{4,0}$ and $A_{4,2}$, $B_{4,2}$ where the non-linearity of the dipole fit becomes apparent and the distributions of best fits are non-Gaussian. The uncolored band results from the union of the result with a cut of three and a cut of four. It appears that excited state uncertainty increases as we head towards larger moments, unsurprisingly as the signal in the data weakens. The obtained results for the dipole fit in terms of value at $t = 0$ and mass are reported in Fig.~\ref{fig:summary}.

\subsection{Radial distributions\label{sec:raddist}}

One of the main physical motivations behind the study of GPDs is their characterization of radial distributions of partonic properties inside hadrons (momentum, energy, pressure, etc.). Such distributions are obtained by a Fourier transform of the $t$-dependence of the GPD, usually at zero skewness \cite{Burkardt:2002hr}. The impact parameter distribution of an unpolarized quark in an unpolarized proton is given by
\begin{equation}
    I(x, \vec{b}_\perp) = \int \frac{\mathrm{d}^2 \vec{\Delta}_\perp}{(2\pi)^2} \,e^{-i \vec{b}_\perp \cdot \vec{\Delta}_\perp} H(x, \xi = 0, t=-\vec{\Delta}^2_\perp)\,,
\end{equation}
whereas the GPD $E$ allows one to characterize the unpolarized quark distribution inside a transversely polarized proton,
\begin{equation}
    I^T(x, \vec{b}_\perp) = \int \frac{\mathrm{d}^2 \vec{\Delta}_\perp}{(2\pi)^2} \,e^{-i \vec{b}_\perp \cdot \vec{\Delta}_\perp} \bigg[H(x, \xi = 0, t=-\vec{\Delta}^2_\perp)+ i\frac{\Delta_y}{2m}E(x, \xi = 0, t = -\vec{\Delta}^2_\perp)\bigg]\,.
\end{equation}
Using the model-dependent extrapolation to any $t$ provided by the previous fits, we can extract the first moments of the isovector component of these distributions,
\begin{align}
I^{u-d}_{n+1}(\vec{b}_\perp) &= \int \frac{\mathrm{d}^2 \vec{\Delta}_\perp}{(2\pi)^2} \,e^{-i \vec{b}_\perp \cdot \vec{\Delta}_\perp} A^{u-d}_{n+1,0}(-\vec{\Delta}^2_\perp)\,,\label{eq:unpolID}\\
I^{u-d, T}_{n+1}(\vec{b}_\perp) &= \int \frac{\mathrm{d}^2 \vec{\Delta}_\perp}{(2\pi)^2} \,e^{-i \vec{b}_\perp \cdot \vec{\Delta}_\perp} \bigg[A^{u-d}_{n+1,0}(-\vec{\Delta}^2_\perp) + i\frac{\Delta_y}{2m}B^{u-d}_{n+1,0}(-\vec{\Delta}^2_\perp)\bigg]\,. \label{eq:transIPD}\\
\end{align}
In the following we use the dipole fits as the $t$-dependence of the $z$ expansion is too unconstrained at large $t$ to give rise to sensible radial distributions. The extraction presented here is therefore strongly model-dependent. Even then, the $t$-dependence of $A_{4,0}$ and $B_{4,0}$ is very poorly determined and therefore we only present results for $n \in \{0,1,2\}$. 

\begin{figure}[ht!]
    \centering
    \includegraphics[width=0.7\linewidth]{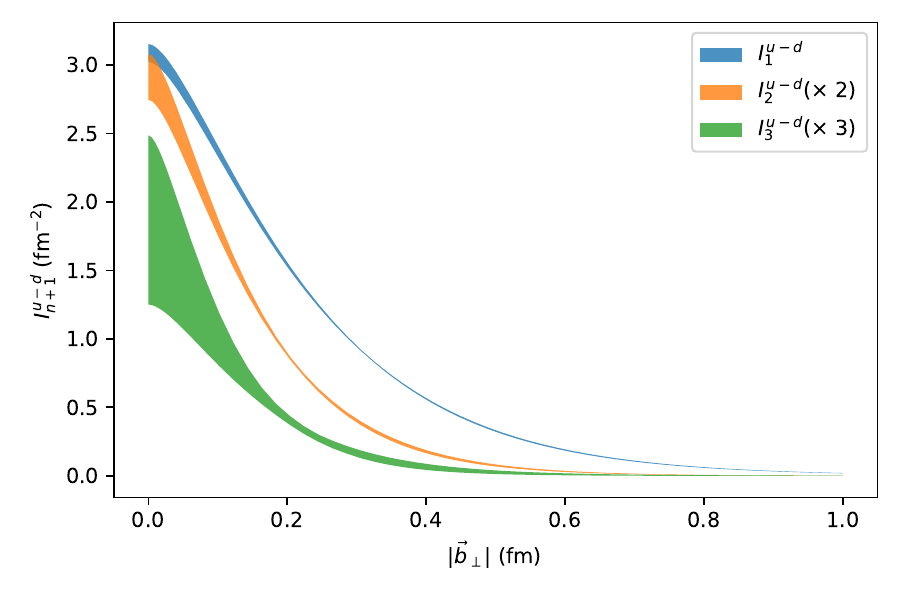}

    \caption{The unpolarized IPD moments as a function of $|\vec{b}_\perp|$, rescaled for visibility. The errorband are obtained from the dipole fits of the data merging both the cuts of 3 and 4 in Euclidean time. \label{fig:rad1d}}
    
\end{figure}

The unpolarized moments \eqref{eq:unpolID} are spherically symmetric in $|\vec{b}_\perp|$. We represent them in Fig.~ \ref{fig:rad1d}. As can already be inferred by the increasing value of the fitted dipole mass (see the summary table of Fig.~ \ref{fig:summary}), when the order of the skewness-independent moments increases, the radial extent of the distribution shrinks. This can be largely understood from a kinematic point of view, as higher-order Mellin moments are increasingly dominated by the large $x$ domain of the distribution. The center of the proton with respect to which our radial coordinate system is defined is the barycenter of the partonic longitudinal momentum. When the longitudinal momentum fraction of the active parton $x$ approaches one, the active parton is by necessity close to the center of the proton, so the impact parameter distribution at large $x$ is very narrow.

\begin{figure}[ht!]
    \centering
    \includegraphics[width=0.45\linewidth]{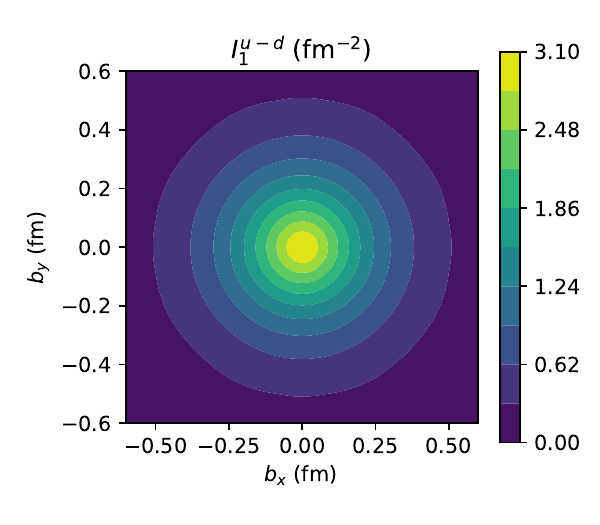}
\includegraphics[width=0.45\linewidth]{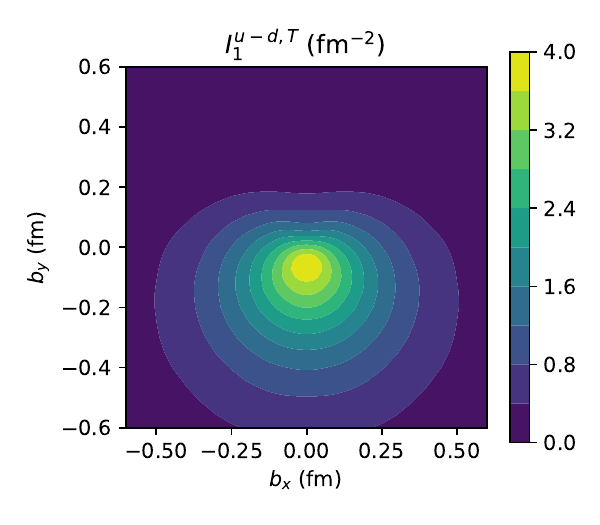}
\includegraphics[width=0.45\linewidth]{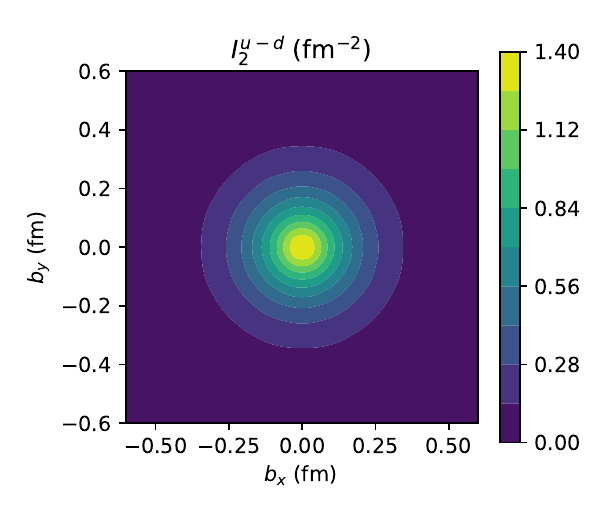}
\includegraphics[width=0.45\linewidth]{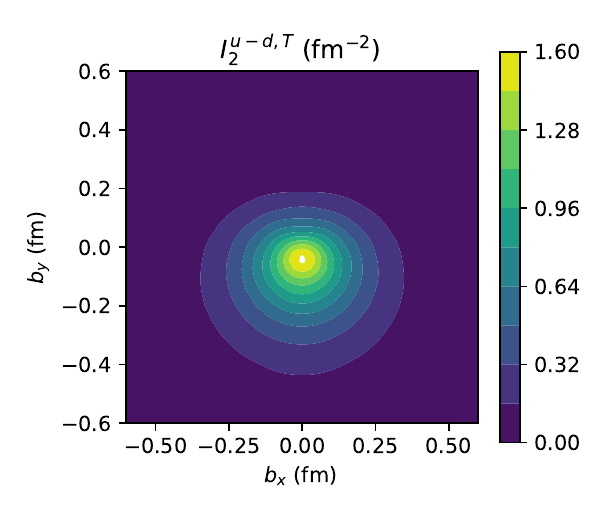}
\includegraphics[width=0.45\linewidth]{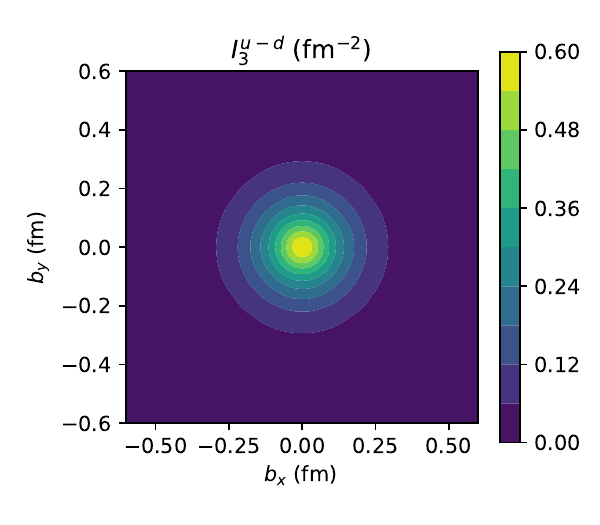}
\includegraphics[width=0.45\linewidth]{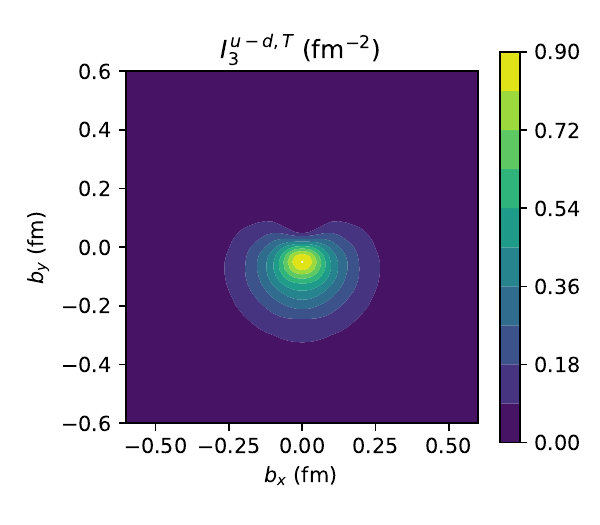}

\caption{The moments $n = 0, 1, 2$ (respectively first, second and third row) of the unpolarized $u-d$ quark combination in an unpolarized proton (left) and in a transversely polarized proton (right). \label{fig:ipd2d}}
    
\end{figure}

The transverse impact parameter distribution moments \eqref{eq:transIPD} are not spherically symmetric. We represent them, along with the unpolarized moments, in Fig.~\ref{fig:ipd2d}. We only use the central value of the dipole fits. A subtle feature of the transverse plots, which can also be observed in \cite{Bhattacharya:2023nmv}, is the fact that moments with $n$ even (extracted from the imaginary part of the Ioffe-time distributions) seem to show a lesser distortion in the transverse impact parameter distribution than the odd moments. In \cite{Bhattacharya:2023nmv}, where the $u+d$ component is extracted as well, it also seems to hold for $n = 3$. Here again, the distribution shrinks at large $n$. One can notice that perturbation theory predicts the GPD to be independent of $t$ when $x \rightarrow 1$ \cite{Yuan:2003fs}:
\begin{equation}
    H^q(x, \xi, t) \sim \frac{(1-x)^3}{(1-\xi^2)^2}\,,\ \ \ E^q(x, \xi, t) \sim \frac{(1-x)^5}{(1-\xi^2)^2}\,.
\end{equation}
This is equivalent to the large $n$ GPD moments becoming independent of $t$, or the impact parameter distribution becoming a simple Dirac peak at $\vec{b}_\perp = 0$.

\subsection{$D$-term \label{Dtermanalysis}}

\begin{figure}[ht!]
\centering
\includegraphics[width=0.9\linewidth]{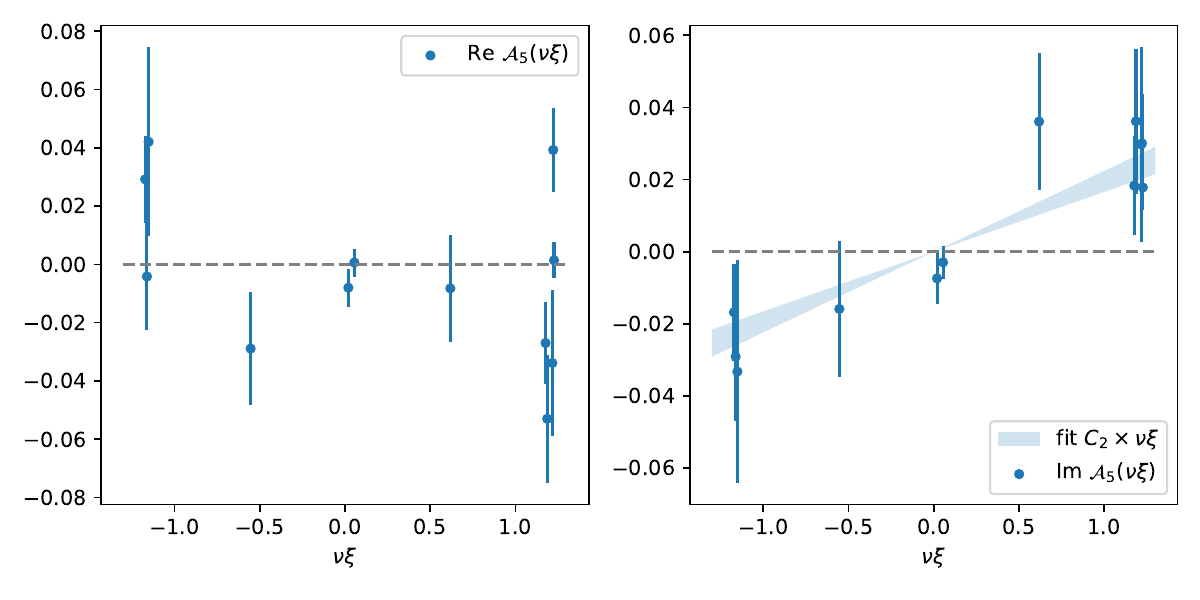}
\caption{The values of $\mathcal{A}_5(\nu, \xi, t \approx -1.36$ GeV$^2, z = 6a)$ with a cut $t_s = 3$ inside a single bin for the real (left) and imaginary (right) parts, along with the results of fits by the functional form of Eq.~\eqref{eq:Dtermfit}. We present the data as a function of $\xi\nu$ instead of the separate $\xi$ and $\nu$.\label{fig:Dtermbin}}    
\end{figure}

For the isovector $D$-term, Eq. \eqref{eq:Dtermdecomp} invites us to fit:
\begin{equation}
\textrm{Im}\,\mathcal{A}_5(\nu, \xi, t, z^2) = \xi\nu C_2(t, z^2) + \mathcal{O}(\nu^3) + \mathcal{O}(\Lambda_{\rm QCD}^2 z^2, t z^2)\,. \label{eq:Dtermfit}
\end{equation} As we have noticed before, there is a trade-off between the value of $\xi$ and $\nu$ as larger $\xi$ requires sacrificing some of the $P_3$ momentum to introduce a momentum transfer. As a result, the range in $\nu \xi$ available to extract the $D$-term does not exceed 1.2 when $z = 6 a$, and we can only achieve sensitivity to the $C_2$ moment of the $D$-term. The situation in one of the bins in $t$ is depicted in Fig. \ref{fig:Dtermbin}. Because of the absence of any discernable trend in $t$, we do not attempt an intrabin $t$ correction in the fashion we used for the rest of the GPD.

In general, $\mathcal{A}_5$ is compatible with 0 and small in magnitude. Some kinematics $(\vec{p}_f, \vec{p}_i)$ show noticeably non-zero values in the real part, a problem we have already identified when discussing the contribution of $\mathcal{A}_5$ in the local matrix element. Therefore, there is a clear sign of potential lattice discretization errors, which we have no means of subtracting from the data within this analysis. We also note that Ref. \cite{Radyushkin:2023ref} highlights the potential presence of ``kinematic'' higher-twists in $\mathcal{A}_5$. When using the full statistics available in some bins, it is possible to fit the imaginary part with the functional form of Eq.~\eqref{eq:Dtermfit} and obtain an extraction for $C_2$ only a few standard deviations away from 0. The imaginary part of $\mathcal{A}_5(\nu, \xi)$ is compatible with being an odd function of the product $\nu\xi$, which mirrors the fact that the $D$-term is an odd function of $x / \xi$. The matching of $C_2$ then proceeds exactly similarly to the one of $A_{2,0}$ and $B_{2,0}$ \eqref{eq:Dtermdecomp}.

\begin{figure}[ht!]
\centering
\includegraphics[width=0.85\linewidth]{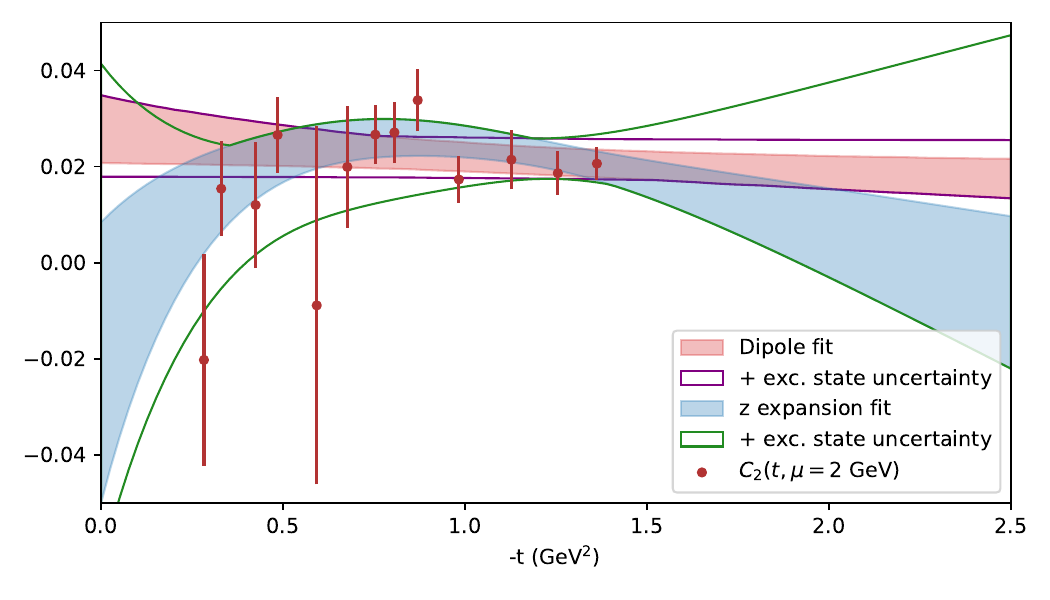}
\caption{The isovector $D$-term form factor $\int d\alpha\,\alpha D(\alpha)$.\label{fig:C2}}    
\end{figure}

The final result on $C_2$ is depicted in Fig. \ref{fig:C2} matched to 2 GeV using the data $z = a$ up to $6a$. The dipole and the $z$-expansion fit produce very different $t$-dependence and extrapolation in the limit where $t \rightarrow 0$, due to the weakness of the constraint imposed by the data. The small magnitude of $C_2$ is consistent with the results of other lattice calculations from local matrix elements, \textit{e.g.} \cite{LHPC:2007blg, Bali:2018zgl,Alexandrou:2019ali}. In the large $N_c$ approximation, one can note that the $u-d$ contribution to the $D$-term is suppressed by a factor $1 / N_c$ compared to the $u+d$ flavor combination \cite{Goeke:2001tz}.

\section{Conclusions\label{sec:conclusion}}
In this work, we present the calculation of off-forward nucleon matrix elements with extensive coverage in initial and final momenta. These matrix elements scan a wide range of $t$ and $\xi$ relevant for understanding the three-dimensional tomography of the nucleon, through a factorization relationship to GPDs analogous to those for DVCS and DVMP. This lattice calculation was made possible by utilizing the technique of distillation to explicitly project onto the momentum states for all operators in the correlation functions. In addition to significantly improving the signal, this procedure provides a natural method for an efficient calculation by allowing many momenta to be calculated, while recycling the components for each operator.

With the current statistical precision, control of excited state contamination is critical for an accurate analysis. We have compared results from two different methods for extracting matrix elements and from different cuts on the data. This step is critical for creating realistic error estimates for the matrix elements. The data with different cuts in Euclidean time can lead to significant deviations in the final results, as was seen in the extraction of generalized form factors. In future studies, we will use the distillation to implement distinct nucleon operators for each momentum. This allows for a GEVP analysis to explicitly remove contributions from excited states. The excited state contamination will only become more difficult to control in the chiral limit, so this full control of excited state contamination will be crucial for physical extrapolations. 

By using this Lorentz-invariant approach, the $\nu$ dependence of this data can be analyzed to study the Mellin moments of the GPDs. In this study, we have utilized relatively small momenta in the direction of the Wilson line, which scans the low $\nu$ region that is dominated by the lowest few moments. We extract the moments of $H$ and $E$ and study the $t$ dependence and, for the first time, the $\xi$ dependence. The $\xi^2$ coefficients of the moments $A_{n,2}$ and $B_{n,2}$ are larger than the $\xi^0$ coefficients $A_{n,0}$ and $B_{n,0}$. We also extracted radial distributions at zero skewness by performing the Fourier transform of the $t$ dependence. 

The use of non-local separations in the matrix elements up to $z = 0.6$ fm gives rise to challenges for the proper evaluation of the systematic uncertainty caused by the perturbative matching, as well as the contamination by higher twist contributions, which we do not have the possibility to properly treat in this study. In view of some clear signal of lattice discretization errors, which we cannot easily differentiate from higher twist contributions, we see the realization of a continuum limit of this study as an important further milestone.

This calculation represents a step on the road to a systematically controlled calculation of nucleon tomography through GPDs. Nucleon tomography requires significantly more matrix elements, to scan the full three-dimensional space of GPDs than determinations of PDFs, particularly for a full study of systematic errors. We have demonstrated a method capable of tackling the problem of kinematical coverage and for studying the significant impact of excited state contamination.

\section{Acknowledgments}
This project was supported by the U.S.~Department of Energy, Office of Science, Contract No.~DE-AC05-06OR23177, under which Jefferson Science Associates, LLC operates Jefferson Lab. KO and HD were supported in part by U.S.~DOE Grant \mbox{\#DE-FG02-04ER41302}. CM is supported in part by U.S.~DOE ECA \mbox{\#DE-SC0023047}.
AR acknowledges support by U.S.~DOE Grant \mbox{\#DE-FG02-97ER41028}. The work of HD and DR was conducted in part under the Laboratory Directed Research and Development Program (LDRD 2412) at Thomas Jefferson National Accelerator Facility for the U.S. Department of Energy. This work has benefited from the collaboration enabled by the Quark-Gluon Tomography (QGT) Topical Collaboration, U.S.~DOE Award \mbox{\#DE-SC0023646}.
This research was funded, in part, by l’Agence Nationale de la Recherche (ANR), project ANR-23-CE31-0019. For the purpose of open access, the authors have applied a CC-BY public copyright licence to any Author Accepted Manuscript (AAM) version arising from this submission.
Computations for this work were carried out in part on facilities of the USQCD Collaboration, which are funded by the Office of Science of the U.S.~Department of Energy. This work was performed in part using computing facilities at William \& Mary which were provided by contributions from the National Science Foundation (MRI grant PHY-1626177), and the Commonwealth of Virginia Equipment Trust Fund. This work used the Extreme Science and Engineering Discovery Environment (XSEDE), which is supported by National Science Foundation grant number ACI-1548562. Specifically, it used the Bridges system, which is supported by NSF award number ACI-1445606, at the Pittsburgh Supercomputing Center (PSC) \cite{6866038, Nystrom:2015:BUF:2792745.2792775}. In addition, this work used resources at NERSC, a DOE Office of Science User Facility supported by the Office of Science of the U.S. Department of Energy under Contract \#DE-AC02-05CH11231, as well as resources of the Oak Ridge Leadership Computing Facility at the Oak Ridge National Laboratory, which is supported by the Office of Science of the U.S. Department of Energy under Contract No. \mbox{\#DE-AC05-00OR22725}. The software codes {\tt Chroma} \cite{Edwards:2004sx}, {\tt QUDA} \cite{Clark:2009wm, Babich:2010mu}, {\tt QPhiX} \cite{QPhiX2}, and {\tt Redstar} \cite{Chen:2023zyy} were used in our work. The authors acknowledge support from the U.S. Department of Energy, Office of Science, Office of Advanced Scientific Computing Research and Office of Nuclear Physics, Scientific Discovery through Advanced Computing (SciDAC) program, and of the U.S. Department of Energy Exascale Computing Project (ECP). The authors also acknowledge the Texas Advanced Computing Center (TACC) at The University of Texas at Austin for providing HPC resources, like Frontera computing system~\cite{10.1145/3311790.3396656} that has contributed to the research results reported within this paper. The authors acknowledge William \& Mary Research Computing for providing computational resources and/or technical support that have contributed to the results reported within this paper.

We acknowledge the EuroHPC Joint Undertaking for awarding this project access to the EuroHPC supercomputer LUMI, hosted by CSC (Finland) and the LUMI consortium through a EuroHPC Extreme Scale access call. This work also benefited from access to the Jean Zay supercomputer at the Institute for Development and Resources in Intensive Scientific Computing (IDRIS) in Orsay, France under project A0080511504.  

\appendix

\section{Full set of Lorentz amplitudes\label{ap:fullset}}

\begin{figure}[ht!]
    \centering
    \includegraphics[width=0.99\linewidth]{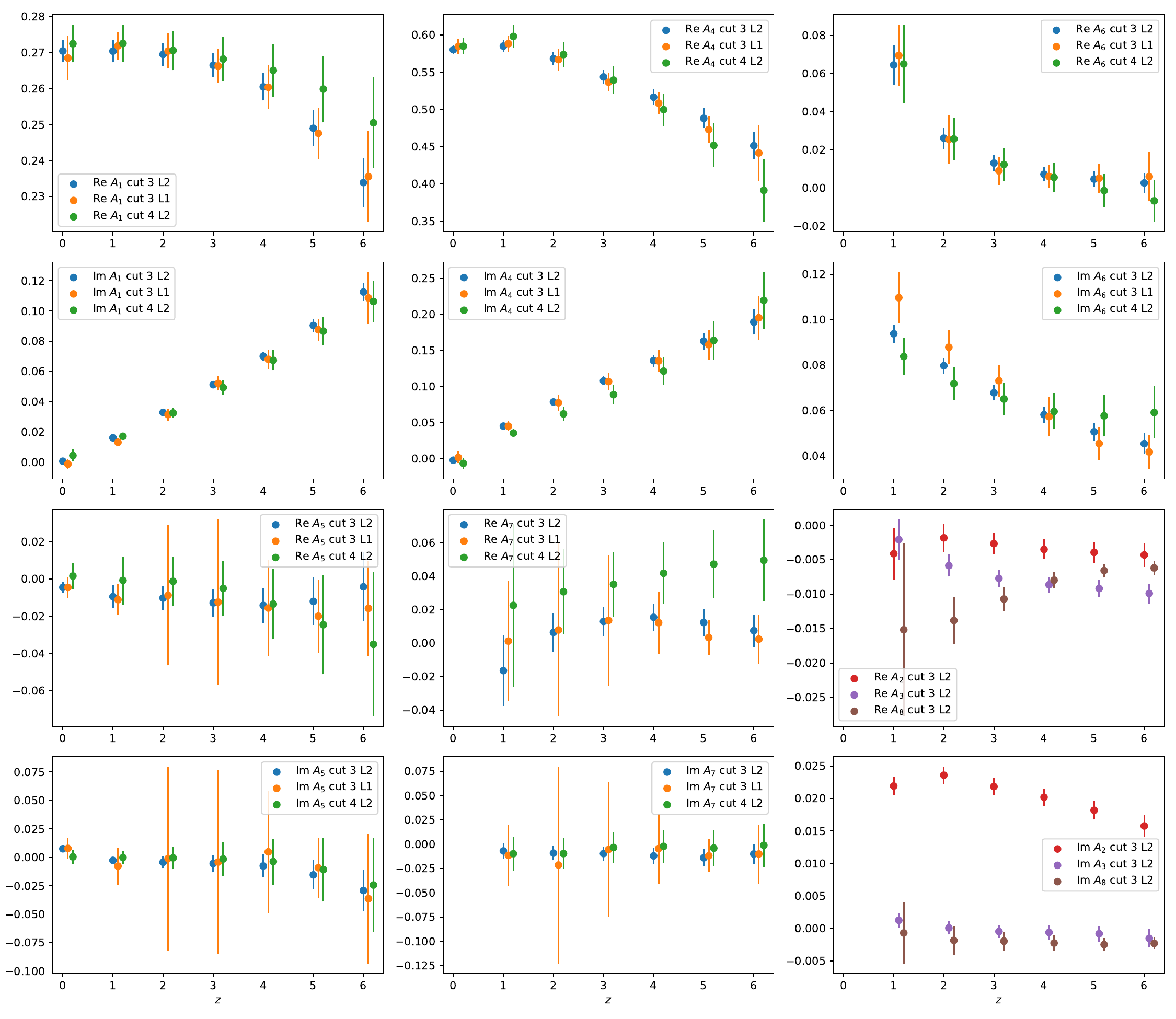}
\caption{Full set of amplitudes for $\vec{p}_f = (2,1,0)$ and $\vec{p}_i = (0, 1, 2)$. See a discussion of the features of the plot in the text. The first row presents the real parts of $\mathcal{A}_1$ (left), $\mathcal{A}_4$ (middle), and $\mathcal{A}_6$ (right), and the second row their imaginary parts. The third row presents the real parts of $\mathcal{A}_5$ (left), $\mathcal{A}_7$ (middle), and $\mathcal{A}_{2,3,8}$ at once (right), and the fourth row their imaginary parts. In all cases but the grouped plot of $\mathcal{A}_{2,3,8}$, we show three extractions of the amplitudes: using a cut of $t_s = 3$ (blue, see section \ref{sec:isolateMats}) and an $l^2$ minimization (or SVD, see section \ref{sec:SVD}), using a cut of $t_s = 4$ and the $l^2$ minimization (green), or using a cut of 3 and an $l^1$ minimization (orange). In the last plot, we only use a cut of 3 with the $l^2$ minimization for clarity. \label{fig:pres1}}
    
\end{figure}

We present here the full set of Lorentz amplitudes $\mathcal{A}_k$ for $k \in \{1, ..., 8\}$ following the decomposition of Eq. \eqref{eq:decomp} for two pairs of momentum: $\vec{p}_f = (2,1,0)$ and $\vec{p}_i = (0, 1, 2)$ (lattice units) on Fig. \ref{fig:pres1}, and $\vec{p}_f = (1,-1,0)$ and $\vec{p}_i = (-1, 1, 0)$ on Fig. \ref{fig:pres2}. The data is presented as a function of $z$ in lattice units. We only show positive $z$ as we average over $\pm z$ using the symmetries discussed in section \ref{sec:spacetime}. We remind the reader that this means that each data point exists therefore at a different ``scale'' of the order of $-1/z^2$. The amplitudes $\mathcal{A}_{2,3,8}$ which do not enter the construction of light-cone GPDs are represented together in the last plots, whereas the other five get their individual plot. Let us comment on some interesting features of both figures.

\paragraph{Figure \ref{fig:pres1}}

\begin{figure}[ht!]
    \centering
    \includegraphics[width=0.7\linewidth]{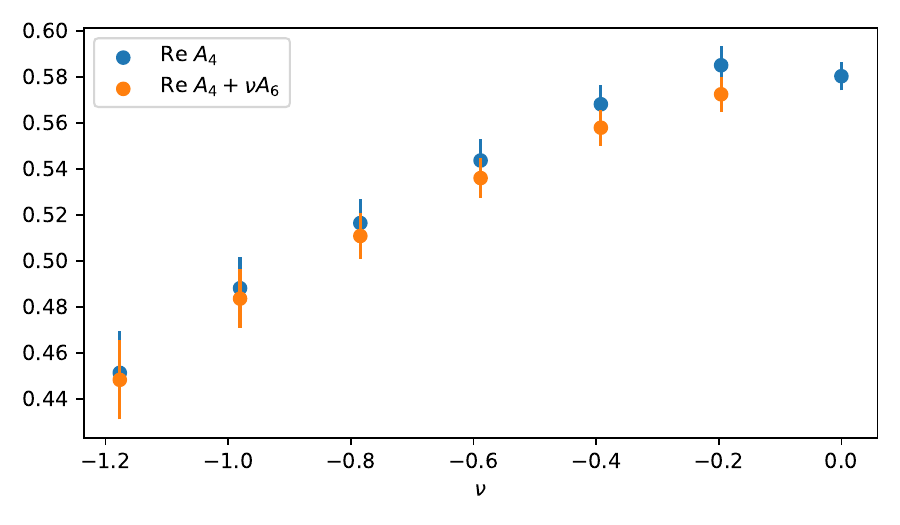}
\caption{Effect of adding $\nu \mathcal{A}_6$ to $\mathcal{A}_4$ for $\vec{p}_f = (2,1,0)$ and $\vec{p}_i = (0, 1, 2)$. \label{fig:discontinuity}}
    
\end{figure}

\begin{figure}[ht!]
    \centering
    \includegraphics[width=0.99\linewidth]{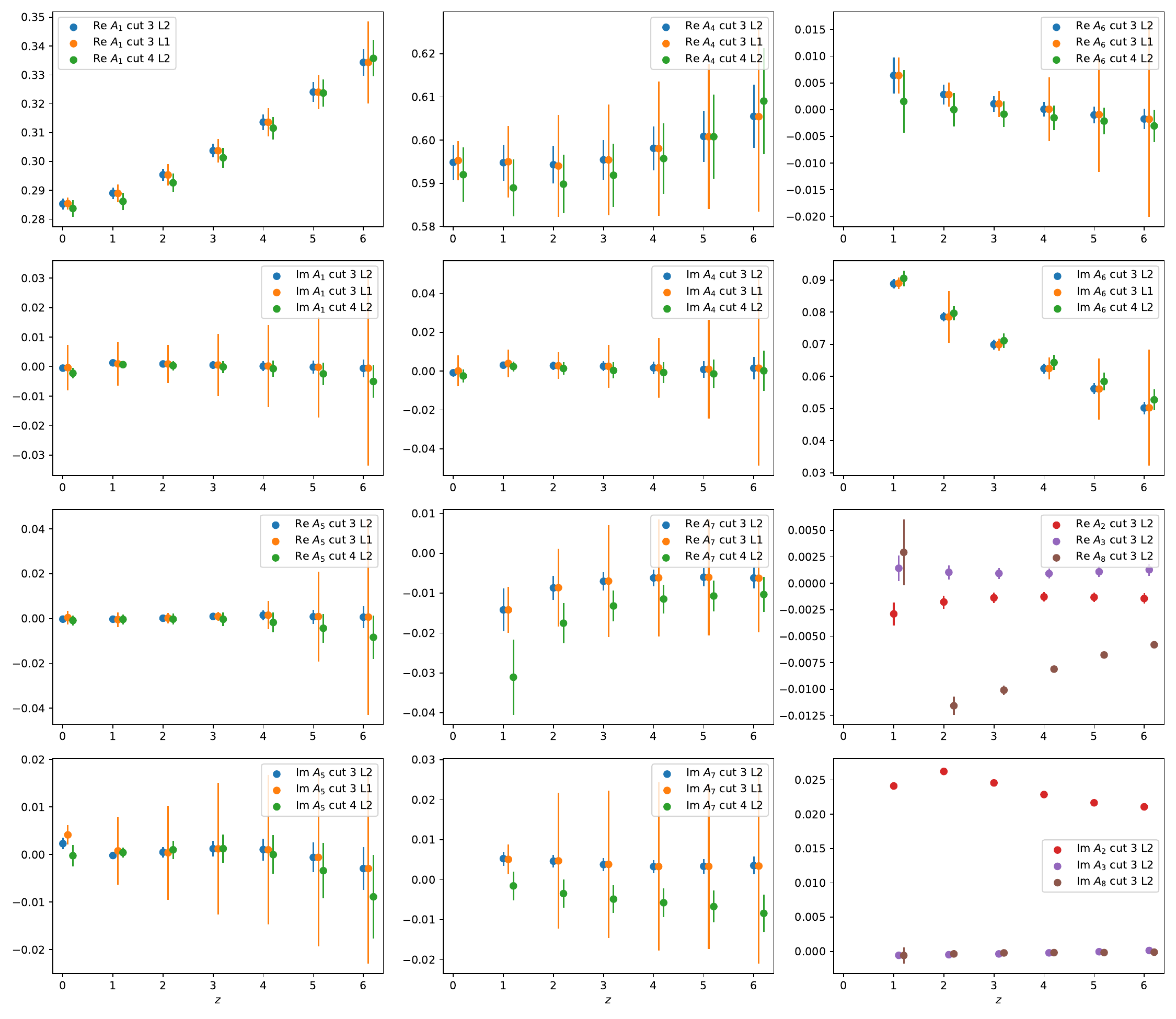}
\caption{Full set of amplitudes for $\vec{p}_f = (1,-1,0)$ and $\vec{p}_i = (-1, 1, 0)$. \label{fig:pres2}}
    
\end{figure}

\begin{itemize}
    \item For the five amplitudes that enter the construction of light-cone GPDs, we compare a solution for the extraction of the amplitudes from the matrix elements using either the SVD ($l^2$ minimization) or an $l^1$ minimization. We find that the $l^1$ minimization produces generally larger uncertainties but with central values in excellent agreement with the $l^2$ extraction. We consider this as a validation of the robustness of the SVD method.
    \item For those five amplitudes, we also show the comparison between the SVD on the data with a cut in lattice time separation of 3 and 4. The larger cut produces more uncertainty, but also in a good consistency with the smaller cut. The results of this paper are jointly extracted using the cut in 3 and 4 to give an account of excited state contribution.
    \item $\mathcal{A}_5$ and $\mathcal{A}_7$ do not show a strong deviation from 0, a situation generally observed in the dataset. $\mathcal{A}_5$ contains signal of the $D$-term, the kinematic higher-twist contamination $Y$ discussed in \cite{Radyushkin:2023ref}, and lattice systematic errors. As for $\mathcal{A}_7$, it enters the construction of the light-cone GPD $E$, but is generally small in magnitude and barely different from 0.
\item The kinematic factors of the amplitudes $\mathcal{A}_{2,3,6,7,8}$ cancel in the limit $z = 0$. Therefore, we cannot evaluate the amplitudes in this limit. $\mathcal{A}_6$ grows noticeably at small $z$. It is in fact necessary to counter-act the discontinuity of $\mathcal{A}_4$ between the value computed at $z = 0$ (the form factor $F_2$), and the value at $z = 1$. Adding $\nu \mathcal{A}_6$ to $\mathcal{A}_4$ to form the light-cone GPD $E$ alleviates the discontinuity, as we demonstrate in Figure \ref{fig:discontinuity}.
\end{itemize}

\paragraph{Figure \ref{fig:pres2}}

The choice of momentum pair $\vec{p}_f = (1,-1,0)$ and $\vec{p}_i = (-1, 1, 0)$ means that $\nu$ and $\nu\xi$ are both 0 for every value of $z$. Therefore, we should have $\mathcal{A}_1 = F_1 + \mathcal{O}(z^2)$ and $\mathcal{A}_4 = F_2 + \mathcal{O}(z^2)$. $\mathcal{A}_1$ presents a distinctive evolution with $z$, which we discuss in section \ref{sec:EFFs}. $\mathcal{A}_4$ on the other hand is compatible with a constant within its uncertainty. The general features identified in the previous figure remain valid.

\section{To $\gamma_3$ or not to $\gamma_3$}\label{app:to_gammaz}
In the lattice calculation of PDFs, the choice of matrix elements with Dirac structure $\gamma^3$ has been considered disfavored. In the Lorentz decomposition, $M^\mu= p^\mu \mathcal{M} + z^\mu \mathcal{N}$, choosing $\mu$ parallel to the Wilson line will introduce a contribution from $\mathcal{N}$ which does not appear in the light cone definition of the PDF~\cite{Radyushkin:2017cyf}. In the work of~\cite{Constantinou:2017sej}, the matrix element of the operator $\bar{\psi}(z) \Gamma W(z;0)\psi(0)$ is calculated at 1-loop lattice perturbation theory. The matrix elements contain terms proportional to $\bar{u} \Gamma u$ and $\bar{u}\{\Gamma, \slashed{z}\}u$, describing a finite mixing with operators which occurs due to the breaking of chiral symmetry. As discussed in this appendix for the case of the vector structures, this pattern can be consolidated with another contribution we typically neglect or discard. For the vector current, the mixing comes from $\bar{u}\{\gamma^\mu, \slashed{z}\}u=-2z^\mu\bar{u}u$ which can only impact the $\mathcal{N}(\nu,z^2)$ or $A_2(\nu,t,z^2)$ for PDFs and GPDs respectively. Since these do not contain the leading order contributions we are interested in, this contamination is benign. Similarly for axial $\Gamma=\gamma^\mu\gamma^5$ structure in the case of the pion DA~\cite{Kovner:2024pwl}, the mixing tensor structure can only generate effects of $O(z^2)$ which will not ultimately effect the asymptotic short distance limit of the relevant amplitude. For the vector case, the inclusion of $\gamma_3$ data has a negligible effect on the amplitudes $A_{1,4,5}$ when $z/a>0$. 

\begin{figure}
    \centering
\includegraphics[width=0.95\textwidth]
{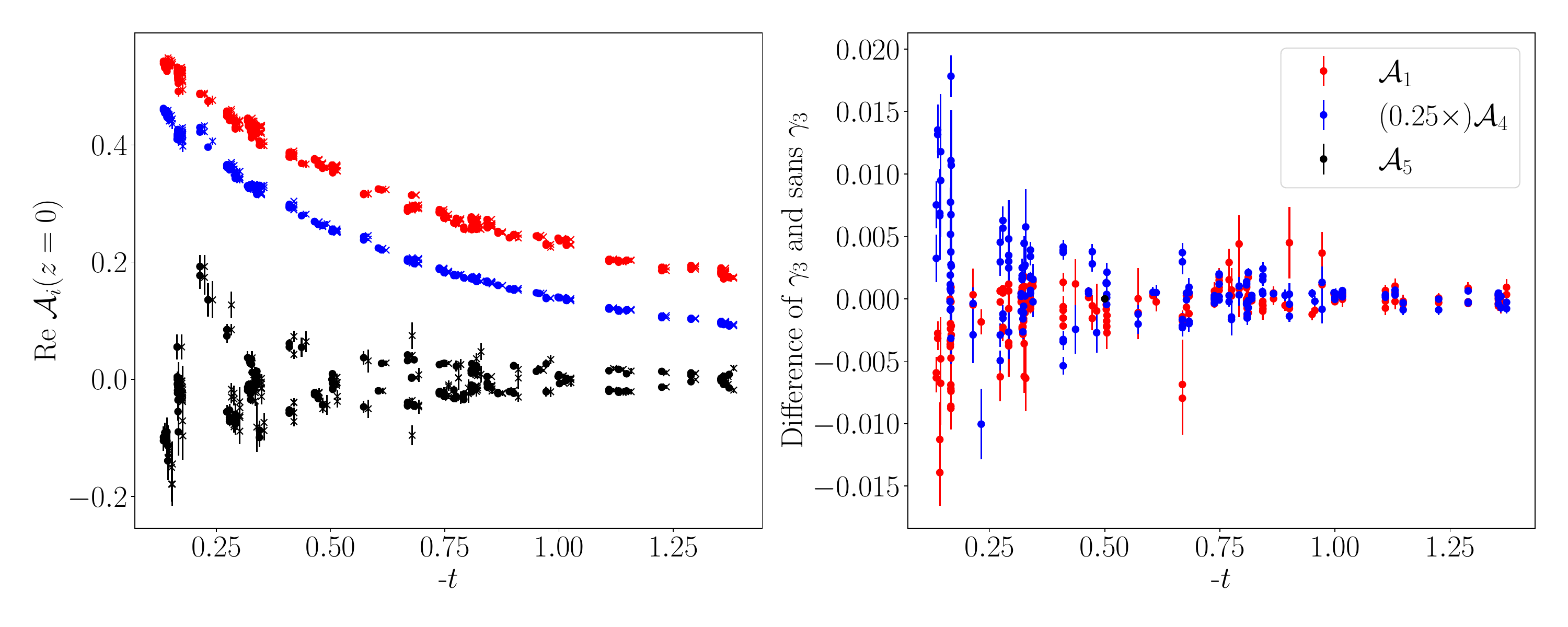}
    \caption{(Left) The amplitudes $\amp{1,4,5}$ at $z=0$ calculated by using data with and without $\gamma_3$. The sans $\gamma_3$ data (circles) are shifted to the right slightly from the with $\gamma_3$ data (crosses). (Right) The difference of the amplitudes calculated with and without $\gamma_3$. The difference of $\amp{5}$ which has significant changes is not shown.}
    \label{fig:togammaz0}
\end{figure}

First, for $z=0$, the first $\amp{1,4}$ amplitudes correspond to the form factors $F_{1,2}$ while $\amp{5}$ should be 0 in the continuum by the Ward Identity $q_\mu M^\mu=0$. The rest of the amplitudes have a factor of $z$ in the kinematic factor and do not belong here. Using the SVD procedure in section~\ref{sec:pGITD-extraction}, $\amp{1,4,5}$ were extracted while including and while neglecting the $M^3$ matrix element, shown in Fig.~\ref{fig:togammaz0}. The amplitudes remain quite similar with a few percent deviations in $A_{1,4}$, relevant at our statistical precision while the quality of $\amp{5}=0$ changes slightly. After binning of the $t$-dependence (see section \ref{sec:EFFs}), the elastic form factors are indiscernible whether $M^z$ is used or not. Remembering the duplicity relating half helicity combinations to the other half, for $z=0$, there were either 8 or 6 independent matrix elements used to obtain 3 unknown functions. Adding in more data without increasing the number of unknowns means there are new constraints being applied when $M^3$ is added.

\begin{figure}
    \centering
\includegraphics[width=0.95\textwidth]
{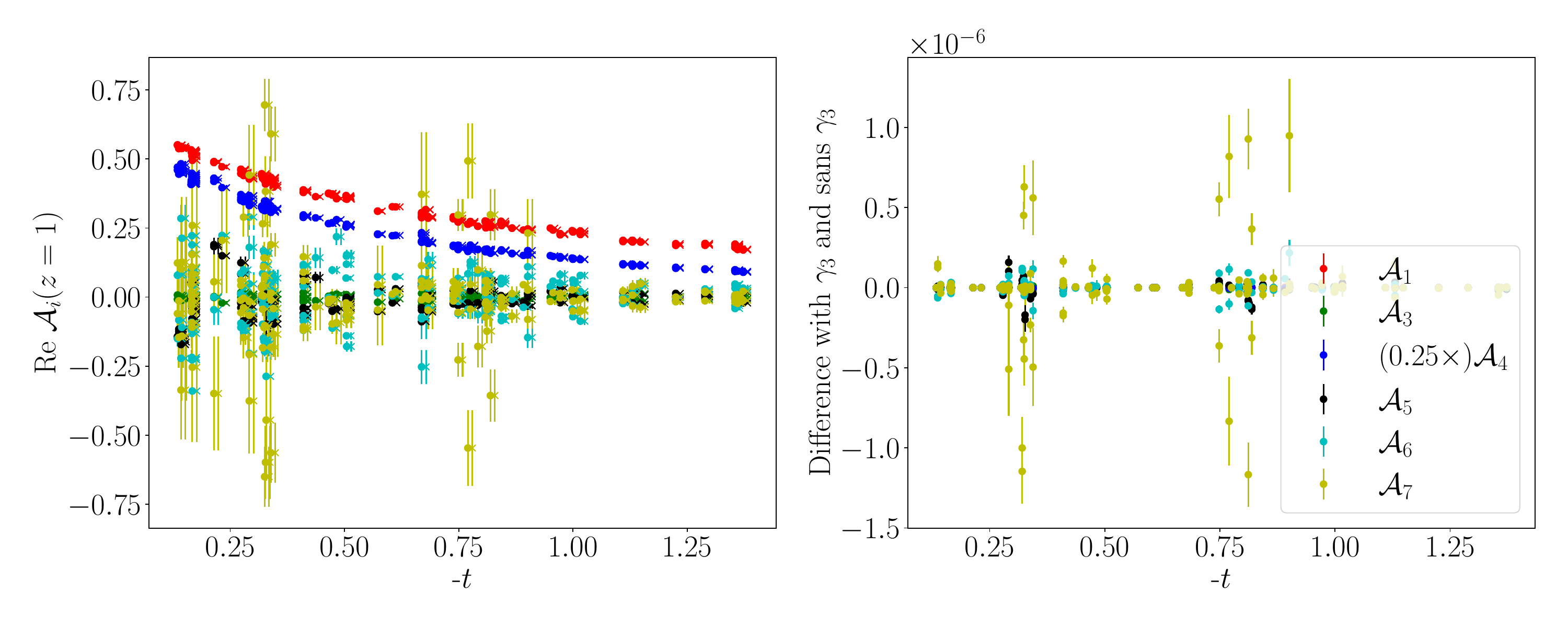}
    \caption{Same as Fig.~\ref{fig:togammaz0} but for amplitudes $\amp{1,3,4,5,6,7}$ at $z=1$.}
    \label{fig:togammaz1}
\end{figure}

While studying the real component of $z=a$ data, which shows the same patterns as larger $z$ and with the imaginary component, if the $M^3$ data is neglected then the $\amp{2,8}$ terms are dropped since they always contribute 0 to $M^{t,x,y}$. The two sets of results are remarkably identical, shown in Fig.~\ref{fig:togammaz1}. In fact, as we have already mentioned in section \ref{sec:SVD}, at $z \neq 0$, the $M^3$ data brings exactly two constraints for two new unknown amplitudes, so it exactly serves to determine $\mathcal{A}_{2,8}$ without any incidence on the other amplitudes. The only deviations on the other amplitudes come from numerical noise, at a level of $10^{-6}$ corresponding to the accuracy at which we computed the kinematic matrix.

\section{Effect of the range of separations $z$ \label{differentz}}

\begin{figure}[ht!]
\centering
\includegraphics[width=0.8\linewidth]{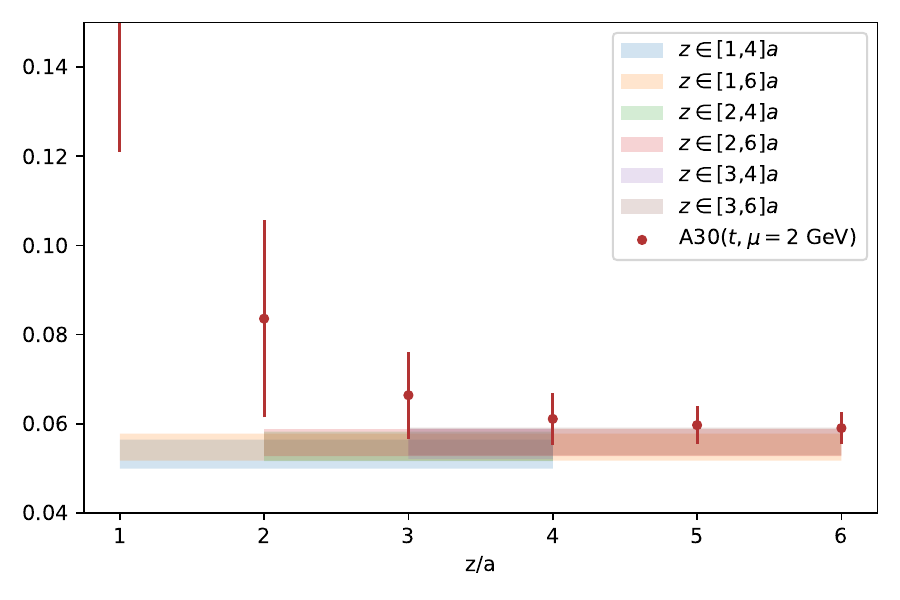}
\caption{The red points are the matched $A_{3,0}(t = -0.33 \textrm{ GeV}^2, \mu = 2$ GeV) depending on the non-local separation $z$ for a cut $t_s = 4$. The bands represent the correlated fit by a constant of a subset of the red points. It appears that the correlated fit result is very stable whereas an uncorrelated fit would show a significant variation. We observed a similar robustness of the correlated constant fits for the non-local extraction of the EFF.  \label{fig:constantmatched}}    
\end{figure}

We compare in Fig. \ref{fig:constantmatched} the extraction of the moment $A_{3,0}$ in one bin in $t$ depending on the range in $z$ one chooses to consider. The standard analysis in the main body of the article is performed with $z \in [a, 6a]$. We observe that varying this range gives statistically similar results, not least because of the very high correlation between the values in $z$. The very inaccurate result at small $z$ stems from the fact that $A_{3,0}$ is a subdominant moment in the real part of the Ioffe-time distribution, which can only be captured accurately when the Ioffe time is large enough.

\begin{figure}[ht!]
\centering
\includegraphics[width=0.45\linewidth]{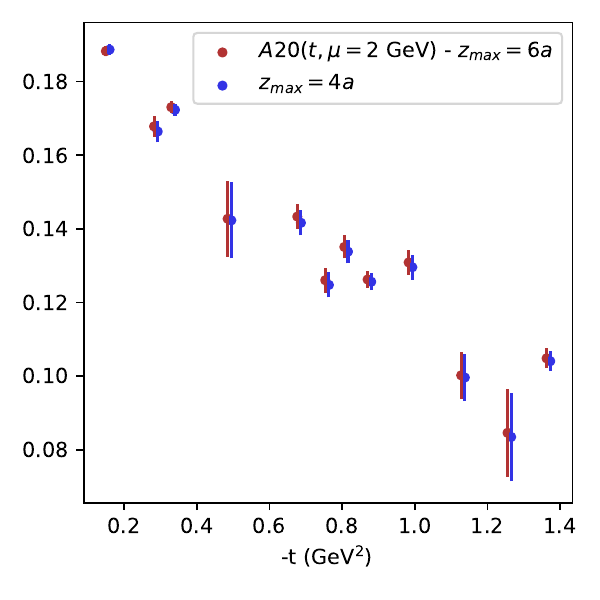} \includegraphics[width=0.45\linewidth]{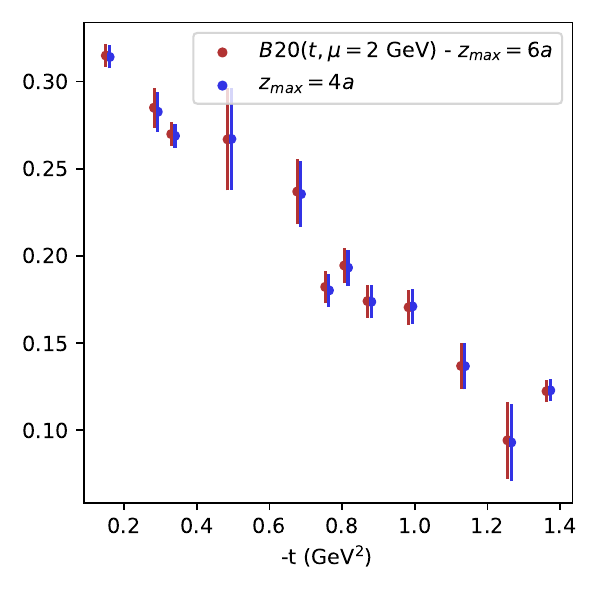}
\caption{$A_{2,0}(t, \mu = 2$ GeV) (left) and $B_{2,0}(t, \mu = 2$ GeV) (right) extracted from the lattice cut of 3 by fitting the matched values for $z \in [a,6a]$ (red) or $z \in [a,4a]$. \label{fig:testHTinit}}    
\end{figure}

\begin{figure}[ht!]
\centering
\includegraphics[width=0.45\linewidth]{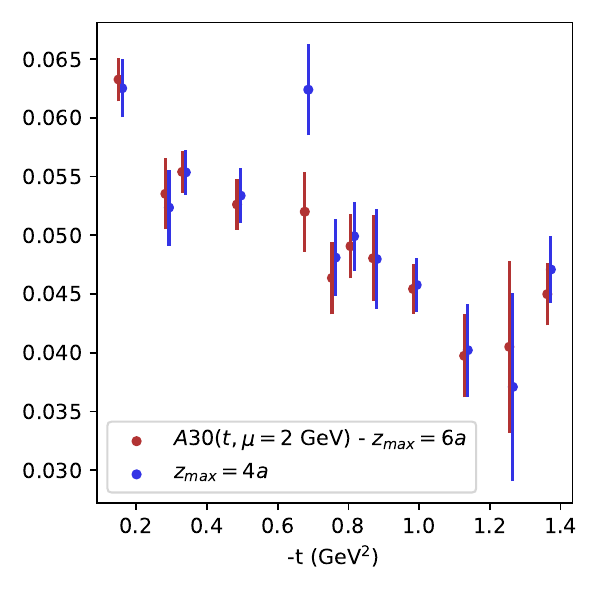} \includegraphics[width=0.45\linewidth]{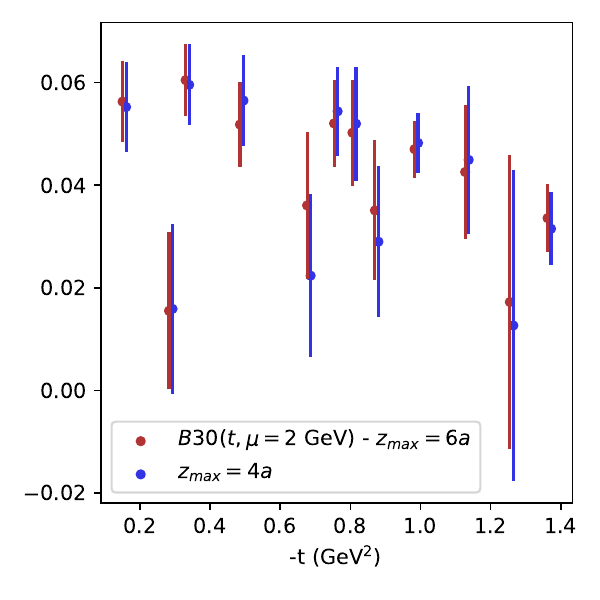}
\caption{$A_{3,0}(t, \mu = 2$ GeV) (left) and $B_{3,0}(t, \mu = 2$ GeV) (right) extracted from the lattice cut of 3 by fitting the matched values for $z \in [a,6a]$ (red) or $z \in [a,4a]$.}    
\end{figure}

In Figs. \ref{fig:testHTinit} to \ref{fig:testHTfinal}, we show systematically the extraction of the moments using $z \in [a, 6a]$ and $z \in [a, 4a]$. It appears that there is no significant statistical difference between using the two methods, and no clear sign of higher twist contributions within the precision of this study.

\begin{figure}[ht!]
\centering
\includegraphics[width=0.45\linewidth]{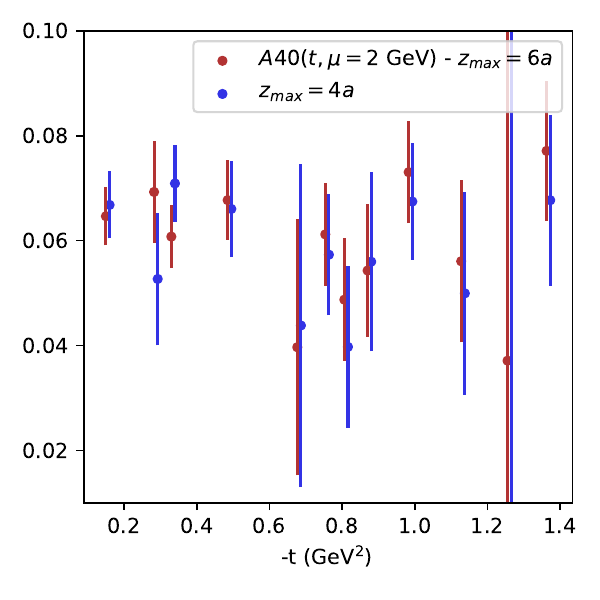} \includegraphics[width=0.45\linewidth]{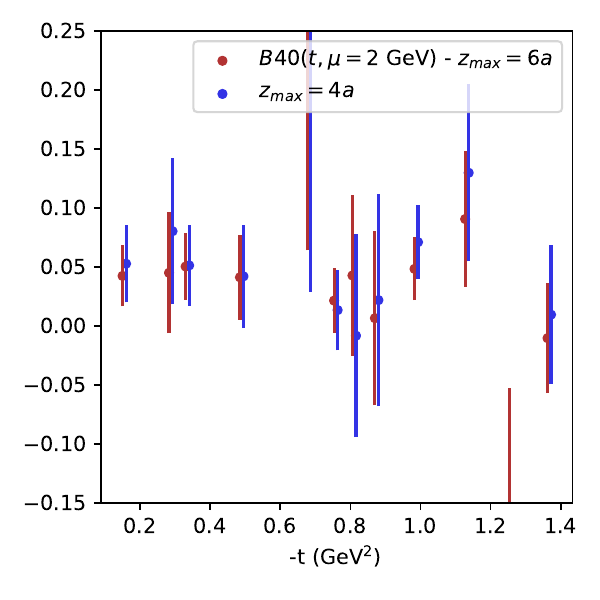}
\caption{$A_{4,0}(t, \mu = 2$ GeV) (left) and $B_{4,0}(t, \mu = 2$ GeV) (right) extracted from the lattice cut of 3 by fitting the matched values for $z \in [a,6a]$ (red) or $z \in [a,4a]$.}    
\end{figure}

\begin{figure}[ht!]
\centering
\includegraphics[width=0.45\linewidth]{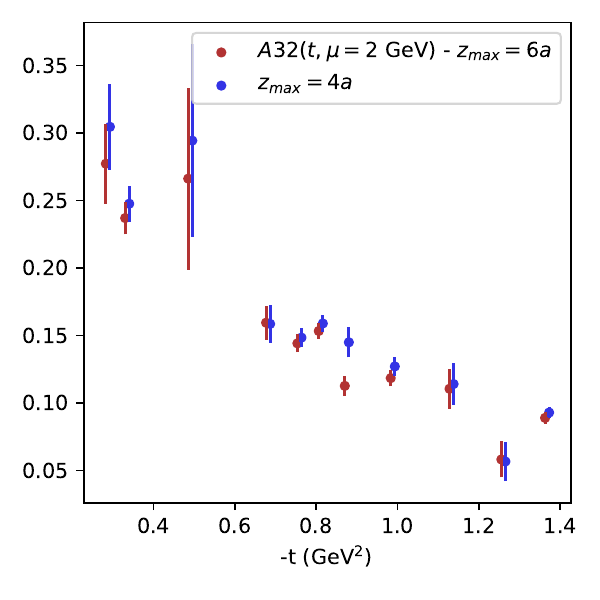} \includegraphics[width=0.45\linewidth]{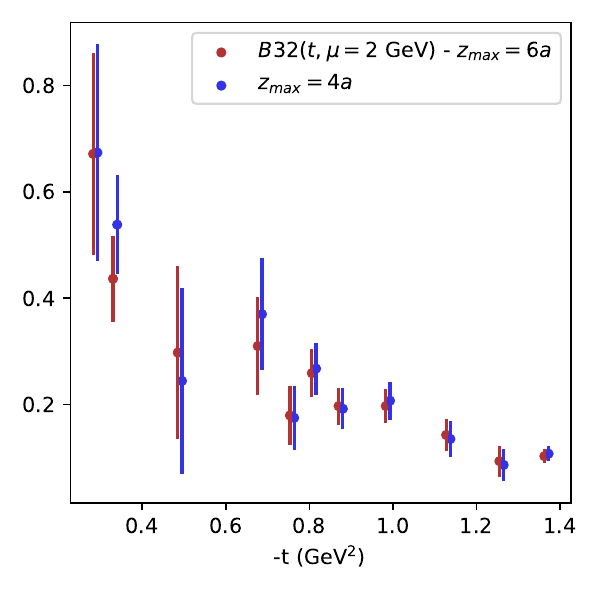}
\caption{$A_{3,2}(t, \mu = 2$ GeV) (left) and $B_{3,2}(t, \mu = 2$ GeV) (right) extracted from the lattice cut of 3 by fitting the matched values for $z \in [a,6a]$ (red) or $z \in [a,4a]$.}    
\end{figure}

\begin{figure}[ht!]
\centering
\includegraphics[width=0.45\linewidth]{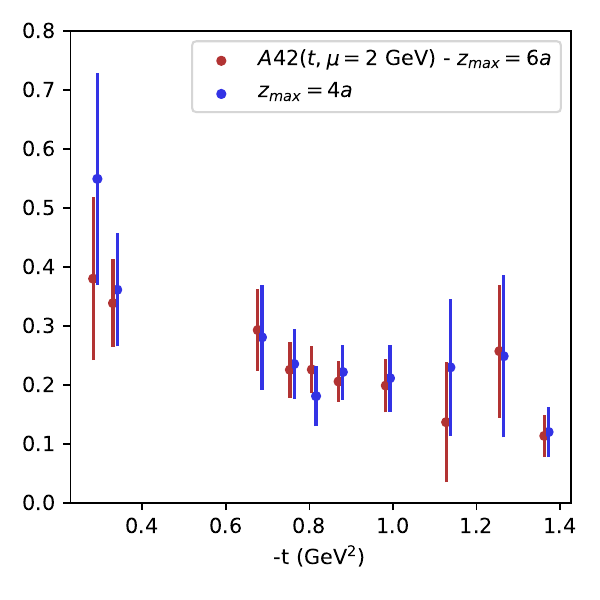} \includegraphics[width=0.45\linewidth]{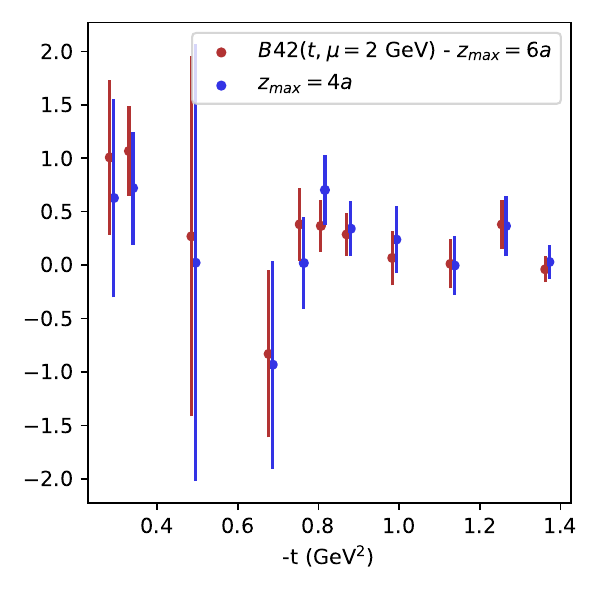}
\caption{$A_{4,2}(t, \mu = 2$ GeV) (left) and $B_{4,2}(t, \mu = 2$ GeV) (right) extracted from the lattice cut of 3 by fitting the matched values for $z \in [a,6a]$ (red) or $z \in [a,4a]$.}    
\end{figure}

\begin{figure}[ht!]
\centering
\includegraphics[width=0.8\linewidth]{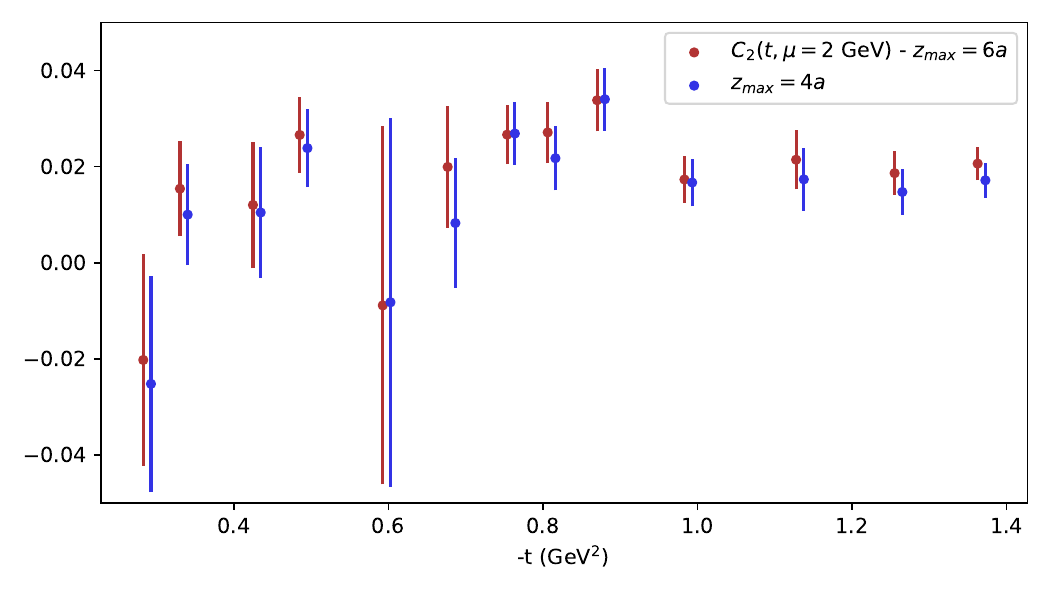}
\caption{$C_{2}(t, \mu = 2$ GeV) extracted from the lattice cut of 3 by fitting the matched values for $z \in [a,6a]$ (red) or $z \in [a,4a]$. \label{fig:testHTfinal}}    
\end{figure}

\FloatBarrier

\bibliography{srcs.bib}

\end{document}